\documentclass[times]{nmeauth}
\pdfoutput=1
 \usepackage{color}
 \usepackage[section]{placeins}
 \usepackage{pict2e}

\usepackage{moreverb}

\usepackage[colorlinks,bookmarksopen,bookmarksnumbered,citecolor=red,urlcolor=red]{hyperref}

\newcommand\BibTeX{{\rmfamily B\kern-.05em \textsc{i\kern-.025em b}\kern-.08em
T\kern-.1667em\lower.7ex\hbox{E}\kern-.125emX}}

\usepackage{stmaryrd}
\usepackage{xcolor,graphicx}

\usepackage{algorithm}
\usepackage{algorithmic}

\usepackage{amssymb,amsmath,amstext, amsfonts, mathrsfs, mathtools}
\usepackage{arydshln}
\newtheorem{remark}{Remark}

\usepackage{cite}

\usepackage{anyfontsize}

\graphicspath{{fig/}}
\DeclareGraphicsExtensions{.pdf}

\begin{document}

\runningheads{C.~Ager et al.}{A consistent and comprehensive approach for FSCI}

\title{A consistent and comprehensive computational approach for general Fluid-Structure-Contact Interaction problems}

\author{Christoph~Ager\corrauth, Alexander~Seitz, Wolfgang~A.~Wall}
\address{Institute for Computational Mechanics ,
Technical University of Munich,\linebreak
Boltzmannstr. 15, 85747 Garching b. M\"u{}nchen
}

\corraddr{Christoph~Ager, Institute for Computational Mechanics, Technical University of Munich, Boltzmannstra{\ss}e 15, D-85747 Garching, Germany. E-mail:~ager@lnm.mw.tum.de}
 \begin{abstract}
 We present a consistent approach that allows to solve challenging general nonlinear 
fluid-structure-contact interaction (FSCI) problems.
The underlying continuous formulation includes both "no-slip" fluid-structure interaction as well as frictionless contact between multiple elastic bodies.
The respective interface conditions in normal and tangential orientation and especially the role of the fluid stress within the region of closed contact are discussed for the general problem of FSCI. 
To ensure continuity of the tangential constraints from no-slip to frictionless contact, a transition is enabled by using the general Navier condition with varying slip length.
Moreover, the fluid stress in the contact zone is obtained by an extension approach as it plays a crucial role for the lift-off behavior of contacting bodies.
With the given continuity of the spatially continuous formulation, continuity of the discrete problem (which is essential for the convergence of Newton's method) is reached naturally.
As topological changes of the fluid domain are an inherent challenge in FSCI configurations, a non-interface fitted Cut Finite Element Method (CutFEM) is applied to discretize the fluid domain.
All interface conditions, that is the ``no-slip'' FSI, the general Navier condition, and frictionless contact are incorporated using Nitsche based methods, thus retaining the continuity and consistency of the model. 
To account for the strong interaction between the fluid and solid discretization, the overall coupled discrete system is solved monolithically.
Numerical examples of varying complexity are presented to corroborate the developments. 
In a first example, the fundamental properties of the presented formulation such as the contacting and lift-off behavior, the mass conservation, and the influence of the slip length for the general Navier interface condition are analyzed. Beyond that, two more general examples demonstrate challenging aspects such as topological changes of the fluid domain, large contacting areas, and underline the general applicability of the presented method.
 \end{abstract}


\maketitle

\vspace{1cm}

\definecolor{darkgreen}{rgb}{0,0.5,0}
\definecolor{darkblue}{rgb}{0,0,0.5}
\definecolor{green}{rgb}{0,1,0}
\definecolor{orange}{rgb}{1.0,0.5,0.0}
\definecolor{darkorange}{rgb}{0.6,0.3,0.0}

\newcommand{\textS}[1]{\textcolor{orange}{#1}}
\newcommand{\textA}[1]{\textcolor{red}{#1}}

\newcommand{\remS}[1]{\textcolor{blue}{#1}}
\newcommand{\remA}[1]{\textcolor{darkgreen}{#1}}

\newcommand{\tns}[1]{\underline{\boldsymbol{#1}}}
\newcommand{\hattns}[1]{\hat{\underline{\boldsymbol{#1}}}}
\newcommand{\disctns}[1]{\underline{\mathfrak{#1}}}
\newcommand{\mat}[1]{\underline{\boldsymbol{#1}}}
\newcommand{\partiald}[2]{\dfrac{ \partial #1}{\partial #2}}
\newcommand{\partialdt}[2]{\dfrac{ \partial^2 #1}{\partial {#2}^2}}
\newcommand{\ddx}[2]{\dfrac{\text{d} {#1}}{\text{d} #2}}
\newcommand{\ddt}[1]{\dfrac{\text{d} {#1}}{\text{d} t}}
\newcommand{\pddt}[1]{\dfrac{\partial {#1}}{\partial t}}
\newcommand{\pddtX}[1]{\left.\dfrac{\partial {#1}}{\partial t}\right|_{\tns{X}}}
\newcommand{\pddtx}[1]{\left.\dfrac{\partial {#1}}{\partial t}\right|_{\tns{x}}}
\newcommand{\derivn}[2]{\partial^{#1}_{\normal}#2}
\newcommand{\derivdir}[2]{\mathcal{D}_{#1}\left[#2\right]}

\newcommand{\innerp}[3]{\left(#1, #2\right)_{#3}}
\newcommand{\innerpb}[3]{\left\langle #1, #2 \right\rangle_{#3}}
\newcommand{\jump}[1]{\ensuremath{\left[\!\left[#1\right]\!\right]} }
\newcommand{\grad}   { \boldsymbol{\nabla}    }
\newcommand{\gradRef}   { \boldsymbol{\nabla}_0    }
\newcommand{\laplace}   { \boldsymbol{\Delta}}
\renewcommand{\div} { \grad \! \cdot \!} 
\newcommand{\divRef} { \gradRef \! \cdot \!} 
\newcommand{\tr}   {\rm{tr}\,}
\newcommand{\ccprod}   { \boldsymbol{:}   }
\newcommand{\zerovec}{\tns{0}}
\newcommand{\dzerovec}{\disctns{0}}
\newcommand{\unity}{\tns{I}}
\newcommand{\abs}[1]{\ensuremath{\left|#1\right|}}
\newcommand{\averg}[1]{\ensuremath{\left\lbrace#1\right\rbrace} }
\newcommand{\avergt}[1]{\ensuremath{<#1> }}
\newcommand{\norminf}[1]{\ensuremath{\left|\left|#1\right|\right|_{\infty}}}
\newcommand{\normL}[2]{\ensuremath{\left|\left|#1\ensuremath\right|\right|_{L^2\left(#2\right)}}}
\newcommand{\norm}[2]{\ensuremath{\left|\left|#1\ensuremath\right|\right|_{#2}}}
\newcommand{\extensionsymbol}{\ensuremath{\mathcal{E}}}
\newcommand{\extension}[2]{\ensuremath{\extensionsymbol_{#2}\left[#1\right]}}
\newcommand{\extensionsym}[1]{\ensuremath{\extensionsymbol_{#1}}}

\newcommand{\testfkt}{\delta \tns{\velocityf}}
\newcommand{\stress}{\tns{\sigma}}
\newcommand{\normal}{\tns{n}}
\newcommand{\tangent}{\tns{t}}
\newcommand{\geninterface}{\interface^*}
\newcommand{\residual}{\underline{\mathcal{R}}}
\newcommand{\btime}{t_0}
\newcommand{\etime}{t_\mathrm{E}}
\newcommand{\timesfulltime}{\times [\btime, \etime]}
\newcommand{\Pnormal}{\tns P_n}
\newcommand{\Ptangent}{\tns P_t}
\newcommand{\WeakB}[1][]{B_{#1}}
\newcommand{\WeakC}[1][]{C_{#1}}
\newcommand{\WeakD}[1][]{D_{#1}}
\newcommand{\Weakc}[1][]{c_{#1}}
\newcommand{\Weakd}[1][]{d_{#1}}
\newcommand{\Weakf}[1][]{f_{#1}}
\newcommand{\strainenergy}[1]{\psi^{#1}}

\newcommand{\domain}{\Omega}
\newcommand{\fullbound}{\partial \domain}
\newcommand{\refdomain}{\Omega_0}
\newcommand{\fullrefbound}{\partial \refdomain}

\newcommand{\fluidletter}{{\mathrm{}}}
\newcommand{\fluidletterr}{{\mathrm{F}}}
\newcommand{\domainf}{\Omega^\fluidletterr}
\newcommand{\velocityf}{v}
\newcommand{\displacementf}{u}
\newcommand{\pressuref}{p}
\newcommand{\densityf}{\rho^\fluidletterr}
\newcommand{\velf}[1][]{\tns{\velocityf}^{\fluidletter #1}}
\newcommand{\velfn}{\velf\tns{\velocityf}^{\fluidletter,n}}
\newcommand{\velfnp}{\tns{\velocityf}^{\fluidletter,n+1}}
\newcommand{\velgridf}{\tns{\velocityf}^G}
\newcommand{\velfD}{\tns{\hat{\velocityf}}^\fluidletter}
\newcommand{\velfB}{\tns{\mathring{\velocityf}}^\fluidletter}
\newcommand{\pf}{\pressuref^\fluidletter}
\newcommand{\pfn}{\pressuref^{\fluidletter,n}}
\newcommand{\pfnp}{\pressuref^{\fluidletter,n+1}}
\newcommand{\viscf}{\mu^\fluidletter}
\newcommand{\epsf}{\tns{\epsilon}^\fluidletter}
\newcommand{\stressf}[1][]{\tns{\sigma}^\fluidletterr_{#1}}
\newcommand{\stressfn}{\tns{\sigma}^{\fluidletterr,n}}
\newcommand{\stressfnp}{\tns{\sigma}^{\fluidletterr,n+1}}
\newcommand{\bodyff}{\hat{\tns b}^\fluidletterr}
\newcommand{\bodyffn}{\hat{\tns b}^{\fluidletterr,n}}
\newcommand{\bodyffnp}{\hat{\tns b}^{\fluidletterr,n+1}}
\newcommand{\normalf}{\tns n^\fluidletterr}
\newcommand{\tangf}{\tns t^\fluidletterr}
\newcommand{\tractionf}{\tns{h}^{\fluidletterr}}
\newcommand{\tractionfN}{\tns{\hat{h}}^{\fluidletterr,\mathrm{N}}}
\newcommand{\tractionfNn}{\tns{\hat{h}}^{\fluidletterr,\mathrm{N},n}}
\newcommand{\tractionfNnp}{\tns{\hat{h}}^{\fluidletterr,\mathrm{N},n+1}}
\newcommand{\timef}{t}
\newcommand{\testvelf}[1][]{\delta \tns{\velocityf}^{\fluidletter #1}}
\newcommand{\testpf}{\delta \pressuref^\fluidletter}
\newcommand{\thetaf}{\theta}
\newcommand{\timestepf}{\Delta t}
\newcommand{\nboundf}{\interface^{\fluidletterr,\mathrm{N}}}
\newcommand{\dboundf}{\interface^{\fluidletterr,\mathrm{D}}}
\newcommand{\fullboundf}{\partial \domainf}
\newcommand{\restboundf}{\interface^{\fluidletterr,\mathrm{I}}}
\newcommand{\restboundfh}{\restboundf_{\mathrm{h}}}
\newcommand{\incvelf}{\Delta \disctns{\velocityf}^\fluidletterr}
\newcommand{\incdispf}{\Delta \disctns{\displacementf}^A}
\newcommand{\dispf}{\tns{\displacementf}^\fluidletterr}
\newcommand{\velfg}{\partiald{\tns{\displacementf}^\fluidletterr}{\timep}}
\newcommand{\testfktspacef}{V^F}

\newcommand{\aleletter}{{\mathrm{G}}}
\newcommand{\dispg}{\tns{\displacements}^{\aleletter}}
\newcommand{\velg}{ \partiald{\dispg}{\timep}}
\newcommand{\velgh}{\partialddisni{\dispg_{\mathrm{h}}}}

\newcommand{\poroletter}{{\mathrm{P}}}
\newcommand{\poroletterr}{{\mathrm{P}}}
\newcommand{\porofluidletter}{{\poroletter^\fluidletter}}
\newcommand{\porofluidletterr}{{\poroletterr^\fluidletterr}}
\newcommand{\porosolidletter}{\mathrm{P^S}}
\newcommand{\porosolidletterr}{\mathrm{P^S}}
\newcommand{\timep}{t}
\newcommand{\domainp}{\Omega^\poroletterr}
\newcommand{\refdomainp}{\Omega^\poroletterr_0}
\newcommand{\displacementp}{u}
\newcommand{\velocityp}{v}
\newcommand{\pressurep}{p}
\newcommand{\porosity}{\phi}
\newcommand{\porosityB}{\mathring{\porosity}}
\newcommand{\velp}{\tns{\velocityp}^{\poroletterr}}
\newcommand{\dispp}{\tns{\displacementp}^{\poroletterr}}
\newcommand{\disppB}{\tns{\mathring{\displacementp}}^\poroletterr}
\newcommand{\velpsB}{\tns{\mathring{\velocityp}}^{\porosolidletterr}}
\newcommand{\velpB}{\tns{\mathring{\velocityp}}^{\poroletterr}}
\newcommand{\velps}{ \partiald{\dispp}{\timep}}
\newcommand{\velpss}{ \dot{\dispp}}
\newcommand{\accps}{ \partialdt{\dispp}{\timep}}
\newcommand{\pp}{\pressurep^{\poroletterr}}
\newcommand{\densitypf}{\densityf}
\newcommand{\refdensitypf}{\densityf}
\newcommand{\refdensityps}{\rho^{\porosolidletterr}_0}
\newcommand{\densityps}{\rho^{\porosolidletterr}}
\newcommand{\refmassps}{m^{\porosolidletterr}_0}
\newcommand{\refavdensityps}{\tilde{\rho}_0^{\porosolidletterr}}
\newcommand{\avdensityps}{\tilde{\rho}^{\porosolidletterr}}
\newcommand{\stresspkp}{\tns{S}^\poroletterr}
\newcommand{\stressfpkp}{\tns{P}^\poroletterr}
\newcommand{\stressp}{\tns{\sigma}^\poroletterr}
\newcommand{\bodyfpf}{\hat{\tns b}^{\porofluidletterr}}
\newcommand{\bodyfp}{\hat{\tns b}^{\poroletterr}}
\newcommand{\refbodyfp}{\bodyfp_0}
\newcommand{\tractionpf}{\tns{h}^{\porofluidletterr}}
\newcommand{\tractionps}{\tns{h}^{\porosolidletterr}}
\newcommand{\reftractionps}{\tractionps_0}
\newcommand{\tractionp}{\tns{h}^{\poroletterr}}
\newcommand{\reftractionp}{\tractionp_0}
\newcommand{\tractionpfN}{\hat{h}^{\porofluidletterr,\mathrm{N}}}
\newcommand{\tractionpN}{\tns{\hat{h}}^{\poroletterr,\mathrm{N}}}
\newcommand{\reftractionpN}{\tractionpN_0}
\newcommand{\viscp}{\viscf}
\newcommand{\permeabpscalar}{k}
\newcommand{\permeabp}{\tns{\permeabpscalar}}
\newcommand{\matpermeabp}{\tns{\matpermeabpscalar}}
\newcommand{\matpermeabpscalar}{K}
\newcommand{\initmatpermeabpscalar}{\mathring{\matpermeabpscalar}}
\newcommand{\initmatpermeabp}{\mathring{\matpermeabp}}
\newcommand{\Jp}{J^\poroletterr}
\newcommand{\strainenergyp}{\strainenergy{\poroletterr}}
\newcommand{\strainenergyps}{\psi^{\porosolidletterr}}
\newcommand{\strainglp}{\tns{E}^\poroletterr}
\newcommand{\straincgp}{\tns{C}^\poroletterr}
\newcommand{\defgradp}{\tns{F}^\poroletterr}
\newcommand{\testvelp}{\delta \tns{\velocityp}^\poroletterr}
\newcommand{\testdispp}{\delta \tns{\displacementp}^\poroletterr}
\newcommand{\testvelpss}{\delta  \dot{\dispp}}
\newcommand{\testpp}{\delta \pressurep^\poroletterr}
\newcommand{\dboundpf}{\interface^{\porofluidletterr,\mathrm{D}}}
\newcommand{\nboundpf}{\interface^{\porofluidletterr,\mathrm{N}}}
\newcommand{\dboundp}{\interface^{\poroletterr,\mathrm{D}}}
\newcommand{\nboundp}{\interface^{\poroletterr,\mathrm{N}}}
\newcommand{\refdboundp}{\dboundp_0}
\newcommand{\refnboundp}{\nboundp_0}
\newcommand{\restboundp}{\interface^{\poroletterr,\mathrm{I}}}
\newcommand{\refrestboundp}{\restboundp_0}
\newcommand{\restboundpf}{\interface^{\porofluidletterr,\mathrm{I}}}
\newcommand{\fullboundp}{\partial \domainp}
\newcommand{\reffullboundp}{\interface^\poroletterr_0}
\newcommand{\normalp}{\tns n^\poroletterr}
\newcommand{\tangp}{\tns t^\poroletterr}
\newcommand{\snormalp}{\tilde{\tns n}}
\newcommand{\stangp}{\tilde{\tns t}}
\newcommand{\refnormalp}{\normalp_0}
\newcommand{\velpnD}{\hat{\velocityp}^\poroletterr_n}
\newcommand{\disppD}{\tns{\hat{\displacementp}}^\poroletterr}
\newcommand{\coordsymbol}{x}
\newcommand{\coord}{\tns \coordsymbol}
\newcommand{\refcoord}{\tns X}
\newcommand{\coordp}{\coord^\poroletterr}
\newcommand{\refcoordp}{\refcoord^\poroletterr}
\newcommand{\incvelp}{\Delta \disctns{\velocityp}^\poroletterr}
\newcommand{\incdispp}{\Delta \disctns{\displacementp}^\poroletterr}

\newcommand{\structletter}{{\mathrm{}}}
\newcommand{\structletterr}{{\mathrm{S}}}
\newcommand{\displacements}{u}
\newcommand{\velocitys}{v}
\newcommand{\domains}[1][]{\Omega^{\structletterr #1}}
\newcommand{\refdomains}{\domains_0}
\newcommand{\refdomainsh}{\domains_{0,\mathrm{h}}}
\newcommand{\refdomainph}{\domainp_{0,\mathrm{h}}}
\newcommand{\disps}{\tns{\displacements}^{\structletter}}
\newcommand{\vels}{ \partiald{\disps}{\timep}}
\newcommand{\accs}{ \partialdt{\disps}{\timep}}
\newcommand{\defgrads}{\tns{F}^{\structletter}}
\newcommand{\Js}{J^{\structletter}}
\newcommand{\stresspks}{\tns{S}^\structletterr}
\newcommand{\stressfpks}{\tns{P}^\structletterr}
\newcommand{\stresss}{\tns{\sigma}^\structletterr}
\newcommand{\strainenergyNHs}{\strainenergys_{NH}}
\newcommand{\strainenergys}{\psi^{\structletter}}
\newcommand{\straingls}{\tns{E}^{\structletter}}
\newcommand{\strainrcg}{\tns{C}^{\structletter}}
\newcommand{\densitys}{\rho^{\structletterr}}
\newcommand{\refdensitys}{\rho^{\structletterr}_0}
\newcommand{\refbodyfs}{\hat{\tns b}^{\structletterr}_0}
\newcommand{\refbodyfsnt}{\hat{\tns b}^{\structletterr}_{0,n+\theta}}
\newcommand{\dispsD}{\tns{\hat{\displacements}}^\structletter}
\newcommand{\dispsB}{\tns{\mathring{\displacements}}^\structletter}
\newcommand{\velsB}{\tns{\mathring{\velocitys}}^\structletterr}
\newcommand{\tractions}{\tns{h}^{\structletterr}}
\newcommand{\reftractions}{\tractions_0}
\newcommand{\tractionsN}{\tns{\hat{h}}^{\structletterr,\mathrm{N}}}
\newcommand{\reftractionsN}{\tractionsN_0}
\newcommand{\reftractionsNnt}{\tractionsN_{0,n+\theta}}
\newcommand{\nbounds}{\interface^{\structletterr,\mathrm{N}}}
\newcommand{\dbounds}{\interface^{\structletterr,\mathrm{D}}}
\newcommand{\refnbounds}{\nbounds_0}
\newcommand{\refnboundsh}{\nbounds_{0,\mathrm{h}}}
\newcommand{\refnboundph}{\nboundp_{0,\mathrm{h}}}
\newcommand{\refdbounds}{\dbounds_0}
\newcommand{\refdboundsh}{\dbounds_{0,\mathrm{h}}}
\newcommand{\refdboundph}{\dboundp_{0,\mathrm{h}}}
\newcommand{\restbounds}{\interface^{\structletterr,\mathrm{I}}}
\newcommand{\refrestbounds}{\restbounds_0}
\newcommand{\refrestboundsh}{\restbounds_{0,\mathrm{h}}}
\newcommand{\restboundsh}{\restbounds_{\mathrm{h}}}
\newcommand{\refrestboundph}{\restboundp_{0,\mathrm{h}}}
\newcommand{\fullbounds}{\partial \domains}
\newcommand{\fullrefbounds}{\partial \refdomains}
\newcommand{\refnormals}{\normals_0}
\newcommand{\testdisps}{\delta \tns{\displacements}^\structletter}
\newcommand{\coords}{\tns \coordsymbol^\structletterr}
\newcommand{\refcoords}{\tns X^\structletterr}
\newcommand{\incdisps}{\Delta \disctns{\displacements}^\structletter}
\newcommand{\normals}{\tns n^\structletterr}

\newcommand{\timeso}{t}
\newcommand{\slaveletter}{{1\text{}}}
\newcommand{\masterletter}{{2\text{}}}
\newcommand{\coordsgone}{\tns x^\structletterr_{\interface_\slaveletter}}
\newcommand{\coordsgtwo}{\tns x^\structletterr_{\interface_\masterletter}}
\newcommand{\normalgone}{\normal_{\interface_\slaveletter}}
\newcommand{\normalgtwo}{\normal_{\interface_\masterletter}}
\newcommand{\stresssone}{\tns{\sigma}^{\structletterr_\slaveletter}}
\newcommand{\stressstwo}{\tns{\sigma}^{\structletterr_\masterletter}}
\newcommand{\domainsone}{\Omega^{\structletterr_\slaveletter}}
\newcommand{\domainstwo}{\Omega^{\structletterr_\masterletter}}
\newcommand{\fsiinterfaceone}{\interface^{\fluidletterr\structletterr_\slaveletter}}
\newcommand{\fsiinterfacetwo}{\interface^{\fluidletterr\structletterr_\masterletter}}
\newcommand{\sciinterface}{\interface^{\structletterr,c}}
\newcommand{\refsciinterface}{\sciinterface_0}
\newcommand{\sciinterfaceone}{\interface^{\structletterr_\slaveletter,c}}
\newcommand{\sciinterfacetwo}{\interface^{\structletterr_\masterletter,c}}
\newcommand{\interfaceone}{\interface_\slaveletter}
\newcommand{\interfacetwo}{\interface_\masterletter}
\newcommand{\Pnormalgone}{\left(\normalgone \otimes \normalgone\right)}
\newcommand{\Pnormalgtwo}{\left(\normalgtwo \otimes \normalgtwo\right)}
\newcommand{\gap}{\left( \project{\coord}{\normal} - \coord \right) \cdot \normal}
\newcommand{\pensolid}{\gamma^{\structletterr}}
\newcommand{\stresssnn}{\sigma^{\structletterr}_{nn}}
\newcommand{\stresssnnone}{\sigma^{\structletterr_\slaveletter}_{nn}}
\newcommand{\stresssnntwo}{\sigma^{\structletterr_\masterletter}_{nn}}

\newcommand{\structWeight}{\omega}
\newcommand{\InterfaceStressn}{\overline{\tns{\sigma}}_{n}}
\newcommand{\normalInterfaceStress}{\overline{\sigma}_{nn}}
\newcommand{\normalAVstructStress}{\overline{\sigma}_{nn}^{\structletterr}}
\newcommand{\Gap}{g_n}
\newcommand{\proj}[1]{\check{#1}}

\newcommand{\penfluid}{\gamma^{\fluidletterr}}
\newcommand{\penscalingfluid}{\phi^F_{v} h_{\Gamma}^{-1}}
\newcommand{\penscalingfluidn}{\phi^F_{n} h_{\Gamma}^{-1}}
\newcommand{\penscalingfluidt}{\penscalingfluidtnh h_{\Gamma}^{-1}}
\newcommand{\penscalingfluidtnh}{\phi^F_{t}}
\newcommand{\penscalingsolid}{\phi^S}
\newcommand{\stressfnn}{\sigma^{\fluidletterr}_{nn}}
\newcommand{\stressfnnext}{\sigma^{\fluidletterr}_{nn,\mathcal{E}}}
\newcommand{\velreln}{\velocityf^{rel}_{n}}
\newcommand{\velrelnext}{\velocityf^{rel}_{n,\mathcal{E}}}
\newcommand{\penfluidext}{\gamma^{\fluidletterr}_\mathcal{E}}
\newcommand{\testvelfempty}{\testvelf_{\emptyset}}
\newcommand{\testpfempty}{\testpf_{\emptyset}}

\newcommand{\BJfac}{\beta_{BJ}}
\newcommand{\scaleFt}[1][]{{\color{darkgreen}c^{F #1}_t}}
\newcommand{\scalePt}[1][]{{\color{darkgreen}c^{P #1}_t}}
\newcommand{\scaleSt}[1][]{{\color{darkgreen}c^{S #1}_t}}
\newcommand{\scaleFn}[1][]{{\color{darkgreen}c^{F #1}_n}}
\newcommand{\scalePn}[1][]{{\color{darkgreen}c^{P #1}_n}}
\newcommand{\scaleSn}[1][]{{\color{darkgreen}c^{S #1}_n}}
\newcommand{\interface}{\Gamma}
\newcommand{\refinterface}{\interface_0}
\newcommand{\fsiinterface}{\interface^{FS}}
\newcommand{\fsiinterfaceh}{\interface^{FS}_\mathrm{h}}
\newcommand{\reffsiinterface}{\fsiinterface_0}
\newcommand{\fpiinterface}{\interface^{FP}}
\newcommand{\reffpiinterface}{\fpiinterface_0}
\newcommand{\psiinterface}{\interface^{PS}}
\newcommand{\psciinterface}{\interface^{PS,c}}
\newcommand{\pscpiinterfaceh}{\interface^{P,c}_\mathrm{h}}
\newcommand{\pscsiinterfaceh}{\interface^{S,c}_\mathrm{h}}
\newcommand{\pscpiinterface}{\interface^{P,c}}
\newcommand{\pscsiinterface}{\interface^{S,c}}
\newcommand{\refpsiinterface}{\psiinterface_0}
\newcommand{\refpsciinterface}{\psciinterface_0}
\newcommand{\sliplengh}{\kappa}
\newcommand{\lagmultpss}{\tns \lambda}
\newcommand{\lagmultpssnormal}{\lambda_n}
\newcommand{\lagmultpsp}{\tns \lambda_2}
\newcommand{\testnlagmultpss}{\delta \lambda_{n}}
\newcommand{\testtlagmultpss}{\delta \lambda_{t_i}}
\newcommand{\testlagmultpss}{\delta \tns{\lambda}}
\newcommand{\testnlagmultpsp}{\delta \lambda_{2,n}}
\newcommand{\testtlagmultpsp}{\delta \lambda_{2,t_i}}
\newcommand{\dlagmultpss}{\tns{\lambda}}
\newcommand{\dlagmultpsp}{\disctns{\lambda}_2}
\newcommand{\testdualpss}{\Phi}
\newcommand{\dmatrix}{\mat{D}}
\newcommand{\mmatrix}{\mat{M}}
\newcommand{\pmat}{\mat{P}}
\newcommand{\timefac}{\theta}
\newcommand{\prescontact}{p^{PS,c}}

\newcommand{\adjointsign}{\zeta}

\newcommand{\identity}{\mat{I}}
\newcommand{\itraction}{\tns h^\mathrm{I}}
\newcommand{\itractionf}{\tns h^{\fluidletter,\mathrm{I}}}
\newcommand{\itractionfh}{\itractionf_{\mathrm{h}}}
\newcommand{\itractions}{\tns h^{\structletterr,\mathrm{I}}}
\newcommand{\refitractions}{\itractions_0}
\newcommand{\itractionsh}{\itractions_{\mathrm{h}}}
\newcommand{\refitractionsh}{\itractions_{0,\mathrm{h}}}
\newcommand{\refitractionshnt}{\itractions_{0,\mathrm{h},n+\theta}}
\newcommand{\itractionpf}{h^{\porofluidletterr,\mathrm{I}}}
\newcommand{\itractionp}{\tns h^{P,I}}
\newcommand{\refitractionp}{\tns h^{P,I}_0}

\newcommand{\dummycite}[1]{{\color{red}[:-)#1] }\color{black}}
\newcommand{\todo}[1]{{\color{blue}(TODO: #1)}\color{black}}

\newcommand{\mycos}[1]{\text{cos}\left(#1\right)}
\newcommand{\mysin}[1]{\text{sin}\left(#1\right)}

\newcommand{\bjcoeff}{\alpha_{BJ}}

\newcommand{\jumpvaln}{\tns{\hat{g}}_n}
\newcommand{\jumpvalt}{\tns{\hat{g}}_t}
\newcommand{\jumpvals}{\tns{\hat{g}}_{\sigma}}
\newcommand{\jumpvalsn}{{\hat{g}}_{\sigma^n}}

\newcommand{\pfa}{\pf_{\mathcal{A}}}
\newcommand{\velfa}{\velf_{\mathcal{A}}}
\newcommand{\ppa}{\pp_{\mathcal{A}}}
\newcommand{\velpa}{\velp_{\mathcal{A}}}
\newcommand{\disppa}{\dispp_{\mathcal{A}}}
\newcommand{\velpsa}{ \partiald{\disppa}{\timep}}
\newcommand{\accpsa}{ \partialdt{\disppa}{\timep}}
\newcommand{\permeabpa}{ \permeabp_{\mathcal{A}}}
\newcommand{\stresspkpa}{ \stresspkp_{\mathcal{A}}}
\newcommand{\stresspa}{ \stressp_{\mathcal{A}}}
\newcommand{\stressfa}{ \stressf_{\mathcal{A}}}
\newcommand{\Jpa}{ \Jp_{\mathcal{A}}}
\newcommand{\defgradpa}{ \defgradp_{\mathcal{A}}}

\newcommand{\bimapi}{\Phi}
\newcommand{\bimapii}{\Psi}
\newcommand{\bimapiii}{\Xi}
\newcommand{\referencecoord}{\tns \chi}
\newcommand{\dimensions}{d}
\newcommand{\solutionspace}{\mathcal{S}}
\newcommand{\testspace}{\mathcal{T}}
\newcommand{\discretespace}{\mathcal{N}}
\newcommand{\solds}{\solutionspace_{\disps}}
\newcommand{\soldp}{\solutionspace_{\dispp}}
\newcommand{\soldsh}{\solutionspace_{\disps,\mathrm{h},n+1}}
\newcommand{\soldph}{\solutionspace_{\dispp,\mathrm{h},n+1}}
\newcommand{\stestds}{\testspace_{\testdisps}}
\newcommand{\stestdp}{\testspace_{\testdispp}}
\newcommand{\stestdsh}{\testspace_{\testdisps,\mathrm{h}}}
\newcommand{\stestdph}{\testspace_{\testdispp,\mathrm{h}}}
\newcommand{\solvf}{\solutionspace_{\velf}}
\newcommand{\solvp}{\solutionspace_{\velp}}
\newcommand{\solvfhn}{\solutionspace_{\velf,\mathrm{h},n+1}}
\newcommand{\solvphn}{\solutionspace_{\velp,\mathrm{h},n+1}}
\newcommand{\solvfhnm}{\solutionspace_{\velf,\mathrm{h},n}}
\newcommand{\solvfh}{\solutionspace_{\velf,\mathrm{h}}}
\newcommand{\stestvf}{\testspace_{\testvelf}}
\newcommand{\stestvp}{\testspace_{\testvelp}}
\newcommand{\stestvfh}{\testspace_{\testvelf,\mathrm{h}}}
\newcommand{\stestvph}{\testspace_{\testvelp,\mathrm{h}}}
\newcommand{\stestvfhn}{\testspace_{\testvelf,\mathrm{h},n+1}}
\newcommand{\solpf}{\solutionspace_{\pf}}
\newcommand{\solpp}{\solutionspace_{\pp}}
\newcommand{\solpfh}{\solutionspace_{\pf,\mathrm{h}}}
\newcommand{\solpph}{\solutionspace_{\pp,\mathrm{h}}}
\newcommand{\solpfhn}{\solutionspace_{\pf,\mathrm{h},n+1}}
\newcommand{\solpfhnm}{\solutionspace_{\pf,\mathrm{h},n}}
\newcommand{\stestpf}{\testspace_{\testpf}}
\newcommand{\stestpp}{\testspace_{\testpp}}
\newcommand{\stestpfh}{\testspace_{\testpf,\mathrm{h}}}
\newcommand{\stestpfhn}{\testspace_{\testpf,\mathrm{h},n+1}}
\newcommand{\stestpph}{\testspace_{\testpp,\mathrm{h}}}
\newcommand{\stestpphn}{\testspace_{\testpp,\mathrm{h},n+1}}
\newcommand{\Czero}{\mathcal{C}^0}
\newcommand{\Hone}{\mathcal{H}^1}
\newcommand{\Ltwo}{\mathcal{L}^2}

\newcommand{\nele}{n_{\mathrm{ele}}}
\newcommand{\nnodeele}[1]{n_{{\mathrm{nod},#1}}}
\newcommand{\nnode}{n_{\mathrm{nod}}}
\newcommand{\nsnode}{n_{\mathrm{nod}}^{\structletterr}}
\newcommand{\nfnode}{n_{\mathrm{nod}}^{\fluidletterr}}
\newcommand{\npnode}{n_{\mathrm{nod}}^{\poroletterr}}
\newcommand{\nsele}{\nele^{\structletterr}}
\newcommand{\nfele}{\nele^{\fluidletterr}}
\newcommand{\npele}{\nele^{\poroletterr}}
\newcommand{\ele}[1]{{\domain}_{\mathcal{T}_\mathrm{h},#1}}
\newcommand{\sele}[1]{\ele{#1}^{\structletterr}}
\newcommand{\fele}[1]{\ele{#1}^{\fluidletterr}}
\newcommand{\pele}[1]{\ele{#1}^{\poroletterr}}
\newcommand{\eleset}{\breve{\domain}_{\mathcal{T}_\mathrm{h}}}
\newcommand{\eleseta}{\breve{\domain}_{\mathcal{T}_\mathrm{h},\mathcal{A}}}
\newcommand{\elesetan}{\breve{\domain}_{\mathcal{T}_\mathrm{h},\mathcal{A},n+1}}
\newcommand{\eleseti}{\breve{\domain}_{\mathcal{T}_\mathrm{h},\mathcal{I}}}
\newcommand{\elesetc}{\breve{\domain}_{\mathcal{T}_\mathrm{h},\restboundf}}
\newcommand{\elesetuc}{\breve{\domain}_{\mathcal{T}_\mathrm{h},\setminus\restboundf}}
\newcommand{\nt}{n_\mathrm{t}}
\newcommand{\hh}{_\mathrm{h}}

\global\long\def\transpose#1{#1^{\mathsf{T}}}
\global\long\def\inverse#1{#1^{\mathsf{-1}}}
\global\long\def\invtrans#1{#1^{\mathsf{-T}}}

\newcommand{\shapef}{N}
\newcommand{\shapefl}{\tilde{\shapef}}
\newcommand{\localcoord}{\tns \xi}

\newcommand{\ncoord}{\tns{\mathsf{\coordsymbol}}}

\newcommand{\ndisps}{\tns{\mathsf{\displacements}}^{\structletter}}
\newcommand{\nvelf}{\tns{\mathsf{\velocityf}}^{\fluidletterr}}
\newcommand{\ndispg}{\tns{\mathsf{\displacements}}^{\aleletter}}
\newcommand{\npf}{\tns{\mathsf{\pressuref}}^{\fluidletterr}}
\newcommand{\ndispp}{\tns{\mathsf{\displacementp}}^{\poroletterr}}
\newcommand{\nvelp}{\tns{\mathsf{\velocityp}}^{\poroletterr}}
\newcommand{\npp}{\tns{\mathsf{\pressurep}}^{\poroletterr}}
\newcommand{\ndispsi}[1]{\tns{\mathsf{\displacements}}^{\structletter,#1}}
\newcommand{\nvelfi}[1]{\tns{\mathsf{\velocityf}}^{\fluidletterr,#1}}
\newcommand{\ndispgi}[1]{\tns{\mathsf{\displacements}}^{\aleletter,#1}}
\newcommand{\npfi}[1]{\tns{\mathsf{\pressuref}}^{\fluidletterr,#1}}
\newcommand{\ndisppi}[1]{\tns{\mathsf{\displacementp}}^{\poroletterr,#1}}
\newcommand{\nvelpi}[1]{\tns{\mathsf{\velocityp}}^{\poroletterr,#1}}
\newcommand{\nppi}[1]{\tns{\mathsf{\pressurep}}^{\poroletterr,#1}}
\newcommand{\nbound}{\interface^{*,\mathrm{N}}}
\newcommand{\dbound}{\interface^{*,\mathrm{D}}}
\newcommand{\timestep}{\Delta t}
\newcommand{\OST}{One-Step-$\theta$ }

\newcommand{\partialddis}[2]{\tilde{\partial}_t \left[#1\right]_{#2}}
\newcommand{\partialdtdis}[2]{\tilde{\partial}_t^2 \left[#1\right]_{#2}}
\newcommand{\partialddisni}[1]{\tilde{\partial}_t \left[#1\right]}
\newcommand{\partialdtdisni}[1]{\tilde{\partial}_t^2 \left[#1\right]}

\newcommand{\Real}{\mathbb{R}}
\newcommand{\Realdim}{\Real^{\dimensions}}
\newcommand{\Project}[3]{\mathcal{P}^{#3}_{#1}\left[#2\right]}
\newcommand{\Projectsymb}[2]{\mathcal{P}^{#2}_{#1}}

\newcommand{\PG}{Petrov-Galerkin}
\newcommand{\BG}{Bubnov-Galerkin}
\newcommand{\FOS}{FOS}
\newcommand{\EOS}{EOS}
\newcommand{\CIP}{CIP}
\newcommand{\CutFEM}{CutFEM}
\newcommand{\FSI}{FSI}

\newcommand{\damping}{\omega}
\newcommand{\iletter}{{\mathrm{I}}}

\section{Introduction}
\label{sec:intro}
The development of a consistent and comprehensive computational approach that allows to investigate
fluid-structure interaction (FSI) including contact\footnote{The terms ``contact'', ``solid-solid interaction'', or ``solid-solid contact'' are exclusively used for the interaction of two solid bodies/domains in this work.
Other phase boundaries, which are sometimes also referred to as ``contact'', are not considered in this contribution.
}
of submersed elastic bodies is the focus of this contribution.
Applications, ranging e.g.\ from the dynamic behavior of biological or mechanical valves to hydrodynamic bearings and tire/wet road contact, often require reliable formulations to solve the 
fluid-structure-contact interaction (FSCI) problem.
As motivated by these examples,
examinations will be carried out at system scale, at which the relevant physics 
can be well represented by means of the continuum mechanics theory.
Challenges for corresponding numerical methods, among others, include the handling of the occurring topological changes in the fluid domain,
the numerical stability of the formulation,
the representation of a physical solution close to the interface,
and the transition between fluid-structure interaction and contact.
For a limited range of problems, the explicit numerical treatment of contact can be avoided by smartly chosen boundary conditions in the setup of the numerical problem and then solved in existing classical FSI frameworks.
These strategies of circumventing the general problem of FSCI, which causes quite a limitation with respect to many applications, are not the focus of this work.

The physical processes involved in contacting bodies submersed in fluid modeled by continuum mechanics theory did not get sufficient attention in previous developments to solve FSCI numerically.
FSCI on a macroscopic length scale is characterized by topological changes of the fluid domain.
On this scale it is important how potential fluid dynamics effects between bodies in the zone of contacting solid interfaces are modeled.
Thus, we first provide a comprehensible clarification of the involved phenomena and discuss resulting implications for the physical modeling for a macroscopic description of FSCI.
The purely geometric classical contact condition that solid bodies are not allowed to penetrate each other directly applies also in the presence of fluid.
In contrast to that, the usual condition from dry contact that no tensile forces can be transmitted between the contacting bodies requires deeper insight into the contacting process.
\begin{figure}[tbp]
\centering
\def\svgwidth{1\textwidth}
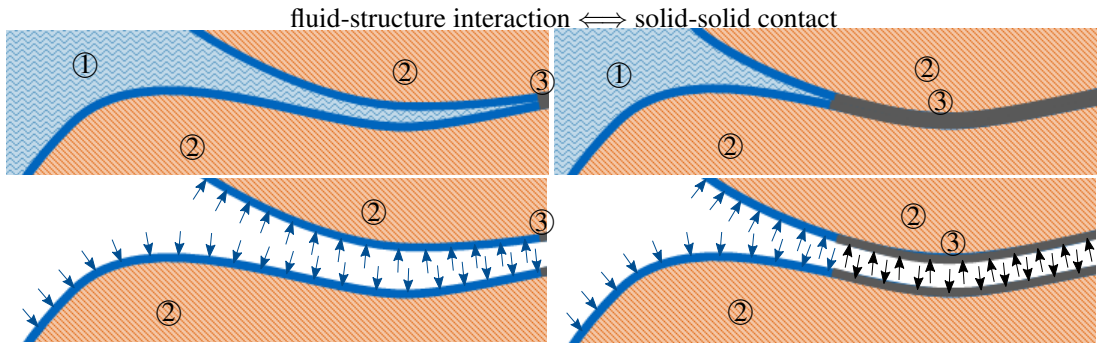
\caption{Detailed view of solid-solid contact with surrounding fluid in two only slightly different states of contact (top). Visualization of the acting force on the solid boundary by arrows. Blue arrows indicate the traction due to the fluid-structure interaction and black arrows indicate the traction due to solid-solid contact (bottom).
The following symbols are used:
\raisebox{.5pt}{\textcircled{\raisebox{-.9pt} {1}}}, fluid domain resolved by macroscopic model; \raisebox{.5pt}{\textcircled{\raisebox{-.9pt} {2}}}, solid domain; \raisebox{.5pt}{\textcircled{\raisebox{-.9pt} {3}}}, solid-solid contact zone.}
\label{fig:fsci_principles}
\end{figure}
In Figure~\ref{fig:fsci_principles} (top), two configurations for contact of two elastic bodies submersed in fluid are shown. 
Both configurations only differ by a slightly different vertical position of the upper solid body, i.e. already small displacements allow to transfer one into the other.
For a continuous formulation, small changes in the position of solid bodies should result only in small changes of the 
interface traction.
This aspect is highlighted in Figure~\ref{fig:fsci_principles} (bottom), where the fluid is replaced by the force acting on the respective solid boundaries.
Therein, the FSI traction (indicated by blue arrows) transfers into the contact traction (indicated by black arrows) and vice versa.
For a continuous problem, the magnitude of the acting traction is not allowed to 
jump at the transition between interface regions with
these two types of condition.
To ensure that, 
we define the FSI traction on the entire interface. 
This
applies especially for the solid-solid contact zone which is located between the two contacting bodies (see Figure~\ref{fig:fsci_principles}, interface zone {\textcircled{\raisebox{-.9pt} {3}}}), even if the enclosed fluid domain is not represented by the macroscopic fluid model. 
This enables a continuous transition from the contact to the FSI traction which otherwise would not be possible.
It is worth mentioning that this is not unphysical or only a ``numerical 'trick'', as in reality fluid will still remain at this interface but would only be visible on a smaller length scale.
The no-adhesive force condition of classical contact mechanics can thereby be reformulated based on the difference between contact and fluid traction.

As a result of this setup increasing the fluid pressure can lift-off two contacting bodies not only in the physical world but also in the numerical model.
This effect demonstrates the importance of a defined fluid stress state at every point in the contact zone.
Depending on different criteria, fluid models of variable complexity and accuracy on the reduced dimensional manifold of the contact zone can be considered.
These criteria include the microstructure of the contacting surfaces (smooth vs.\ rough), the quantities/processes of interest (e.g.\ leakage of a sealing), and the macroscopic problem configuration (closed vs.\ open fluid domain).
As this work intends to provide a general FSCI framework,
a simple extension based approach to model the fluid in the contact zone is applied herein. 
The general framework then allows to include also various types of alternative and more sophisticated fluid models in the contact zone in the future.
For a model with greater physical depth, the reader is e.g.\ referred to \cite{ager2018}, where a homogenized poroelastic layer formulation is applied to model the fluid flow through contacting rough surfaces.

When looking at the available literature one can note that a 
large portion of literature on FSCI formulations is either interested in the analysis of heart valves 
\cite{dehart2003,loon2006,santos2008,astorino2009,borazjani2013,espino2014,espino2015,kamensky2015,laadhari2016,meschini2018}
and only a smaller set in solving a more general problem setup of FSCI \cite{sathe2008,mayer20103,wick2014,bogaers2016,frei2018,ager2018}.
However, while most of those formulations work well for certain problem setups, they suffer from some restrictions preventing their application to more general complex problem classes.
In the following we try to give a brief overview on existing formulations but will not describe the different approaches in detail but rather point to certain features and especially to assumptions or restrictions. 

In \cite{espino2014,espino2015} contact in surrounding fluid does not need to be considered due to the chosen problem setup with geometrically separated contact and fluid-structure interfaces.
A penetration of the solid bodies is accepted in \cite{dehart2003} since contact is not treated explicitly.
In \cite{borazjani2013}, contact is included, but in the presented computations only the valve opening phase without significant influence of the contact formulation is analyzed.
In \cite{wick2014}, no explicit contact formulation is considered and a minimal distance of one mesh cell still remains between two flaps.
Using reduced modeling with included contact of the heart valve, \cite{laadhari2016} avoids the requirement for a general FSCI formulation.
In \cite{ager2018}, a very general FSCI formulation for contact of rough surfaces is presented, where the interface is modeled via a homogenized poroelastic layer.
Such a formulation is very powerful and also well motivated by the involved physical phenomena but it is also more complex and not always needed. In such frequent cases the approach presented in this paper will be a better alternative.

Explicit treatment of the contact is considered in \cite{loon2006,santos2008,astorino2009,mayer20103,ager2018} by Lagrange multiplier based contact methods, 
in \cite{sathe2008,borazjani2013,kamensky2015, bogaers2016} by methods based on penalty contact contributions,
and in \cite{meschini2018} by an approach based on enforcement of equal structural velocity.
Interface fitted computational meshes for discretization of the fluid domain are enabled by approaches that require to enforce a non-vanishing fluid gap between 
approaching
bodies and therefore avoid topological changes in the fluid domain preventing degenerated elements \cite{sathe2008,bogaers2016}.
Approaches enabling the use of a non-interface fitted discretization, which allow the consideration of ``real'' contact scenarios and the resulting topological changes of the fluid domain directly, are
applied in \cite{dehart2003,loon2006,santos2008,astorino2009,borazjani2013,mayer20103,wick2014,kamensky2015,laadhari2016,meschini2018,ager2018}.
The majority of these formulations consider 
dimensionally reduced structural models (i.e.\ membranes and shells)
\cite{dehart2003,loon2006,santos2008, astorino2009, borazjani2013,kamensky2015,laadhari2016,meschini2018},
whereas bulky structures (i.e.\ structures of significant thickness as compared to the spatial resolution of the computational discretization in the fluid domain)
are only considered in
\cite{mayer20103,wick2014,frei2018,ager2018}. 
The restriction to slender bodies of the non-interface fitted approaches is often related to issues concerning system conditioning and mass conservation errors close to the fluid-structure interface.
This is due to the fact that the discontinuity of the fluid stress between two sides of a submersed solid typically is not represented by the discrete formulation (see e.g.\ \cite{kamensky2015,laadhari2016}), which prevents the analysis of configurations including large pressure jumps.
This issue is not a fundamental limitation for non-interface fitted FSI as shown e.g.\ in \cite{burman2014,alauzet2016} (without contact), but increases the complexity of such a formulation including the underlying algorithm.

It is surprising that, except for \cite{ager2018,frei2018}, none of the referenced works includes a substantial discussion concerning the requirement for the fluid state in the contact zone as elaborated earlier.
Most works include contact just as a constraint additional to the incorporation of FSI conditions, which is 
still enforced in closed contact.
If such an approach is carried out properly, it can result in a continuous FSCI formulation. Still, the strategy to recover the fluid state in the contact zone, which is required to enforce the FSI conditions, remains an open question.
For formulations which circumvent topological changes of the fluid domain  \cite{sathe2008,bogaers2016} this issue does not arise directly, as there is always a numerically motivated fluid domain 
in between the contacting bodies.
For all other formulations which support the actual contact of surfaces, the different approaches for the numerical solution of the fluid problem provide a nonphysical fluid state outside the fluid domain in the contact zone by default, which builds the basis to incorporate the FSI interface conditions.
Depending on the underlying numerical method and the respective FSCI problem, this can provide, but not necessarily is, a sufficiently good approximation of the fluid state in the contact zone.
We would like to point out that this fluid state in the contact zone has an essential influence for example to the detaching behavior of contacting solid bodies, e.g. a high/low fluid pressure supports/prohibits the separation of two bodies.
Another important aspect which should be considered when applying such a strategy of incorporating contact on top of the FSI conditions is that 
the FSI traction includes a tangential component.
In contrast to the normal component of the FSI traction, which serves merely as an offset to normal contact traction, the tangential component directly acts on the contacting surfaces potentially deteriorating the solution accuracy.

In this paragraph, we would like to comment briefly on the so called ``collision paradox'', which states that for incompressible, viscous fluid with ``no-slip'' condition on smooth boundaries, contact between submersed bodies cannot occur in finite time (see e.g. \cite{hillairet2009,gerard2015}).
This is in contrast to the observation of contacting solid bodies which is also observed when bodies are submersed in fluid.
A physical explanation for this paradox can be found in the missing consideration of non-smooth surfaces (analyzed in e.g.\ \cite{cawthorn2010,gerard2010})
arising from the rough microstructure, or other effects on the micro scale, not considered in the macroscopic fluid model of the ``no-slip" boundary condition \cite{neto2005}.
Nevertheless, even for the computational analysis of physical models where these effects are not considered, contact still has to be considered. 
This is a result of the numerical solution approaches which are always accompanied by approximations of the underlying physical model when considering general configurations.
If there is no explicit contact treatment within the FSI formulation only fluid forces in the gap keep the bodies apart. But when an artificial collision of solid bodies occurs, e.g. during an iterative nonlinear solution procedure, there is no separation force acting as there is no remaining fluid between these bodies.
This is shown in \cite{dehart2003,astorino2009}, where a penetration of surfaces can be observed as no contact formulation is considered.

In this contribution, we present a general FSCI formulation considering flows based on the incompressible Navier-Stokes equations interacting with nonlinear elastic solids that are not restricted to slender structural bodies.
Therein, contact is not considered as a mere additional constraint on the FSI problem, but focus is rather put on the mutual exclusiveness of fluid-solid and solid-solid coupling.
Thus, the application of fluid forces on the interface in the zone of active contact, where typically no good representation of the fluid solution is available, is automatically eliminated.
The approach is applied to non-interface fitted discretizations for the fluid domain by the Cut Finite Element Method (CutFEM), due to its ability to represent 
sharp discontinuities of the solution at the interface.
This is of essential importance for the discrete representation of the prevalent discontinuity of the fluid stress between opposite sides of (potentially thin) structural bodies.
Hence, non-physically high gradients arising from a continuous fluid solution representation can be avoided (see remark in \cite[page 1753]{santos2008}).
Crucial for a continuous discrete form is the continuity of the transition from fluid-solid to solid-solid interaction, which is achieved by the use of Nitsche-based methods for both constraints.

The CutFEM in general enables the use of non-interface fitted, fixed Eulerian meshes for the fluid discretization in complex and deforming domains.
This method is perfectly suited for handling of large interface motion and topological changes of the computational domain, typically occurring for FSCI problems,
and therefore is applied here.
To enable a determined, continuous transition from the ``no-slip'' 
condition to frictionless contact, a relaxation of the tangential constraint is proposed, while retaining the mass-balance in normal direction.
This is enabled by a flexible formulation capable of handling the ``no-slip'' and the ``full-slip'' limit on the fluid-solid interface.
CutFEM has seen great progress in recent years and meanwhile enjoys a solid mathematical base.
Initial analysis was performed for the Poisson equation \cite{burman2012}, extended to the Stokes equation \cite{burman2014_2,massing2014}, 
and finally, including advection, on the Oseen equation \cite{schott2014,massing2016}.
Therein, a so-called ``ghost penalty'' stabilization \cite{burman2010} guarantees a well-conditioned formulation for arbitrary interface positions. 
Successful applications of the CutFEM on two-phase flow and fluid-structure interaction 
are presented in \cite{gross2007,hansbo2014,schott2015} and  \cite{burman2014,massing2015,schott2017}, respectively.
Therein, the ``no-slip'' interface condition is applied weakly by a Nitsche-based method.
The basis for the general Navier interface condition applied in this work was presented for the Poisson equation in \cite{Juntunen2009}, extended to the general Navier boundary condition for the Oseen equation in \cite{winter2017},
and applied to enforce the tangential coupling condition on the interface of an poroelastic solid and a viscous fluid in \cite{ager2018,ager2018b}.

To obtain a continuous transition of the discrete formulation from fluid-structure to contact interaction, both FSI and the treatment of contact are enforced via Nitsche's method.
A first application of Nitsche's method to contact problems was presented in \cite{Wriggers2008}. More recently, the development of Nitsche-type methods for contact problems gained more attention due to the mathematical analysis of symmetric and skew-symmetric Nitsche methods provided by \cite{Chouly2015sym,Chouly2013,Chouly2014tresca} for small deformation frictionless and frictional contact problems.
In addition, \cite{Burman2017} analyzed a penalty-free variant for the Signiorini-problem.
Based on these works, \cite{Mlika2017} extended Nitsche's method to nonlinear elasticity at finite deformation and \cite{Seitz2018} to nonlinear thermomechanical problems.
Most classical contact formulations employ a so-called slave-master concept introducing an inherent bias to the formulation by a (user-defined) choice of the slave and master surface.
In the context of Nitsche methods, \cite{Mlika2017,Chouly2018} introduced an unbiased variant.
The method proposed in \cite{Seitz2018} is based on a harmonic weighting of the contact stress resulting in an almost unbiased approach as the only bias is introduced by the applied integration rule.
In this work, we will extend the method of \cite{Seitz2018} to a completely unbiased form by integrating not only on the slave but also on the master surface similar to so-called two-half-pass algorithms \cite{Sauer2015}.
Finally, the transition between active and inactive contact has to be balanced carefully with the ambient fluid traction to achieve continuity of the discrete formulation.

The resulting formulation is discretized in time by the one-step-$\theta$ scheme.
Finally, the nonlinear system of equations is solved for all unknowns, i.e.\ nodal structural displacements, fluid velocities and pressures, by a Newton-Raphson based procedure.
Due to the strong interaction of all involved physical domains for the FSCI problem this is done simultaneously, i.e.\ a monolithic procedure
is applied (see e.g. \cite{verdugo2016}). 

Recently, \cite{frei2018} presented a Nitsche-based formulation for FSCI similar to the one derived in this paper.
Therein, linear Stokes flow and linear elastic solids based on a fully Eulerian description in combination with contact to a rigid, straight obstacle at the fluid boundary is analyzed and stability results for this formulation are shown. 
In contrast to the extrapolation based strategy proposed in this contribution, which allows for complete topological changes, two strategies based on a thin remaining fluid film are presented in \cite{frei2018} to obtain the fluid stress in the contacting zone.

The paper is organized as follows.
In Section~\ref{sec:goveq} the governing equations, comprised of the structural and fluid mechanics model as well as conditions on the interface in normal an tangential direction of the FSCI problem, are given.
This is followed by a presentation of the discrete formulation, including all volume and interface contributions, and the solution strategy in Section~\ref{sec:FE}.
Different numerical examples, capable of analyzing different aspects of the formulation, are presented in Section~\ref{sec:examples}.
Finally, in Section~\ref{sec:conclusion} a short summary and an outlook are given.
\section{Governing equations}
\label{sec:goveq}
\begin{figure}[t]
\centering
\def\svgwidth{0.5\textwidth}
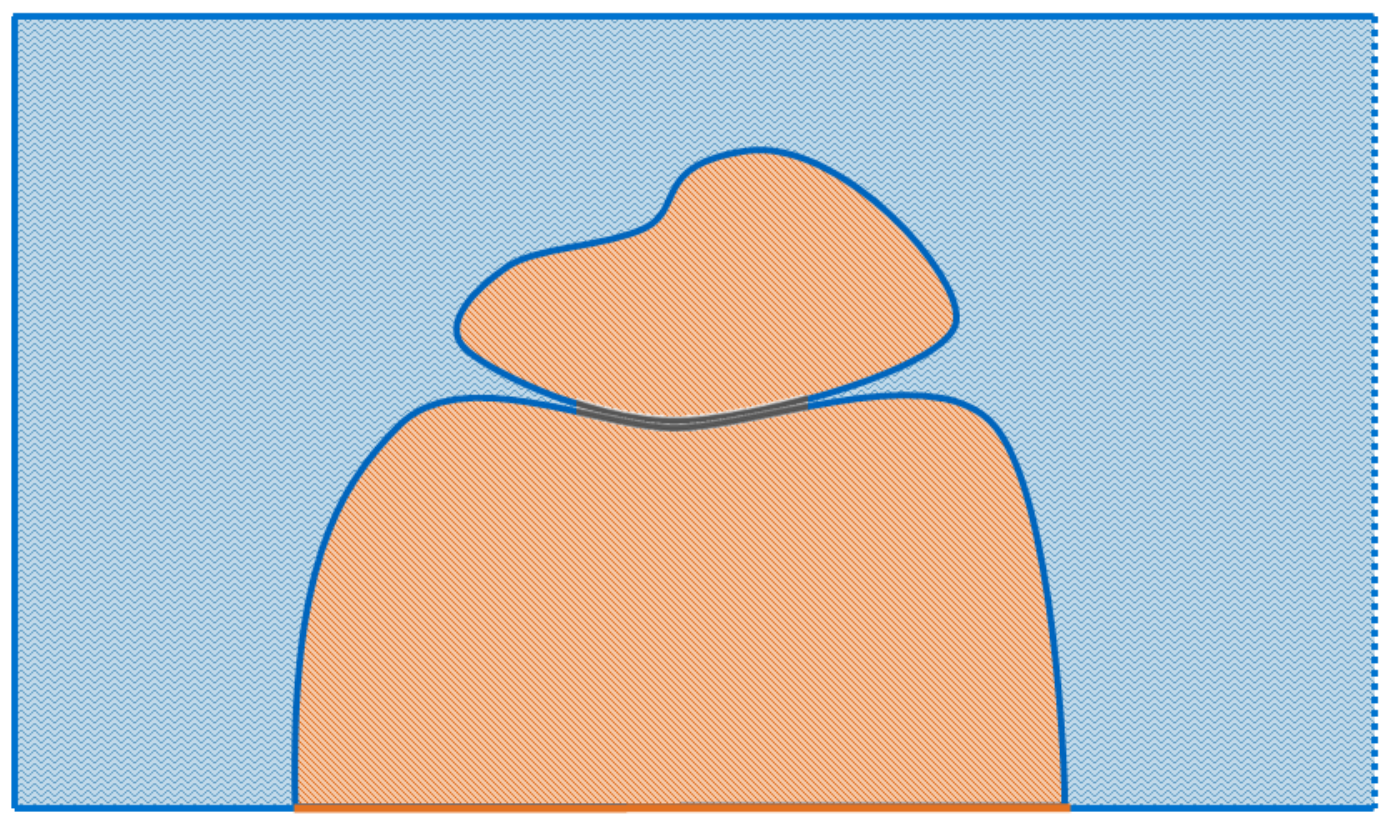
\caption{Fluid-structure-contact interaction (FSCI) problem setup for two contacting bodies ``$1$'' and ``$2$'': fluid domain $\domainf$, solid domain $\domains = \domainsone \cup \domainstwo$, 
fluid-structure interface $\fsiinterface = \fsiinterfaceone \cup \fsiinterfacetwo$, the active (closed) contact interface $\sciinterface = \sciinterfaceone \cup \sciinterfacetwo$,
overall coupling interface $\interface = \interfaceone \cup \interfacetwo$, and outer boundaries $\dboundf,\nboundf,\dbounds, \nbounds$.}
\label{fig:setup}
\end{figure}

In this section, we discuss the governing equations and conditions for all physical domains and interfaces of the FSCI problem.
A typical configuration for such a problem is shown in Figure \ref{fig:setup}.
The domain $\domain$ of the overall FSCI problem includes the fluid domain $\domainf$ and the solid domain $\domains$.
The overall coupling interface $\interface$ consists of the fluid-structure interface $\fsiinterface$ and the active (closed) contact interface $\sciinterface$.
The different boundaries on the outer boundary $\partial \domain$ are denoted by $\dboundf,\nboundf,\dbounds,$ and $\nbounds$.
In the following, all quantities $* \, , \tns *$ with additional ``zero''-index $*_0\, , \tns *_0$  are described in the undeformed reference/material configuration,
whereas a missing index indicates the current configuration (see \cite{Holzapfel2000} for details).
An additional ``hat''-symbol $\hat{*}\, , \tns{\hat{*}}$ indicates time-dependent prescribed quantities at the boundaries and in the domains.
Prescribed quantities at the initial point in time $\btime$ are indicated by the ``ring''-symbol  $\mathring{*}\, , \tns{\mathring{*}}$.
The outer boundary of a domain $\domain^*$ is specified by $\partial \domain^*$.

\subsection{Structural domain $\domains$}
\label{sec:struct}
The displacements of every point in the hyperelastic structural domain are governed by the transient balance of linear momentum:
\begin{align}
 \refdensitys \accs - \divRef \left(\defgrads \cdot \stresspks \right) - \refdensitys \refbodyfs = \zerovec  \quad \text{in} \;  \refdomains \times [\btime,\etime],
 \label{eq:structure_eq}
\end{align}
\begin{align}
\quad \stresspks = \partiald{\strainenergys}{\straingls}, \quad \straingls = \frac{1}{2}\left[\left(\defgrads\right)^T\cdot\defgrads - \identity\right], \quad \defgrads = \identity + \partiald{\disps}{\refcoords}.
 \label{eq:structure_constitutive}
\end{align}
Therein, the displacement vector $\disps = \coords - \refcoords$ describes the motion of a material point (with position $\refcoords$ at initial time $\timeso = \btime$), 
due to deformation of the elastic body, to the current position $\coords$.
The structural density in the undeformed configuration is denoted by $\refdensitys$, the material divergence operator by $\divRef \tns *$, the deformation gradient by $\defgrads$,
the second Piola-Kirchhoff stress tensor by $\stresspks$, and the body force per unit mass by $\refbodyfs$.
A hyperelastic strain energy function $\strainenergys$ characterizes the nonlinear material behavior and hence provides the stress-strain relation.
Therein, the strain is quantified by the Green-Lagrange strain tensor $\straingls$.
The Cauchy stress can be expressed by $\stresss = \frac{1}{\Js} \defgrads \cdot \stresspks \cdot \left( \defgrads \right)^T$, with $\Js$ being the determinant of the deformation gradient $\defgrads$.
This representation of solid stress $\stresss$ in the current configuration will be required for coupling of the solid domain and the fluid domain on their common interface.
Additional initial conditions for the displacement field $\dispsB$ and velocity field $\velsB$ are required:
\begin{align}
\disps= \dispsB \quad \text{in} \;  \refdomains \times \left\lbrace t_0 \right\rbrace, \qquad
\vels = \velsB \quad \text{in} \;  \refdomains \times \left\lbrace t_0 \right\rbrace.
\label{eq:structure_init}
\end{align}
Finally, to complete the description of the initial boundary value problem for nonlinear elastodynamics,
adequate boundary conditions on the outer boundary $\fullrefbound\cap\fullrefbounds$ have to be specified with the predefined displacement $\dispsD$ on Dirichlet boundaries $\refdbounds$ and the
given traction $\reftractionsN$ on Neumann boundaries $\refnbounds$:
\begin{align}
\disps = \dispsD \quad\text{on} \;  \refdbounds \times [\btime,\etime], \qquad
\left( \defgrads \cdot \stresspks \right) \cdot \refnormals = \reftractionsN \quad \text{on} \;  \refnbounds \times [\btime,\etime].
\label{eq:structure_bcs}
\end{align}
The outward-pointing reference unit normal vector on the boundary $\partial \refdomains$ is specified by $\refnormals$.
Conditions on the remaining subset of the structural boundary $\reffsiinterface \cup \refsciinterface = \fullrefbounds \setminus \left( \refdbounds \cup \refnbounds \right)$,
where the structural domain is coupled to the fluid domain or contact occurs will be discussed in Sections \ref{sec:interfacen} and \ref{sec:interfacet}.
This remaining subset is not part of the outer boundary of the FSCI problem $\fullrefbound\cap\left(\reffsiinterface \cup \refsciinterface\right)=\emptyset$.

\subsection{Fluid domain $\domainf$}
In the fluid domain transient, incompressible, viscous flow is considered.
Therefore, the governing equations are the incompressible Navier-Stokes equations which include the balance of mass and linear momentum:

\begin {align}
\densityf \partiald{\velf}{\timef} + \densityf \velf \cdot \grad \velf + \grad \pf - \div( 2 \viscf \epsf (\velf)) - \densityf\bodyff  &= \zerovec   \quad \text{in} \;  \domainf \times [\btime,\etime],\label{eq:fluidm_eq}\\
\div \velf &= 0   \quad \text{in} \;  \domainf \times [\btime,\etime].\label{eq:fluidc_eq}
\end{align}
Therein, the velocity and the pressure of the fluid continuum at a specific point in space is denoted by $\velf$ and $\pf$, respectively.
The constant fluid density is denoted by $\densityf$, the constant dynamic viscosity by $\viscf$, and the prescribed body force per unit mass by $\bodyff$.
Further, the symmetric strain-rate tensor is defined by $\epsf (\velf) = \frac{1}{2}\left[\grad \velf + \left(\grad \velf\right)^T\right]$.
Due to the present derivative of the velocity in time, the initial velocity field $\velfB$ has to be prescribed:
\begin{align}
\velf = \velfB \quad \text{in} \;  \domainf \times \left\lbrace t_0 \right\rbrace.
\label{eq:fluid_init}
\end{align}
By prescribing adequate boundary conditions on the outer boundary $\fullbound\cap\fullboundf$, the description of the fluid problem is completed.
Thereby the fluid velocity $\velfD$ on Dirichlet boundaries $\dboundf$, 
or the fluid traction $\tractionfN$ on Neumann boundaries $\nboundf$ is predefined:
\begin{align}
\velf = \velfD \quad \text{on} \;  \dboundf \times [\btime,\etime], \qquad
\stressf \cdot \normalf = \tractionfN \quad \text{on} \;  \nboundf\times [\btime,\etime].
\label{eq:fluid_bcs}
\end{align}
Herein, the Cauchy stress $\stressf = -\pf \identity + 2 \viscf \epsf (\velf)$ and the outward unit normal $\normalf$ of the fluid domain is utilized.
Again, conditions on the remaining subset of the fluid boundary $\fsiinterface = \fullboundf\setminus\left(\dboundf\cup\nboundf\right)$, which equals the common interface of fluid and structural domain,
will be discussed in Sections \ref{sec:interfacen} and \ref{sec:interfacet}.
This remaining subset is not part of the outer boundary of the FSCI problem $\fullbound\cap\fsiinterface=\emptyset$.

\paragraph*{The fluid extension operator}
In order to formulate the interface conditions at any point in space $\coord$ on the overall coupling interface, 
an extension operator $\extensionsym{\coord}:\fsiinterface \longrightarrow \interface$ from the fluid-structure interface $\fsiinterface$ to the overall interface $\interface$ is required.
This extension is applied for all quantities solely defined in the fluid domain $\domainf$ and thus for all quantities on the fluid-structure interface $\fsiinterface$
which are required for the formulation of the interface constraints on $\interface$. In the following, the extension of any quantity $*$ is denoted by an additional index $*_\extensionsymbol$.
Exemplary, the extension of the normal fluid stress $\stressfnn$ to a position $\coord$ on $\interface$ is defined as follows:
\begin{align}
 \stressfnnext\left(\coord\right) =   
 \begin{cases}
  \stressfnn\left(\velf\left(\coord\right),\pf\left(\coord\right)\right) \quad \text{on} \quad \fsiinterface\\
  \extension{\stressfnn\left(\velf\left(\coord_\extensionsymbol\right),\pf\left(\coord_\extensionsymbol\right)\right)}{\coord} \quad \text{on} \quad \sciinterface,
  \end{cases}\nonumber\\
  \quad \text{with} \quad \extension{\stressfnn\left(\velf\left(\coord_\extensionsymbol\right),\pf\left(\coord_\extensionsymbol\right)\right)}{\coord} 
  = \stressfnn\left(\velf\left(\coord\right),\pf\left(\coord\right)\right)
  \quad \text{on} \quad \fsiinterface \cap \sciinterface,
  \label{eq:fextension}
\end{align}
where the extension origin position $\coord_\extensionsymbol$ is properly chosen on $\fsiinterface$.
The last line in \eqref{eq:fextension} represents the continuity of the extension operator. The applied extension operator for all presented numerical examples is discussed in Section \ref{sec:num_extrapol_operator}.
Alternative approaches to obtain fluid quantities on the overall interface $\interface$ are briefly discussed in the Remarks \ref{rem:extension_phys1} and \ref{rem:extension_phys2}.

\subsection{Conditions on the overall coupling interface $\interface$ in normal direction}
\label{sec:interfacen}
For the formulation of the interface constraints, which are splitted in the interface normal direction and in the tangential plane, the solid outward unit normal $\normal=\normals$ will be considered.
The normal component of the respective Cauchy stress is denoted as: $\stresssnn = \stresss \ccprod \Pnormal$ and $\stressfnn=\stressf \ccprod \Pnormal$,
with the normal projection operator being specified as $\Pnormal := \normal \otimes \normal$.

The conditions in the normal direction for purely non-adhesive structural contact configurations are given by the classical Hertz--Signiorini--Moreau (HSM) conditions:
\begin{align}
\Gap := (\proj{\coord}(\coord) -\coord)\cdot\normal&\geq 0 \quad \text{on} \quad \interface \timesfulltime,\label{eq:gapcond_hsm}
\\
\stresssnn &\leq 0 \quad \text{on} \quad  \interface \timesfulltime,\label{eq:contacttractioncond_hsm}
\\
\Gap \stresssnn  &= 0 \quad \text{on} \quad  \interface \timesfulltime,\label{eq:contactexclusioncond_hsm}
\end{align}
which ensure the non-penetration, the absence of adhesive contact forces, and the complementarity between the contact pressure and normal gap $\Gap$.
To obtain the normal gap $\Gap$, the point $\proj{\coord}(\coord)$ is obtained as the projection of  $\coord$ along its normal $\normal$ onto the opposite solid surface; in the case that no such projection exists, we assume $\Gap\to\infty$.
All quantities $*$ evaluated at this projection point will be denoted by a check $\proj{*}$.

In the case contacting bodies are surrounded by fluid, the fluid flow in the contacting zone has to be considered properly as discussed in the introduction.
Applying the classical HSM conditions \eqref{eq:gapcond_hsm}-\eqref{eq:contactexclusioncond_hsm} directly would result in the implicit assumption that fluid does not fill the contact zone.
For such a configuration an instantaneous change from zero traction to the traction arising from the ambient fluid in the contact opening zone on the solid boundary would occur
and thus the formulation of a continuous problem is prohibited.
Considering, on the contrary, the presence of (physically reasonable) fluid in the contact zone (on a smaller length scale and not resolved but just modeled at the current macroscopic scale) leads to modified HSM conditions, 
where a lifting of both bodies occurs for vanishing relative traction of contact (solid) traction and ambient fluid traction.
These conditions result in a continuous problem as the balance of solid and fluid traction is essential on the common interface of a fluid and a solid.
Then, the conditions on the interface $\interface$ formulated for a specific point $\coord$ on $\Gamma$ are:
\begin{align}
\Gap  &\geq 0 \quad \text{on} \quad \interface \timesfulltime,
\label{eq:gapcond}\\
\stresssnn - \stressfnnext\left(\coord\right)&\leq 0 \quad \text{on} \quad  \interface \timesfulltime,
\label{eq:contacttractioncond}\\
\Gap\left[\stresssnn - \stressfnnext\left(\coord\right)\right] &= 0 \quad \text{on} \quad  \interface \timesfulltime.\label{eq:contactexclusioncond}
\end{align}
Condition \eqref{eq:gapcond} enforces a positive or vanishing gap $\Gap$ between two solid bodies.
In condition \eqref{eq:contacttractioncond}, a negative or vanishing relative traction has to be guaranteed, at least in the case without adhesive forces that is considered here.
Finally, in equation \eqref{eq:contactexclusioncond}, either a vanishing gap in the contact case of solid-solid interaction or a vanishing relative traction in the case of fluid-structure interaction
is enforced. Additionally, the dynamic equilibrium between two contacting bodies has to be formulated:
\begin{align}
\stresssnn - \stressfnnext  = 
\proj{\stresssnn} - \proj{\stressfnnext}
\quad &&\text{on} \; \interface \timesfulltime.
\label{eq:InterfaceLinMom}
\end{align}
In the contact case, due to the vanishing gap $\Gap$, the normal fluid traction equals its projection $\stressfnnext=\proj{\stressfnnext}$ and therefore the classical dynamic equilibrium 
between both contacting bodies is recovered. For the fluid-structure interaction case, due to the vanishing relative traction $\stresssnn = \stressfnnext$, both sides of the equilibrium vanish
and as a result equation \eqref{eq:InterfaceLinMom} is automatically fulfilled. Finally, the mass balance for the motion of solid bodies connected to a fluid domain is given as:
\begin{align}
\velreln := \left(\velf - \vels\right) \cdot  \normal = 0 \quad &&\text{on} \; \fsiinterface \timesfulltime.
\label{eq:fsi_noslipn}
\end{align}
Herein, a vanishing normal relative velocity $\velreln$ is enforced solely on the interface $\fsiinterface$, which is part of the fluid outer boundary $\partial \domainf$.
Applying an extension to the normal relative velocity $\velrelnext$, this condition is automatically fulfilled on the remaining subset of the interface $\sciinterface$ and hence on the entire $\interface$.

\begin{remark}[Influence of the fluid extension operator]
\label{rem:fluidsol}
It should be highlighted, that conditions \eqref{eq:contacttractioncond}, \eqref{eq:contactexclusioncond}, and \eqref{eq:InterfaceLinMom} are expressed by an extension of the fluid stress from the fluid-structure interface $\fsiinterface$ to the contact interface $\sciinterface$.
The fluid stress extension has an essential influence only close to the condition changing point/curve $(\fsiinterface \cap \sciinterface)$.
This point is contained in the origin from which the extension is constructed, namely the fluid domain.
Thus, even the application of a simple continuous extension strategy of the fluid stress, which is by definition more accurate close to the fluid domain, provides a sufficiently accurate fluid stress representation for a wide range of problem configurations. 
Still, we would like to emphasize that the continuous extension operator is considered in this work especially to enable a clear presentation due to its simplicity.
In the case that a more accurate physical fluid solution is required in the contact zone, alternative extension based strategies can be considered or appropriate equations to describe the fluid flow  in this zone can be solved.
\end{remark}

\subsection{Conditions on the overall coupling interface $\interface$ in tangential direction}
\label{sec:interfacet}
In the tangential direction, frictionless solid-solid contact in combination with the general Navier boundary condition as a kinematic constraint between solid bodies and the fluid domain is considered for simplicity of presentation.
Then, the following conditions have to be fulfilled on the interface $\interface$:
\begin{align}
\stresss \cdot \normal \cdot \Ptangent = \zerovec \quad &&\text{on} \; \sciinterface \timesfulltime,
\label{eq:c_dyneq_t}\\
\left(\stressf\cdot \normal - \stresss\cdot  \normal\right) \cdot \Ptangent = \zerovec \quad &&\text{on} \; \fsiinterface \timesfulltime,
\label{eq:fsi_dyneq_t}\\
\left(\velf - \vels + \sliplengh \stressf \cdot \normalf \right) \cdot  \Ptangent   = \zerovec  \quad &&\text{on} \; \fsiinterface \timesfulltime.
\label{eq:fsi_navslip}
\end{align}
Herein, the tangential projection operator is specified by $\Ptangent := \unity -\normal \otimes \normal$.
While condition \eqref{eq:c_dyneq_t} represents the vanishing tangential traction component on the contact interface $\sciinterface$, condition \eqref{eq:fsi_dyneq_t} enforces 
the dynamic equilibrium between solid and fluid on interface $\fsiinterface$. 
As these two conditions can coincide at the common point $\sciinterface\cap\fsiinterface$ only in the case of a vanishing tangential fluid traction 
$\left(\stressf\cdot \normal \cdot \Ptangent = \zerovec\right)$, the general Navier boundary condition  \eqref{eq:fsi_navslip} with a varying slip length is applied.
This condition includes the no-slip boundary condition for a vanishing slip length $\sliplengh = 0$, which is the common interface condition, successfully applied for macroscopic problem setups.
Nevertheless, on smaller scales, due to characteristics such as surface roughness or wettability, an interfacial velocity slip can be observed in a large number of experiments \cite{neto2005}.
In this contribution, the main emphasis of applying the general Navier boundary condition 
is to guarantee continuity for transitions between fluid-structure interaction and frictionless contact solid-solid interaction and
to enable a relaxation of the tangential constraint close to the contacting zone.
To obtain these properties, an infinite slip length $\sliplengh=\infty$ is specified close to the common point $\sciinterface\cap\fsiinterface$,
whereas a vanishing slip length still allows the consideration of the no-slip condition for the majority of the fluid-structure interface $\fsiinterface$
representing the macroscopic modeling point of view.
Further details on the specification of the slip length $\sliplengh$ for the presented formulation are given in Section~\ref{sec:num_interface_t}.

\begin{remark}[Continuity of the formulation considering frictional contact]
\label{rem:fric_contact}
It should be pointed out that also for the case when frictional contact is considered, specific treatment of the tangential constraints will be required to result in a continuous problem. 
This issue arises due to the fact, that the fluid wall shear stress on a fluid-structure interface is not automatically equal to the tangential stress resulting from sliding friction of two contacting structures on a macroscopic view. 
In the case of a friction model with vanishing tangential interface traction at the condition changing point/curve $\fsiinterface \cap \sciinterface$, applying the presented strategy directly results in a continuous problem also for frictional contact.
The presented general Navier conditions yields a zero tangential fluid traction at the condition changing point $\fsiinterface \cap \sciinterface$.
Hence, to ensure continuity, a solid contact friction model has to provide a vanishing tangential traction at this point as well.
For instance, this can be achieved using a Coulomb friction law (friction coefficient~$\mathfrak{F}$) based on the relative normal stress with the total friction bound 
$\mathfrak{F}\cdot (\stresssnn - \stressfnnext)$.
\end{remark}
\section{Discrete formulation}
\label{sec:FE}
In this section, the discrete formulation applied to the numerical solution of the FSCI problem is presented.
The spatial discretization of the continuous problem, presented in the previous section, is based on the FEM
and temporal discretization by the one-step-$\theta$ scheme is applied.
First, the semi-discrete weak forms directly derived from the governing equations, including additional fluid stabilization operators, are given.
To account for topological changes in the fluid domain, an elementary feature occurring for the FSCI problems, the CutFEM is applied to the discretization of the fluid equations and is thus discussed in the following.
Therein, details on the determination of a consistent discrete set of fluid domain and fluid-structure interface for the contact case are given.
The interface conditions, which are split in normal and tangential direction, are incorporated by Nitsche-based approaches.
For the normal direction, a single continuous interface traction representation is proposed, automatically incorporating the fluid-structure and contact conditions.
A detailed explanation of the resulting contributions by this normal interface traction is given by analyzing the different cases. 
Further, a Nitsche-based formulation to incorporate the tangential fluid-structure interface condition including potential slip is presented.
The specification of the slip length parameter on the interface to enable a continuous transition from fluid-structure coupling to frictionless contact is discussed.
Finally, all contributions are treated in a single global system of equations and solved monolithically.
Additional details on the solution procedure of the FSCI problem are given at the end of this section.
To shorten the presentation only some aspects that help understanding the approach are discussed here, while many more details can be found 
in the referenced literature 
for the particular building block methods.

In the following, all quantities, including the primary unknowns, the test functions in the weak form, the domains and interfaces as well as derived quantities are discretized in space.
Still, no additional index $h$ is added to these discrete quantities 
for the sake of brevity of presentation.
The expressions $\innerp{*}{*}{\Omega}$ and $\innerpb{*}{*}{\partial \Omega}$ denote the $\Ltwo$-inner products integrated in the domain $\Omega$ and on the boundary/interface $\partial \Omega$, respectively.

\subsection{Weak forms for the domains $\domains, \domainf$}
The weak forms for the structural domain $\mathcal{W}^S$, the fluid domain $\mathcal{W}^F$, and the overall coupled problem $\mathcal{W}^{FS}$ can be derived from equations \eqref{eq:structure_eq} and \eqref{eq:fluidm_eq} - \eqref{eq:fluidc_eq}, respectively.
\begin{align}
\mathcal{W}^S\left[\testdisps,\disps\right] = \innerp{\testdisps}{\refdensitys \accs}{\refdomains} + \innerp{\gradRef \testdisps}{\defgrads \cdot \stresspks}{\refdomains} - \innerp{\testdisps}{\refdensitys \refbodyfs}{\refdomains}
-\innerpb{\testdisps}{\reftractionsN}{\refnbounds}
,\label{eq:w_solid}\\
\mathcal{W}^F\left[\left(\testvelf, \testpf\right),\left(\velf, \pf\right)\right] = \innerp{\testvelf}{\densityf \partiald{\velf}{t}}{\domainf} + \innerp{\testvelf}{\densityf \velf \cdot \grad \velf}{\domainf} 
-\innerp{\div \testvelf}{\pf}{\domainf}\nonumber\\
+\innerp{\epsf (\testvelf)}{2 \viscf \epsf (\velf)}{\domainf}
-\innerp{\testvelf}{\densityf\bodyff}{\domainf} 
-\innerpb{\testvelf}{\tractionfN}{\nboundf}
+\innerp{\testpf}{\div \velf}{\domainf},
\label{eq:w_fluid}\\
\mathcal{W}^{FS}\left[\left(\testdisps, \testvelf, \testpf\right),\left(\disps, \velf, \pf\right)\right] = 
\mathcal{W}^S\left[\testdisps,\disps\right] + \mathcal{W}^F\left[\left(\testvelf, \testpf\right),\left(\velf, \pf\right)\right] \nonumber\\
\underbrace{-\innerpb{\testdisps}{\InterfaceStressn}{\interface} +\innerpb{\testvelfempty}{\InterfaceStressn}{\interface}}_{:=\mathcal{W}^{FS,n}_{\interface}+\mathcal{W}^{FS,t}_{\interface}}.
\label{eq:w_fs}
\end{align}
Herein, $\left(\testdisps,\testvelf, \testpf\right)$ are the corresponding test functions of the primary unknowns $\left(\disps,\velf, \pf\right)$.
The discrete solution space is created by a spatial discretization consisting of elements containing piece-wise polynomials in an element reference coordinate system which are continuous on the inter-element boundaries.
For the pressure $\pf$ and for each component of the vector-valued quantities fluid velocity $\velf$ and solid displacement $\disps$ 
the discrete approximation space is directly constructed by these functions. 
All test functions are discretized by the same space as their corresponding primal unknowns. 
Modifications of these spaces for the incorporation of strong Dirichlet boundary conditions on $\refdbounds$ and $\dboundf$ are performed in the usual way.
For the structural displacements, an interface fitted discretization is applied, meaning that the elements fill the entire domain $\domains$.
Details on the discretization of the fluid domain, which is non-interface fitted, are given in Section~\ref{sec:cutfem}.

Including the unique interface traction $\InterfaceStressn$, which will be discussed in Sections  \ref{sec:num_interface_n} and \ref{sec:num_interface_t}, the respective dynamic equilibrium in normal direction \eqref{eq:contacttractioncond}, \eqref{eq:contactexclusioncond}, and \eqref{eq:InterfaceLinMom}, as well as in tangential direction \eqref{eq:c_dyneq_t}, and  \eqref{eq:fsi_dyneq_t} is incorporated directly into the weak form.
As the interface conditions \eqref{eq:gapcond}-\eqref{eq:fsi_navslip} require a separate treatment of normal and tangential constraints, 
the normal component $\normalInterfaceStress$ and the tangential component $\InterfaceStressn \cdot \Ptangent$ of the interface traction
$\InterfaceStressn = \normalInterfaceStress \cdot \normal + \InterfaceStressn \cdot \Ptangent$
are treated separately in Sections \ref{sec:num_interface_n} and \ref{sec:num_interface_t}.

To extend the interface contribution on $\fsiinterface$ arising from partial integration of the viscous and pressure contributions in domain $\domainf$ 
to the overall interface $\interface$, an additional definition of the fluid test functions $\left(\testvelfempty,\testpfempty\right)$ on the whole interface $\interface$ is consulted.
For the additional interface contributions in \eqref{eq:w_fs} vanishing fluid test functions outside of the fluid domain $\domainf$ are considered:
\begin{align}
\label{eq:ext_testfkt}
 \left(\testvelfempty,\testpfempty\right) =   
 \begin{cases}
  \left(\testvelf,\testpf\right) \quad \text{in} \quad \domainf\\
  0 \quad \text{otherwise}.
  \end{cases}
\end{align}

\subsubsection{Stabilization of the discrete fluid formulation}
In addition to the naturally arising terms of the fluid weak form \eqref{eq:w_fluid}, discrete stabilization operators have to be added
to control convective instabilities, to ensure discrete mass conservation, and to guarantee inf-sup stability for equal order interpolation of velocity and pressure:
\begin{align}
\mathcal{W}^F_{\mathcal{S}}\left[\left(\testvelf, \testpf\right),\left(\velf, \pf\right)\right] &= \mathcal{S}^F_v\left[\testvelf, \left(\velf,\pf\right)\right]+ \mathcal{S}^F_p\left[\testpf, \left(\velf,\pf\right)\right].
\label{eq:w_CIP}
\end{align}
Different realizations of these stabilization operators are possible, including residual-based stabilization and face-oriented stabilization.
In \cite{braack2007} a comparison of various techniques for stabilization of the incompressible flow problem is given.
For the presented numerical examples in Section \ref{sec:examples}, face-oriented stabilization operators are chosen (for details see \cite{massing2016}).
\begin{figure}[t]
\centering
\def\svgwidth{0.65\textwidth}
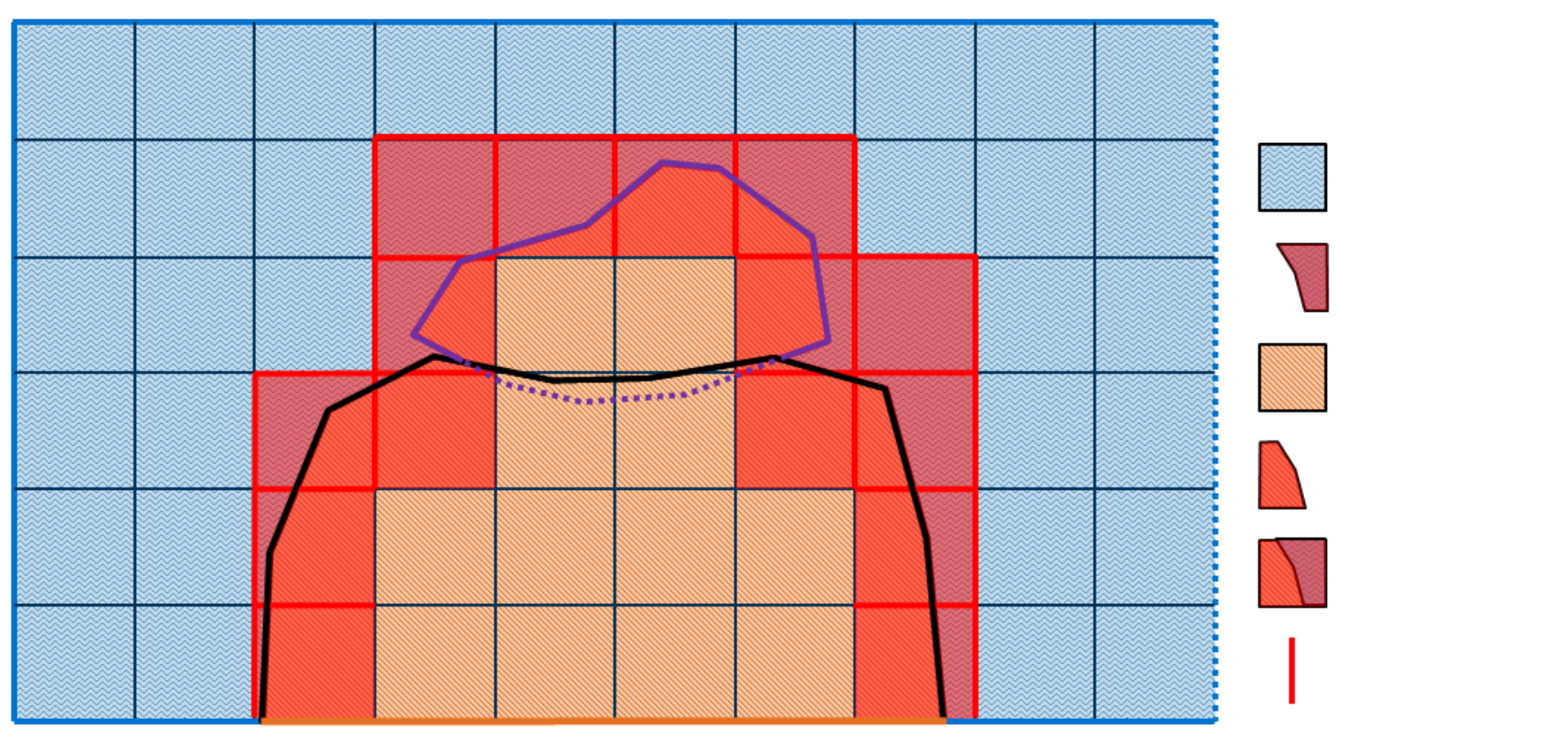
\caption{Basic problem setup for the applied CutFEM, structural domain $\domains = \domainsone \cup \domainstwo$ embedded in the fluid domain $\domainf$. 
A non-interface fitted discretization $\mathcal{T} = \mathcal{T}^F \cup \mathcal{T}_{\fsiinterface} \cup \mathcal{T}^0$
represents the fluid domain $\domainf$ by a set of elements in $\mathcal{T}^F$ and the physical sub-domain $\mathcal{T}^F_{\fsiinterface}$ of the elements in $\mathcal{T}_{\fsiinterface}$.
The non-physical domain, which equals the structural domain $\domains$, consists of a set of elements in $\mathcal{T}^0$ 
and the non-physical sub-domain $\mathcal{T}^0_{\fsiinterface}$ of the elements in $\mathcal{T}_{\fsiinterface}$.
For all inner element faces $\mathcal{F}_{\fsiinterface}$ of $\mathcal{T}^F \cup \mathcal{T}_{\fsiinterface}$, which are connected to one element in $\mathcal{T}_{\fsiinterface}$, the ``ghost penalty'' stabilization is applied.
}
\label{fig:cutfem}
\end{figure}

\subsubsection{The CutFEM utilized for discretization of the fluid domain $\domainf$}
\label{sec:cutfem}
As discussed in the introduction, the CutFEM is applied for the discretization of the fluid domain allowing for a fixed Eulerian computational mesh.
Herein, the boundaries and interfaces of the fluid domain are not required to match the boundary of the computational discretization.
This beneficial feature of the CutFEM allows the direct handling of large motion or deformation of the solid domain $\domains$ and even topological changes of the fluid domain $\domainf$
as it is typically occurring for FSCI problems.
The discretization concept is visualized for an exemplary contacting configuration in Figure~\ref{fig:cutfem}.
The typical small penetration of contacting solid bodies in the discrete solution is visualized exaggerated in this figure.
This aspect is left aside here and is discussed in detail in Section \ref{sec:clarify_interfaces}.

All solid domains $\domainsone$ and $\domainstwo$ are discretized boundary and interface matching.
The fluid discretization is specified to cover the entire fluid domain $\domainf$ and is not matching to the interface $\fsiinterface$. 
As shown by the exemplary configuration in Figure~\ref{fig:cutfem}, the outer boundaries of the fluid domain often match the discretization boundary, which does not necessarily have to be.
Then, the physical fluid domain $\domainf$ results from ``cutting out'' the non-fluid domain which is specified by the boundary of the solid domain $\partial \domains$ and potential non matching outer boundaries.

In the following, a brief overview on the most important aspects for application of the CutFEM to the FSCI problem is given.
The treatment of all interface conditions is not included here, but presented in Sections \ref{sec:num_interface_n} and \ref{sec:num_interface_t}.
A general overview of this method is given in \cite{burman2015cutfem} including references for further details.

The integration of the $\Ltwo$-inner products in the fluid weak form \eqref{eq:w_fluid} has to be performed solely in the physical fluid domain.
This domain is described by the outer fluid boundaries $\dboundf$ and $\nboundf$ as well as the deforming position of the interface $\fsiinterface$ including its solid outward unit normal vector $\normal$.
By separation of the fluid discretization, which is constant in time, in different sets of elements, the numerical integration of \eqref{eq:w_fluid} can be realized.
The computational fluid mesh consists of the sets of elements which are not intersected by the interface $\fsiinterface$ and affiliated to the fluid domain $\mathcal{T}^F$
or affiliated to the non-fluid domain $\mathcal{T}^0$. 
The set of all remaining elements $\mathcal{T}_{\fsiinterface}$ intersected by the interface $\fsiinterface$ is split into the physical fluid part $\mathcal{T}_{\fsiinterface}^F$ and the 
non-fluid part $\mathcal{T}_{\fsiinterface}^0$, which can be identified by the unit solid outward solid normal vector $\normal$.
For the ``non-intersected'' elements in $\mathcal{T}_{\fsiinterface}^F$ standard Gaussian quadrature is applied, whereas no integration has to be performed on elements in $\mathcal{T}_{\fsiinterface}^0$.
For the numerical integration of the physical fluid sub-domain $\mathcal{T}_{\fsiinterface}^F$ of the intersected elements, the method described in \cite{sudhakar2014}, where the divergence theorem is utilized repeatedly, is applied.
No integration has to be performed on the remaining sub-domain $\mathcal{T}_{\fsiinterface}^0$.

Due to the arbitrary relative position of the deformed interface $\fsiinterface$ and the fixed computational fluid mesh, 
any geometric intersection configuration has to be treated properly.
In fact, intersections leading to very small contributions of single discrete degrees of freedom to the weak form \eqref{eq:w_fluid} are critical.
If not handled appropriately, these configurations can lead to an ill-conditioned resulting system of equations or a loss of discrete stability arising from the weak incorporation of interface conditions presented in the 
Sections \ref{sec:num_interface_n} and \ref{sec:num_interface_t}.
These issues can be tackled by additional weakly consistent stabilization operators added to the weak form \eqref{eq:w_fluid}.
Therein, in principle, any non-smoothness of the discrete extension of the solution into the non-fluid domain $\mathcal{T}_{\fsiinterface}^0$ is penalized.
Single degrees of freedom with vanishing contribution in the weak form \eqref{eq:w_fluid} are then still defined by the smooth extension of the solution, even if there is no physical relevance left.
This kind of stabilization is called ``ghost penalty'' stabilization and was first presented in \cite{burman2010} for the Poisson's problem.
The method which is applied here for the stabilization of the fluid equations is analyzed in \cite{massing2016}.
The operators \eqref{eq:w_ghostpenalty} added to the fluid weak form therein penalizes jumps of normal derivatives of the velocity $\velf$ and the pressure $\pf$:
\begin{align}
\mathcal{W}^F_{\mathcal{G}}\left[\left(\testvelf, \testpf\right),\left(\velf, \pf\right)\right] =
\mathcal{G}_v\left(\testvelf,\velf\right)+\mathcal{G}_p\left(\testpf,\pf\right).
\label{eq:w_ghostpenalty}
\end{align}
These operators are integrated on a selected set of inner element faces $\mathcal{F}_{\fsiinterface}$ marked in Figure \ref{fig:cutfem} by red lines.

\begin{remark}[Existence of the discrete fluid test functions in the ghost domain]
\label{rem:discrete_fluidtest}
It should be highlighted that the discrete test functions $\left(\testvelf,\testpf\right)$, in contrary to definition \eqref{eq:ext_testfkt},
do not vanish in the ghost domain $\mathcal{T}^0_{\Gamma^{FS}}$, and are evaluated on the inter-element faces for the face-oriented stabilization and ``ghost penalty'' stabilization \eqref{eq:w_CIP} and \eqref{eq:w_ghostpenalty}
outside of the physical fluid domain $\domainf$.
\end{remark}

\subsubsection{Consistent fluid domain $\domainf$ and fluid-structure interface $\fsiinterface$ representation for the contacting case}
\label{sec:clarify_interfaces}
\begin{figure}[t]
\centering
\def\svgwidth{0.95\textwidth}
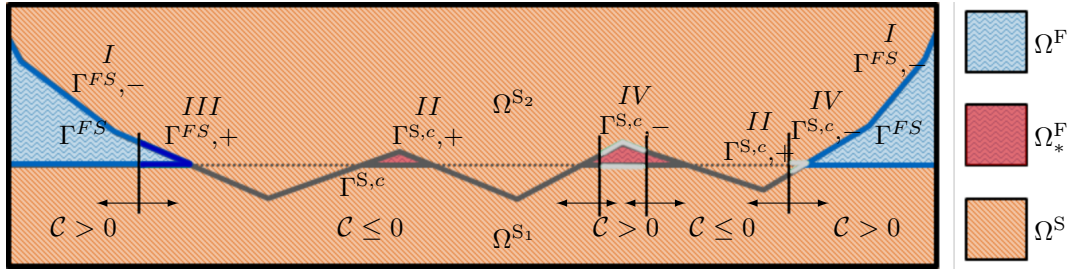
\caption{Detailed (exaggerated) view of the discrete contacting zone of two solid bodies $\domainsone$ and $\domainstwo$. Due to the discrete contact formulation, small fluid fractions $\domainf_*$ can emerge, which are not considered part 
of the fluid domain $\domainf$. The fluid-structure interface $\fsiinterface$ (blue line) is constructed accordingly to this fluid domain $\domainf$.
The remaining part of the interface $\interface$ is the contact interface $\sciinterface$. With the scalar value $\mathcal{C}$ introduced in Section \ref{sec:num_interface_n},
the interface is split into four cases ($I/\fsiinterface,-;II/\sciinterface,+;III/\fsiinterface,+;IV/\sciinterface,-$).
}
\label{fig:domain}
\end{figure}
The weak form \eqref{eq:w_fluid} is solely integrated in the physical domain $\domainf$. This domain is characterized by the non-moving outer boundaries $\dboundf$ and $\nboundf$ as well as the 
moving fluid-structure interface $\fsiinterface$.
The discrete motion of the interface $\fsiinterface$ is given by the general interface $\interface$ and hence by the motion of the solid domain $\domains$.
It is essential to evaluate the overall weak form on a consistent pair of domain $\domainf$ and interface $\fsiinterface$.
This aspect is straight-forward as long as no contact between solid bodies occurs, but should be discussed in detail for the case of contacting discrete bodies.
The contacting scenario illustrated in Figure \ref{fig:domain} results in partial overlap of both solid domain due to the discrete approximation.
Therefore, in a first step all parts of the interface $\interface$ which are overlapping - identified by the solid unit outward solid normal vector $\normal$ - are removed from the 
``intersection'' interface. The corresponding fluid domain $\domainf \cup \domainf_*$ potentially includes small fluid fractions occurring from the discrete contact formulation.
To avoid these ``islands'', the purely numerically occurring segments on the current ``intersection'' interface are removed additionally, leading to the consulted interface $\fsiinterface$.
For sufficiently spatially resolved computational meshes, the identification can be simply performed by a predefined maximal ratio of the element size
compared to the actual size of the bounding box containing a single fluid fraction.
Finally, the intersection of the computational fluid mesh is performed with this interface $\fsiinterface$, resulting in a physical fluid domain $\domainf$ which does not include the domain
$\domainf_*$.
The discrete contact interface is then defined by: $\sciinterface = \interface\setminus \fsiinterface$.

\subsection{Nitsche-based method on the overall coupling interface $\interface$ in normal direction}
\label{sec:num_interface_n}
The representative interface traction $\normalInterfaceStress = \InterfaceStressn \cdot \normal$ in normal direction needs to comply with all interface conditions \eqref{eq:gapcond}-\eqref{eq:fsi_noslipn}.
Defining the normal interface traction to:
\begin{equation}
\normalInterfaceStress=\mathrm{min}\left[ (\stressfnnext +\penfluidext\velrelnext)\;,\;(\normalAVstructStress + \pensolid\Gap) \right],
\label{eq:InterfaceStress}
\end{equation}
with two sufficiently large parameters $\penfluidext>0$ and $\pensolid>0$, allows the fulfillment of these conditions as discussed in the following.
The left-hand side of the minimum corresponds to enforcing the FSI conditions (\eqref{eq:contacttractioncond} in the case equal to zero in combination with \eqref{eq:fsi_noslipn}) and the right-hand side of the minimum enforces the contact no-penetration condition in normal direction (\eqref{eq:gapcond} in the case equal to zero in combination with \eqref{eq:InterfaceLinMom}). As a result, condition \eqref{eq:contactexclusioncond} is fulfilled automatically for both sides of the minimum.
If no feasible projection exists, we assume $\Gap\to\infty$ and as a result the FSI condition is enforced.

In the case that the contact no-penetration condition is active, the balance of linear momentum across the closed contact interface, in which condition \eqref{eq:InterfaceLinMom} reduces to $\stresssnn  = 
\proj{\stresssnn}$, is accommodated for by using the same representative solid stress  $\normalAVstructStress$ on both sides of the potential contact surfaces.
In the most simple case, one of the two potentially contacting solids, e.g.~$\domainsone$ is designated as a so-called slave side and the representative solid stress is chosen as the discrete stress representation of that side $\normalAVstructStress=\stresssnnone$.
An explicit choice of a slave side, however, results in an inherent bias between the two solid sides.
To obtain an unbiased formulation, an arbitrary convex combination $\normalAVstructStress=\structWeight\stresssnnone+(1-\structWeight)\stresssnntwo$ of the stress representations of the two solid sides can be used based on a weight $\structWeight\in [0,1]$.
If this weight is determined independently of the numbering of the contacting solids (i.e.~invariant with respect to flipping the slave and master side), the resulting algorithm is unbiased.
Two possible choices for unbiased method are either choosing $\structWeight=\frac{1}{2}$ \cite{Chouly2018,Mlika2017} or using harmonic weights determined based on material parameters and mesh sizes \cite{Burman2011,Seitz2018}.

In the case that the FSI condition is enforced, the normal interface traction is represented uniquely by the normal fluid traction $\stressfnn$.
Thus, the essential dynamic equilibrium \eqref{eq:contacttractioncond} in the case equal to zero and equilibrium \eqref{eq:InterfaceLinMom} due to vanishing contributions on both sides separately are fulfilled .
For this choice, a properly scaled, consistent penalty contribution $\penfluid\velreln$ is added to guarantee discrete stability of the formulation and to enforce the constraint \eqref{eq:fsi_noslipn}.
In addition to the resulting traction and penalty contribution, a skew-symmetric adjoint consistency term is added to the weak form \eqref{eq:w_fs}:
\begin{align}
\label{eq:num_adj_n}
\mathcal{W}^{\fluidletterr\structletterr,n}_{\interface,\text{Adj}}\left[\left(\testvelf, \testpf\right),\left(\disps,\velf\right)\right] = 
\innerpb{\testpfempty \normal - 2 \viscf \epsf(\testvelfempty) \normal}{\velrelnext \normal}{\interface}.
\end{align}
This term allows the direct balance of the contribution of the fluid pressure in addition to the viscous contribution, when introducing $\stressfnn$ in \eqref{eq:w_fs}.
Due to the inherent constraint \eqref{eq:fsi_noslipn}, this additional contribution does not alter the consistency of the formulation.
When enforcing the FSI conditions, also a representation of the interface traction by the corresponding solid stress would be possible, but is not considered in the following.

\paragraph*{A demonstration of the different resulting interface contributions}
To demonstrate the arising interface contributions from incorporation of the normal interface traction \eqref{eq:InterfaceStress} into the weak form \eqref{eq:w_fs}, the  boundary integral on the interface $\interface$ is split into the solid-solid contact ``$+$'' and the fluid-structure interaction ``$-$'' parts:
\begin{equation}
\begin{aligned}
&\innerpb{*}{*}{\interface,+} = 
  \begin{cases}
  \innerpb{*}{*}{\interface} \quad &\text{if} \quad \mathcal{C} \leq 0\\
  0 \quad &\text{otherwise}
  \end{cases}\;\;, \qquad
  \innerpb{*}{*}{\interface,-} = 
  \begin{cases}
  0 \quad &\text{if} \quad \mathcal{C} \leq 0\\
  \innerpb{*}{*}{\interface} \quad &\text{otherwise}
  \end{cases}\;\;,\\
 &\text{with} \quad \mathcal{C}\left(\disps,\velf, \pf\right)  = 
 \left(\normalAVstructStress + \pensolid \Gap\right)
  - \left(\stressfnnext + \penfluidext\velrelnext\right).
\end{aligned}
\label{eq:num_interfaces}
\end{equation}

\begin{remark}[Relation between the interfaces $\sciinterface$, $\fsiinterface$ and $\interface,+$, $\interface,-$]
\label{rem:discrete_interfaces}
For the continuous problem presented in Section \ref{sec:goveq}, integration on the interface subsets $\interface,+$ and $\interface,-$ coincidences with an integration on
the contact interface $\sciinterface$ and the fluid-structure interface $\fsiinterface$, respectively.
Due to the discrete error this relation does not hold for the discrete formulation, where in general a deviation between these interfaces will occur.
\end{remark}
In definition \eqref{eq:num_interfaces}, the sign of the scalar $\mathcal{C}$ indicates, which side of the $\mathrm{min}[]$ function in \eqref{eq:InterfaceStress} represents the normal interface traction.
In addition to this split of interface $\interface$ in the ``$+$'' and ``$-$'' parts, a purely geometric split into interfaces $\fsiinterface$ and $\sciinterface$ was described in Section \ref{sec:clarify_interfaces}.
As the interface $\fsiinterface$ is part of the outer fluid boundary $\partial \domainf$, the fluid state $(\velf,\pf)$ and the corresponding test functions $(\testvelf,\testpf)$ are directly defined
on this interface without any extension required. 
Combining these two different subdivisions when performing the integration of the normal traction \eqref{eq:w_fs} on the interface $\interface$, 
leads to four cases ($I-IV$) which finally needs to be dealt with:
\begin{equation}
\begin{aligned}
&\innerpb{\testvelfempty-\testdisps}{\normalInterfaceStress \normal}{\interface}=
\innerpb{\testvelfempty-\testdisps}{\normalInterfaceStress \normal}{\interface,+}+\innerpb{\testvelfempty-\testdisps}{\normalInterfaceStress \normal}{\interface,-}=\\
&\underbrace{\innerpb{\testvelfempty-\testdisps}{\normalInterfaceStress \normal}{\sciinterface,+}}_{\text{case } II}
+\underbrace{\innerpb{\testvelfempty-\testdisps}{\normalInterfaceStress \normal}{\sciinterface,-}}_{\text{case } IV}\\
+&\underbrace{\innerpb{\testvelfempty-\testdisps}{\normalInterfaceStress \normal}{\fsiinterface,+}}_{\text{case } III}
+\underbrace{\innerpb{\testvelfempty-\testdisps}{\normalInterfaceStress \normal}{\fsiinterface,-}}_{\text{case } I}.
\end{aligned}
\label{eq:num_cases}
\end{equation}
A visualization of these four cases for a specific discrete contact configuration is given in Figure \ref{fig:domain}.
In the following, the resulting contributions, which have to be evaluated, are depicted.
Vanishing contributions are not included and the extension operator is just applied in the case no 
direct representation of the corresponding quantity is available on the relevant segment.
Further, the skew-symmetric adjoint consistency term introduced in \eqref{eq:num_adj_n} is included, to include all interface contributions evaluated in the normal direction:
\begin{align}
\mathcal{W}^{FS,n}_{\interface}&\left[\left(\testdisps, \testvelf, \testpf\right),\left(\disps, \velf, \pf\right)\right] + \mathcal{W}^{\fluidletterr\structletterr,n}_{\interface,\text{Adj}}\left[\left(\testvelf, \testpf\right),\left(\disps,\velf\right)\right] = I+II+III+IV \qquad \text{with:}\nonumber\\
I &= \innerpb{\testvelf-\testdisps}{\stressfnn \normal}{\fsiinterface,-}
+\innerpb{\testvelf-\testdisps}{\penfluid\velreln \normal}{\fsiinterface,-}\nonumber\\
&+\innerpb{\testpf \normal - 2 \viscf \epsf(\testvelf) \normal}{\velreln \normal}{\fsiinterface,-},\label{eq:num_caseI}\\
II &= \innerpb{-\testdisps}{\normalAVstructStress \normal}{\sciinterface,+} + \innerpb{-\testdisps}{\pensolid\Gap \normal}{\sciinterface,+},\label{eq:num_caseII}\\
III &= \innerpb{\testvelf-\testdisps}{\normalAVstructStress \normal}{\fsiinterface,+} + \innerpb{\testvelf-\testdisps}{\pensolid\Gap \normal}{\fsiinterface,+},\label{eq:num_caseIII}\\
IV &= \innerpb{-\testdisps}{\stressfnnext \normal}{\sciinterface,-}+\innerpb{-\testdisps}{\penfluidext\velrelnext \normal}{\sciinterface,-}.\label{eq:num_caseIV}
\end{align}
Herein, contribution $I$ equals the classical Nitsche-based method for the imposition of the mass conservation on an fluid-structure interface as applied in \cite{burman2014,schott2017}.
This method includes an interface traction representation by the fluid stress, a penalty term which is consistent due to the includes mass conservation \eqref{eq:fsi_noslipn},
and the skew symmetric viscous and pressure adjoint consistency term which also includes \eqref{eq:fsi_noslipn}.

The evaluated terms in summand $II$ coincide in principle with Nitsche-based methods for classical contact problems, e.g.\ applied in \cite{Mlika2017,Chouly2018,Seitz2018}.
Here, the interface traction is represented by a one-sided or two-sided weighted solid stress of both contacting bodies with an additional penalty term including the no penetration condition included in \eqref{eq:gapcond}
and \eqref{eq:contactexclusioncond}. No adjoint consistency terms are applied. Due to the vanishing fluid test functions $(\testvelfempty,\testpfempty)$, no contribution to the fluid weak form occurs.

Finally, contributions $III$ and $IV$ arise solely close to the condition changing point/curve $\mathcal{C}=0$ and the common point/curve of both interface $\sciinterface\cap\fsiinterface$.
The impact of these summands compared to contributions $I$ and $II$ is relatively small and so making use of a simple extension of the fluid quantities in \eqref{eq:num_caseIV} is acceptable.
Still, both contribution have to be applied to ensure a continuous discrete problem and guarantee geometrically fitting interface conditions applied onto the fluid domain.
\begin{remark}[Application of a different representation for contribution $III$]
\label{rem:discrete_interfaces}
For all numerical examples presented in Section \ref{sec:examples}, an alternative formulation of contribution $III$ is applied due to algorithmic reasons.
Therein, the contributions of Nitsche contact $II$ are completed by a fluid-sided interface traction representation for the fluid domain.
\begin{align}
III = 
&\innerpb{\testvelf}{\stressfnn \normal}{\fsiinterface,+}
+\innerpb{\testvelf}{\penfluid\velreln \normal}{\fsiinterface,+}
+\innerpb{\testpf \normal - 2 \viscf \epsf(\testvelf) \normal}{\velreln \normal}{\fsiinterface,+}\nonumber\\
+&\innerpb{-\testdisps}{\normalAVstructStress \normal}{\fsiinterface,+} + \innerpb{-\testdisps}{\pensolid\Gap \normal}{\fsiinterface,+} 
\label{eq:num_caseIII_alt}
\end{align}
By comparison of contributions \eqref{eq:num_caseIII} and \eqref{eq:num_caseIII_alt}, the coincidence of both formulations at the condition-changing point $\mathcal{C}=0$ and for fulfilled 
mass balance \eqref{eq:fsi_noslipn} can be directly seen.
As the impact of contribution $III$ is generally small and arises solely close to $\mathcal{C}=0$, this modification does not have a significant influence onto the performance of the presented formulation.
\end{remark}

\begin{remark}[Determination of the solid penalty parameter $\pensolid$]
\label{rem:solidpen}
For Nitsche's method, the parameter $\pensolid= \pensolid_0 \penscalingsolid$ with a sufficiently large, positive constant $\pensolid_0$ is required to establish discrete stability of the formulation.
Therein, material- and mesh-dependencies of $\pensolid$ are considered in $\penscalingsolid$ by a local generalized eigenvalue problem as presented e.g.\ in \cite{Seitz2018}.
Larger values of $\pensolid_0$ improve the constraint enforcement $(\Gap = 0)$, while smaller values typically reduce the error of the consistent stress representation $(\normalAVstructStress)$.
For the FSCI problem, additional aspects have to be considered. 
The influence of case $IV$ (see \eqref{eq:num_caseIV}) should be minimized, as it incorporates the extended fluid solution and switching between the cases $II$ and $IV$ during the nonlinear solution procedure should be reduced.
A small penalty parameter $\pensolid$ supports this behavior as it turns out to reduce the ratio of $\mathcal{C}\leq 0$ on the interface $\sciinterface$.
As a result, a small but still numerical stable constant $\pensolid_0$ is beneficial for solving the FSCI problem.
This aspect is not critical as the constant $\pensolid_0$ is not problem dependent for a properly defined scaling $\penscalingsolid$ and the same value can be kept for all computation  ($\pensolid_0=1.0$ for all presented numerical examples).
\end{remark}

\begin{remark}[Determination of the fluid penalty parameter $\penfluid$]
\label{rem:fluidpen}
The penalty term in \eqref{eq:InterfaceStress} with the parameter $\penfluid  = \penfluid_0 \penscalingfluidn$ balances viscous, convective and temporal contributions according to \cite{massing2016} and so enables a discrete stable formulation.
Therein, $\penfluid_0$ is a sufficiently large positive constant, $h_{\Gamma}$ an appropriate element volume to interface area ratio, and $\phi^F_{v}$ a scaling 
taking into account the different flow regimes.
For the determination of the constant $\penfluid_0$, constraint enforcement as well as the resulting interface stress error is important.
Additionally, for the computed numerical examples, it was observed, that a small penalty parameter $\penfluid_0$ is beneficial for the FSCI problem as it incorporates an inherent relaxation of the kinematic constraints especially close to the point of changing conditions ($\mathcal{C}= 0$) and hence improves the performance of the nonlinear solution procedure. 
The relevance of this aspect depends highly on the complexity of the considered problem configuration and increases for a reduced accuracy of the applied numerical integration procedure on the interface $\Gamma$.
\end{remark}

\begin{remark}[Applied numerical integration procedure on the interface $\interface$]
\label{rem:interface_int}
For the numerical integration of the contributions \eqref{eq:num_caseI}-\eqref{eq:num_caseIII_alt} on the interface,
non-smooth and non-continuous functions on single solid boundary elements have to be integrated. 
These kinks and jumps potentially occur due to element boundaries of the contact partner or on the intersection of the interface with fluid element boundaries.
To enable an accurate numerical integration, each solid boundary element has to be split by all other element boundaries involved and a numerical integration rule has to be specified, e.g.\ by triangulation, on these segments (see e.g. \cite{farah2015}).
For the numerical examples presented in the following, this most accurate approach was not applied. 
Instead, the numerical integration points on the interface $\Gamma$ are constructed to account for the intersection of the interface with fluid element boundaries solely. To account for the integration of the discontinuous solid stress in the contacting case, an increased number of integration points is applied.
\end{remark}

\subsection{Nitsche-based method on the overall coupling interface $\interface$ in tangential direction}
\label{sec:num_interface_t}
The tangential component of the interface traction $\InterfaceStressn$, has to fulfill the traction free condition \eqref{eq:c_dyneq_t} due to the consideration of frictionless contact
on the contact interface $\sciinterface$:
\begin{align}
\InterfaceStressn \cdot \Ptangent &= \zerovec \quad \text{on} \quad \sciinterface.
\label{eq:num_itraction_tc}
\end{align}
Further, the dynamic equilibrium \eqref{eq:fsi_dyneq_t} and the Navier slip boundary condition \eqref{eq:fsi_navslip} have to be fulfilled on the fluid-structure interface $\fsiinterface$.
A representation of the unique tangential interface traction by:
\begin{align}
\InterfaceStressn \cdot \Ptangent &= \left[-\frac{(\penfluid_{t,\mathrm{0}})^{-1} h_{\Gamma}}{\kappa\viscf+(\penfluid_{t,\mathrm{0}})^{-1} h_{\Gamma}} \stressf  \cdot \normalf 
+ \frac{\viscf}{\kappa\viscf+(\penfluid_{t,\mathrm{0}})^{-1} h_{\Gamma}}\left(\velf - \vels\right)\right]\cdot \Ptangent \quad \text{on} \quad \fsiinterface,
\label{eq:num_itraction_tfsi}
\end{align}
complies with these condition. For the limit cases no-slip (slip length $\sliplengh = 0$) and free-slip (slip length $\sliplengh = \infty$), the tangential interface traction reduces to 
$\InterfaceStressn \cdot \Ptangent = \left[-\stressf  \cdot \normalf 
+ \frac{\viscf}{(\penfluid_{t,\mathrm{0}})^{-1} h_{\Gamma}}\left(\velf - \vels\right)\right]\cdot \Ptangent$
and $\InterfaceStressn \cdot \Ptangent = \zerovec$, respectively.
Incorporating of the tangential interface traction in the weak form \eqref{eq:w_fs} and adding an additional consistent 
skew-symmetric adjoint term results in the contributions:
\begin{align}
&\mathcal{W}^{FS,t}_{\interface}\left[\left(\testdisps,\testvelf\right),\left(\disps,\velf\right)\right]=
-\innerpb{\testvelf-\testdisps}{ \stressf  \cdot \normalf \cdot \Ptangent}{\fsiinterface}\nonumber\\
 &+\frac{\viscf}{\kappa\viscf+(\penfluid_{t,\mathrm{0}})^{-1} h_{\Gamma}}\innerpb{\testvelf - \testdisps}
 {\left[\velf - \vels + {\kappa \stressf \cdot \normalf} \right]\cdot \Ptangent}{\fsiinterface}
 \label{eq:w_fsi_t_nit}\\
 &\mathcal{W}^{FS,t}_{\interface,\text{Adj}}\left[\testvelf,\left(\disps,\velf\right)\right]=\nonumber\\
 &-\frac{(\penfluid_{t,\mathrm{0}})^{-1} h_{\Gamma}}{\kappa\viscf+(\penfluid_{t,\mathrm{0}})^{-1} h_{\Gamma}}\innerpb{-2 \viscf \epsf (\testvelf) \cdot \normalf }
 {\left[\velf - \vels + {\kappa \stressf \cdot \normalf} \right]\cdot \Ptangent}{\fsiinterface} \label{eq:w_fsi_t_nit_adj}.
\end{align}
It can be directly seen that this formulation is consistent,
as the term in the first line includes the naturally arising fluid stress applied on fluid and solid boundary due to the balance \eqref{eq:fsi_dyneq_t} 
and the additional terms include directly the constraint \eqref{eq:fsi_navslip}.
Theses additional terms are present to guarantee a discrete stable formulation and to enforce the kinematic constraint.
They balance the destabilizing effects of the viscous boundary integral occurring in line one and the term of similar structure in line two.
The penalty parameter in tangential direction $\penfluid_{t,\mathrm{0}}$ needs to be a positive and sufficiently large constant.
This Nitsche-based contribution for the general Navier interface condition is based on the formulation presented and analyzed in \cite{Juntunen2009} for the Poisson problem
and \cite{winter2017} for the linearized fluid problem. 
It was successfully applied to enforce the coupling conditions between a poroelastic structure and fluid flow in \cite{ager2018,ager2018b}.

\paragraph*{Definition of the slip length $\sliplengh$}
As motivated already in Section \ref{sec:interfacet} for the overall problem, the no-slip interface condition $\sliplengh = 0$ on $\fsiinterface$ has to be applied.
Solely close to the contacting zone, a relaxation of this constraint is designated.
A continuous transition between the tangential fluid-structure interaction condition \eqref{eq:fsi_dyneq_t}-\eqref{eq:fsi_navslip} and the tangential frictionless contact condition \eqref{eq:c_dyneq_t}
can be guaranteed for an infinite slip length $\sliplengh = \infty$ on $\fsiinterface\cap\sciinterface$.
The applied interpolation between these limiting points is given by
\begin{align}
\label{eq:def_sliplength}
\sliplengh =
   \begin{cases}
    0 \quad &\text{if} \quad \Gap > h\\
    \sliplengh_0 h \left[\frac{h}{\Gap}-1\right] \quad &\text{if} \quad h \geq \Gap > 0.\\
    \infty \quad &\text{otherwise}
  \end{cases}
\end{align}
Herein, the minimal value of the gap $\Gap$ between two solid interfaces to apply the no-slip interface condition is specified by the fluid element size $h$.
The interpolation function can be specified by the constant reference slip length $\sliplengh_0$.
It should be pointed out that for a reduction of the fluid element size $h$, also the range of influence for this modification compared to a pure no-slip condition is reduced.
For small scales, an alternative formulation for the slip length in relation \eqref{eq:def_sliplength} due to the underlying physical slip can improve the accuracy of the 
interface condition.
A second advantage of allowing a certain amount of slip on the interface is to avoid blockage of single fluid elements between approaching surfaces due to the insufficient discrete solution space.
This aspect is less essential for a weak enforcement of the interface condition by Nitsche's method with an appropriately chosen penalty parameter $\penfluid_{t,\mathrm{0}}$ 
than for a strong enforcement of interface conditions (see \cite{Bazilevs2007} for a comparison of strong enforcement and weak imposition of fluid boundary conditions).
To give an example, in \cite{loon2006} this issue is resolved by a modification of the fluid-structure interface constraint close to contact.

\subsection{Overall formulation for the coupled FSCI problem}
\label{sec:num_globalsystem}
Finally, by making use of the corresponding interface traction representation in normal \eqref{eq:InterfaceStress} and in tangential \eqref{eq:num_itraction_tc} - \eqref{eq:num_itraction_tfsi} direction in the weak form \eqref{eq:w_fs} of the overall coupled problem,
and summing up all additional contributions including interface adjoint consistency terms and discrete stabilization contributions,
the following semi-discrete weak form of the FSCI problem has to be solved. 
Find $\left(\disps,\velf, \pf\right)$ such that $\forall \left(\testdisps,\testvelf, \testpf\right)$:
\begin{equation}
\begin{aligned}
&\mathcal{W}^{FS}\left[\left(\testdisps, \testvelf, \testpf\right),\left(\disps, \velf, \pf\right)\right]
+\mathcal{W}^F_{\mathcal{S}}\left[\left(\testvelf, \testpf\right),\left(\velf, \pf\right)\right]
+\mathcal{W}^F_{\mathcal{G}}\left[\left(\testvelf, \testpf\right),\left(\velf, \pf\right)\right]\\
+&\mathcal{W}^{\fluidletterr\structletterr,n}_{\interface,\text{Adj}}\left[\left(\testvelf, \testpf\right),\left(\disps,\velf\right)\right]
+\mathcal{W}^{\fluidletterr\structletterr,t}_{\interface,\text{Adj}}\left[\testvelf,\left(\disps,\velf\right)\right]
= 0.
\end{aligned}
\label{eq:num_wfinal}
\end{equation}
For discretization of the weak form \eqref{eq:num_wfinal} in time, the one-step-$\theta$ scheme is applied to the occurring time derivatives in solid \eqref{eq:w_solid} and 
fluid weak form \eqref{eq:w_fluid} with an equal time integration factor $\theta$.
This procedure leads to a nonlinear system of equations of form $\tns{\mathcal{R}}_{n} = \zerovec$ for each discrete instance in time (index $n$) in the interval $[\btime,\etime]$.
An iterative solution scheme based on a Newton-Raphson procedure is applied to solve this nonlinear system:
\begin{align}
\label{eq:linearizations}
\tns{C}_{n}^i\cdot \Delta  \disctns x_{n}^{i+1} = -\tns{\mathcal{R}}_{n}^i, \quad
\disctns x_{n}^{i+1} = \disctns x_{n}^{i} + \Delta  \disctns x_{n}^{i}, \quad
\tns{C}_{n} \approx \partiald{\tns{\mathcal{R}}_{n}}{\disctns x_{n}}.
\end{align}
Herein, all equations arising from the overall weak form \eqref{eq:num_wfinal} are included in $\tns{\mathcal{R}}_{n}$ and
and all unknowns $\disctns x_{n} =  \left[\disctns \displacements_{n}, \disctns \velocityf_{n}, \disctns \pressuref_{n}\right]$ (including the structural displacement, fluid velocity and pressure) are solved and updated in every iteration step simultaneously.
The matrix $\tns C_{n}$ includes the essential linearizations of the residual vector $\tns{\mathcal{R}}_{n}$ with respect to the unknowns $\disctns x_{n}$.
For a sufficiently small value of a residual norm $||\tns{\mathcal{R}}^i_n||<\epsilon$, the current iteration state
approximates the solution state for this timestep $\disctns x_{n}^{i+1} = \disctns x_{n}$.
Based on the previously computed solution state, the solution at the next discrete instance in time is computed by another Newton-Raphson iteration procedure.
It should be highlighted, that the discrete solution space of the fluid pressure and velocity potentially changes between different iteration steps or time steps.
As a consequence, a procedure to transfer the previously evaluated solution state to the current solution space is required for the nonlinear solution procedure and the discrete time integration.
Details on this procedure and the overall solution algorithm for the CutFEM-based fluid-structure interaction, which builds the basis for the algorithm applied herein, are presented in \cite{schott2017}.
From an algorithmic point of view, solely the evaluation of different contributions on the interface varies from the presentation therein.

In the following, we present algorithmic details for the monolithic solution algorithm, applied for the computation of the subsequently presented numerical examples.
The configurations of the presented examples (except for the first validation example) are chosen to challenge the presented FSCI formulation.
Thus, it is ensured that the fluid-structure coupling, the solid-solid contact, as well as the change between these conditions have an essential impact to the overall problem.
To guarantee a strong interaction between the structures and the fluid, equal initial densities are considered within all domains. Specific strategies to enable a robust solution of the resulting highly nonlinear problems are discussed in the following.

\subsubsection{Damping strategy for the update procedure}
\label{sec:ex_damping}
A simple damped update procedure of the vector of unknowns $\disctns x_{n}^{i+1} = \disctns x_{n}^{i} + \omega^{i}_n \Delta  \disctns x_{n}^{i}$ in \eqref{eq:linearizations}
with parameter $\omega^{i}_{n}=[0.0,1.0]$ turned out to be beneficial for the convergence behavior.
The damping parameter at the initial iteration of each timestep step is set to $\omega^{0}_n = 1.0$. 
Based on the relative change of the residual norm $||\tns{\mathcal{R}}^{i}_n||/||\tns{\mathcal{R}}^{i-1}_n||$ between single iterations, the damping parameter is reduced for an increasing residual norm and vice versa.

\subsubsection{Update strategy for geometric intersection}
\label{sec:ex_ffg}
A simple procedure to avoid deterioration of the convergence behavior in the Newton-Raphson procedure due to 
``algorithmic''-discontinuities arising from geometric tolerances in the algorithm intersecting the interface $\fsiinterface$
and the computational fluid domain $\domainf$ is applied. 
Herein, the geometric intersection 
(includes the creation of numerical integration points in the physical fluid domain $\domainf$ and on the fluid-structure interface $\fsiinterface$)
is just updated as long as the maximal displacement increment $||\Delta \disctns \displacements^{i}_n||_{\infty}>\epsilon_{\text{geom}}$ in an iterations step exceeds a specified valued.
For the remaining iteration steps, the intersection information of the previous iteration step is applied.

\subsubsection{Solution space update strategy}
\label{sec:ex_fds}
As explained in detail in \cite{schott2017}, the solution space is updated dynamically within the iterative solution procedure for solving the system of nonlinear equations.
For classical FSI computations without structural contact, applying this procedure typically results in a constant solution space after few iterations.
Nevertheless, including contact increases the sensitivity of the formulation with respect to changing solution spaces.
This aspect can result in periodically repeating changes of the solution space within the iterative solution procedure for specific geometric configurations and so prohibits the convergence of the scheme.
The reason for this behavior is a discontinuity in the discrete formulation,
which arises due to the change in the considered set of faces in the weakly consistent ``ghost penalty'' stabilization \eqref{eq:w_ghostpenalty}, when changing the discrete fluid solution spaces.
The influence of this effect onto the convergence of the Newton-Raphson based procedure is especially relevant in the case when two physical fluid domains are merged or separated.
Thus, when exceeding a maximum number of iterations in the nonlinear solution procedure,
no reduction in the computational nodes carrying fluid degrees of freedom is performed anymore.
The fluid solution space is then just enlarged within the actual timestep.
To retain a solvable system of equations, the ``ghost penalty'' stabilization has to include the faces connecting all additional degrees of freedom to the physical domain.
This strategy leads to a constant set of faces considered for the stabilization during the Newton-Raphson based procedure and as a result avoids the occurring discontinuity in the discrete formulation.
With this modification, the consistency of the formulation is not touched.
Only some additional fluid degrees of freedom, which represent an extension of the solution in the non-physical domain, are appended to the system.
To ease the use of this strategy, the discrete fluid solution space is constructed by a maximum of one set of fluid unknowns on each node in the presented computations in the following.
As long as no slender solid bodies are considered, this restriction still results in an appropriate discrete fluid solution space (for more details on multiple sets of fluid unknowns on single computational nodes see \cite{schott2014}).

\subsubsection{Applied extension operator}
\label{sec:num_extrapol_operator}
In Section \ref{sec:interfacen} and Section \ref{sec:num_interface_n}, an extension operator $\extensionsym{\coord}$ is required to extend the scalar fluid quantities, normal stress, relative normal velocity,
and fluid Nitsche penalty parameter from the fluid-structure interface to the contact interface $\fsiinterface \longrightarrow \sciinterface$.
In the numerical examples presented, a very simple operator is applied. 
Herein, starting from the coordinate $\coord$ on interface $\sciinterface$, the closest point $\coord_\extensionsymbol$ to $\Gamma^{FS}\cap\Gamma^c$ is computed.
In this point, the scalar quantity is evaluated.
Then a constant extension is applied and as a result the computed value of the scalar quantity equals the extension.

\begin{remark}[Alternatives to the proposed extension strategy for fluid filled contact zones]
\label{rem:extension_phys1}
This kind of extension includes the modeling assumption that the contacting zone is filled with fluid.
As long as the influence of this extension on the computational model is limited to the neighborhood of the fluid domain, 
this approximation of the fluid solution seems sufficient.
If a better fluid solution on the contact interface is required, a physical model has to be solved to avoid the extension.
Depending on the requirements for this solution, potential models are based on the Reynolds equation \cite{reynolds1886} or a poroelastic layer \cite{ager2018}.

\end{remark}
\begin{remark}[Alternative to the proposed extension strategy for vanishing fluid in the contact zones]
\label{rem:extension_phys2}
If vanishing fluid in the contact zone is modeled, a continuous extension from the physical fluid solution to a vanishing fluid solution (zero ambient pressure) depending on the distance to the fluid domain can be applied alternatively.
When making use of this approach, it has to be guaranteed that gaps emerging from opening contact in this zone of vanishing fluid solution are not considered as part of the fluid domain $\domainf$ to avoid a non-physical model.
Such a configuration equals classical structural contact mechanics and therefore is not considered in the following.
\end{remark}

\section{Numerical examples}
\label{sec:examples}
In the following section, three numerical examples with focus on different aspects of computationally solving FSCI problems are presented. To start with, the falling, contacting, and lifting of a rounded stamp is analyzed to verify  the principal processes present in all FSCI configurations. The examination of an elastic pump proves the applicability of the framework to handle topological changes of the fluid domain including significantly different fluid solutions between the separated domains. Finally, a flow-driven squeezed elastic structure is analyzed, which includes highly dynamic mechanisms, large contact areas, and numerous contacting and lift-off processes. 
For all examples presented in this section, four-noded bi-linear quadrilateral elements are applied for the spatial discretization of all solid domains and fluid domains.

\subsection{Falling, contacting, and lifting of a rounded stamp}
\label{sec:ex1}
The first presented numerical example, a simple configuration including the falling, contacting, and lifting of a rounded stamp, is considered to analyze basic properties of the presented formulation.
Due to the symmetry of this configuration, just the half rounded stamp $\domainstwo$ and fluid domain $\domainf$ are considered.

\paragraph{Problem description}
\begin{figure}[htbp]
\hspace{-1.7cm}
\begin{minipage}[hbt]{0.5\textwidth}
\input{fig/ex1}
\end{minipage}
\hspace{1.2cm}
\begin{minipage}[hbt]{0.5\textwidth}
\raggedleft
\def\figscaling{0.5}
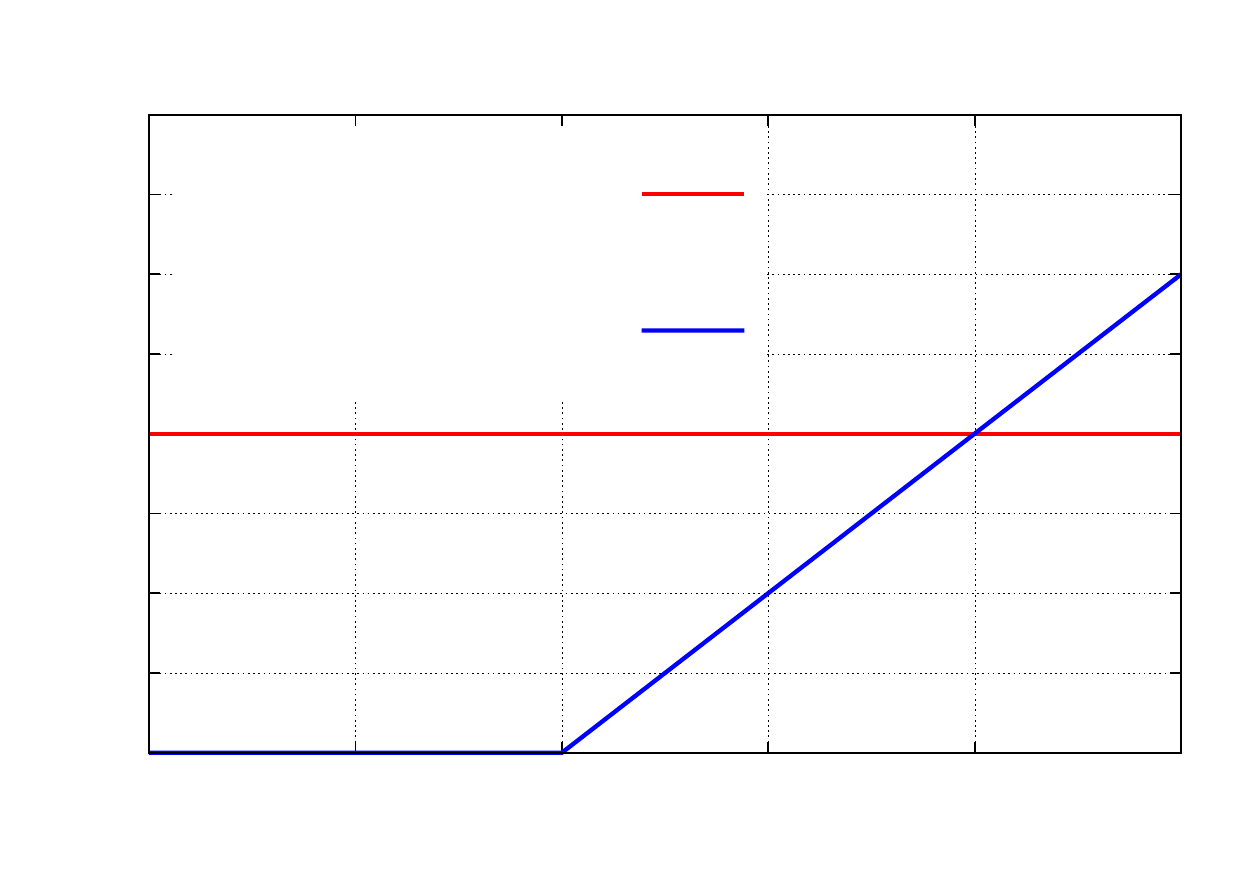
\end{minipage}
\caption{Geometry and boundary conditions for the falling, contacting, and lifting of a rounded stamp. Due to the symmetry of the configuration,  only the part with $x\leq0$ is considered (left). The prescribed, time-dependent external loads $\reftractionsN$ and $\tractionfN$ are given in the diagram (right).}
\label{fig:ex1_setup}
\end{figure}

The geometric setup and basic boundary conditions are visualized in Figure \ref{fig:ex1_setup}.
The solid domain $\domainsone$ is rigid and fixed in space by a Dirichlet boundary condition on the overall domain.
In the initial phase, the stamp is exposed solely to a prescribed constant-in-time Neumann load on the boundary $\nbounds$
in negative $y$-direction
(see Figure \ref{fig:ex1_setup} (right)),  which induces the falling motion.
After a certain time, contact between the solid domains $\domainsone$ and $\domainstwo$ will occur and a stationary state will be established subsequently.
Finally, after $t=1000$ a Neumann fluid load is prescribed in the normal direction of the boundary $\nboundf$. 
This load increases linearly in time as indicated in Figure \ref{fig:ex1_setup} (right). 
The fluid material parameters are specified as density $\densityf=10^{-3}$ and dynamic viscosity $\viscf = 1.0$.
The solid density in the undeformed configuration is equal to the fluid density $\refdensitys=10^{-3}$.
A Neo-Hookean model with the hyperelastic strain energy function 
\begin{align}
\label{ex1:strainenergysp}
\strainenergys = c \left[\text{tr}\left(\left(\defgrads\right)^T\cdot\defgrads\right)-3\right]+\frac{c}{\beta}\left(\left(\Js\right)^{-2 \beta}-1\right), \quad
c = \frac{E}{4(1+\nu)}, \quad \beta = \frac{\nu}{1-2 \nu}
\end{align}
describes the material behavior of the solid domain $\domainstwo$, with with Young's modulus $E = 100$ and Poisson's ratio $\nu = 0.0$.
To analyze the presented formulation, two different spatial resolutions are applied.
For the ``coarse'' variant, the fluid mesh consists of $16 \times 24 = 384$ elements and the solid mesh of domain $\domainstwo$ is created by $400$ elements.
In the ``fine'' variant, $64 \times 96 = 6144$ fluid elements and $6400$ elastic solid elements are used.
The weighting of the solid contact stress is purley based on the domain $\domainstwo$ due to the rigid domain $\domainsone$.
The reference slip length is set to $\sliplengh_0=0.1$ for all compuations including the Navier slip condition.
The discretization in time is performed with the Backward Euler scheme ($\theta = 1.0$), 
with three different sizes of the timestep ($\Delta t = 0.01$ for $t\in\left[0,20\right]$, $\Delta t = 0.2$ for $t\in\left[20,420\right]$, $\Delta t = 2.0$ for $t\in\left[420,2500\right]$)
to account for the varying dynamic of the analyzed system.
\begin{figure}[htbp]
\hspace{-0.02\textwidth}
\begin{minipage}[hbt]{0.5\textwidth}
\centering
\def\figscaling{0.65}
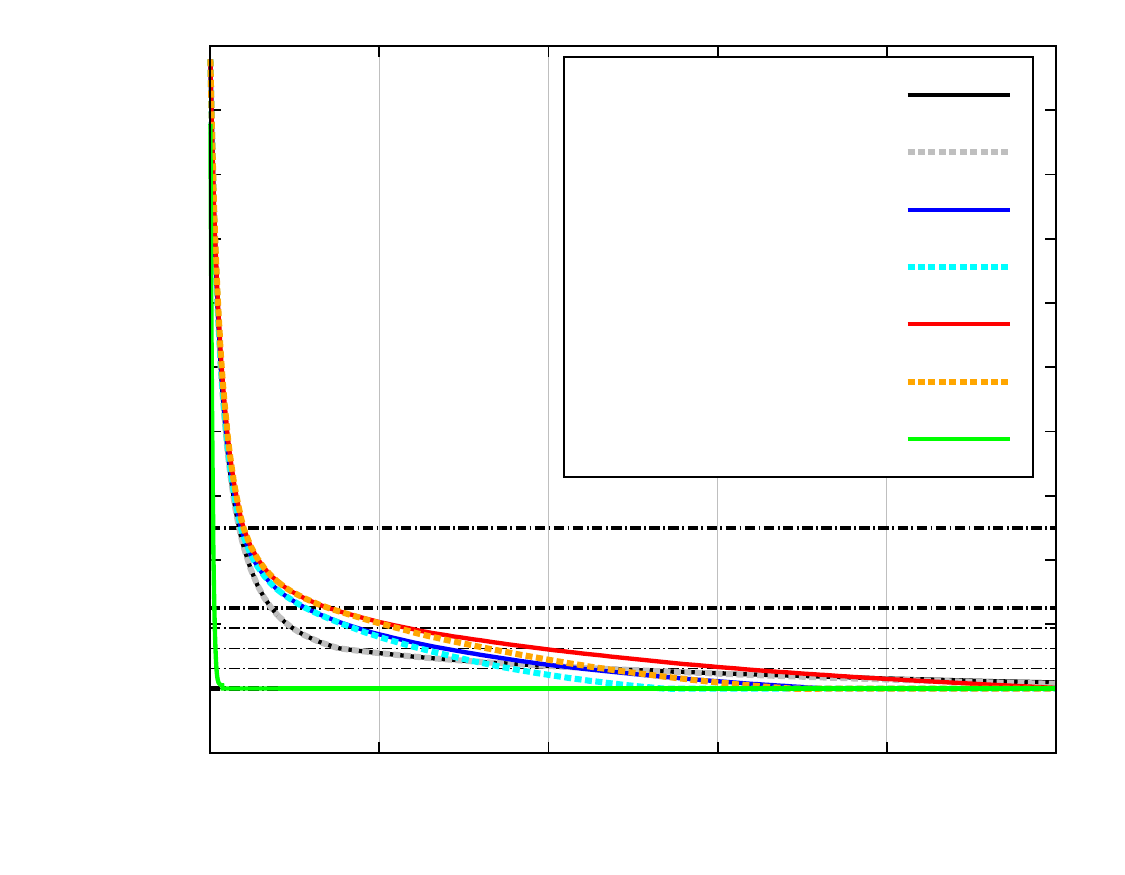
\end{minipage}
\begin{minipage}[htbp]{0.5\textwidth}
\def\figscaling{0.65}
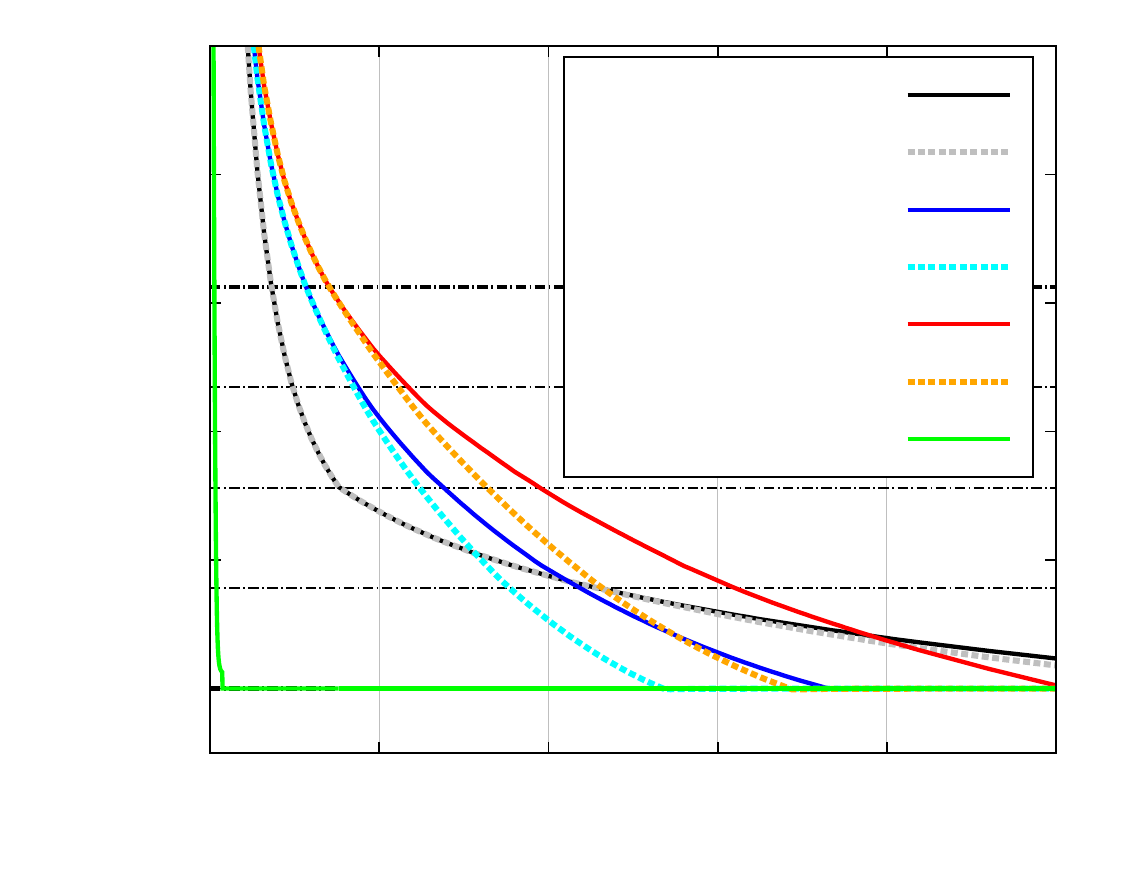
\end{minipage}
\caption{Vertical displacement of the spatial point with initial position $\refcoords = \left(0,a\right)$ of the solid domain $\domainstwo$ for different computed variants over time
(left overview, right detail):
``noslip fine'' (slip length on the interface $\fsiinterface$ specified to $\sliplengh=\infty$, computed with the ``fine'' discretization variant), 
``navslip fine'' (slip length on the interface $\fsiinterface$ as defined in Section \ref{sec:num_interface_t} ($\sliplengh_0=0.1$), computed with the ``fine'' discretization variant),
``noslip'' (slip length on the interface $\fsiinterface$ specified to $\sliplengh=\infty$, computed with the ``coarse'' discretization variant), 
``navslip'' (slip length on the interface $\fsiinterface$ as defined in Section \ref{sec:num_interface_t} ($\sliplengh_0=0.1$), computed with the ``coarse'' discretization variant),
``noslip incpen'' (configuration as ``noslip'' with increased tangential Nitsche penalty constant $\penfluid_{t,0}$ by a factor of $1000$), 
``navslip incpen'' (configuration as ``navslip'' with increased tangential Nitsche penalty constant $\penfluid_{t,0}$ by a factor of $1000$),
``slip'' (slip length on the interface $\fsiinterface$ specified to $\sliplengh=0$, computed with the ``coarse'' discretization variant).
The horizontal black dash-dotted lines (thick lines for coarse mesh) indicate the fluid element boundaries.
}
\label{fig:ex1_disp}
\end{figure}

\paragraph{Numerical results and discussion}
In Figure \ref{fig:ex1_disp}, the vertical displacement of the spatial point with initial position $\refcoords = \left(0,a\right)$ of the solid domain $\domainstwo$ is depicted.
Comparing the ``fine'' and ``coarse'' discretizations shows a good agreement down to a gap of approximately two coarse fluid elements (vertical displacement $\disps_y\left(0,a\right)=-0.375$), where both variants start to deviate significantly.
Solely the ``slip'' variant leads to a fundamentally different impact behavior, which is clear due to the non-physical boundary condition applied to the viscous fluid.
All variants lead to contact in finite time, even though this phenomenon is not expected for the no-slip variants theoretically (see \cite{hillairet2009,gerard2015}).
The explanation for this (realistic) behavior lies in the inherent constraint relaxation arising from the weak imposition by Nitsche's method.
As soon as the solution can no longer be resolved sufficiently, a tangential slip occurs numerically also for the no-slip model.
To substantiate this explanation, a variant with increased tangential penalty parameter $\penfluid_{t,\mathrm{0}}=1000\gamma_{t,\mathrm{0}}^{\fluidletterr,\mathrm{std}}$ (with the parameter of the standard configuration given by $\gamma_{t,\mathrm{0}}^{\fluidletterr,\mathrm{std}}$) is computed, which reduces the numerical slip and thus results as 
expected in a slower approach velocity.

In the following, the difference between the no-slip condition and the Navier slip condition of the computed solution is discussed.
As expected, the Navier slip variant results in an increased velocity, starting from fluid gaps smaller than one fluid element
(see definition of the slip length in Section \ref{sec:num_interface_t}).
Still, the difference between both approaches is not substantial (compared to the error between ``coarse'' and ``fine'' resolution).
While this simple configuration allows to solve the FSCI problem for both interface conditions, applying the Navier slip condition seems to be beneficial for general configurations in two aspects.
Firstly, independent of the approach applied for the imposition of the interface condition, a controlled way of relaxation of the tangential constraint can be incorporated.
Secondly, this type of condition is required to allow for a continuous problem formulation on the interface.

\begin{figure}[htbp]
\begin{minipage}[hbt]{1.0\textwidth}
\centering
\def\figscaling{0.75}
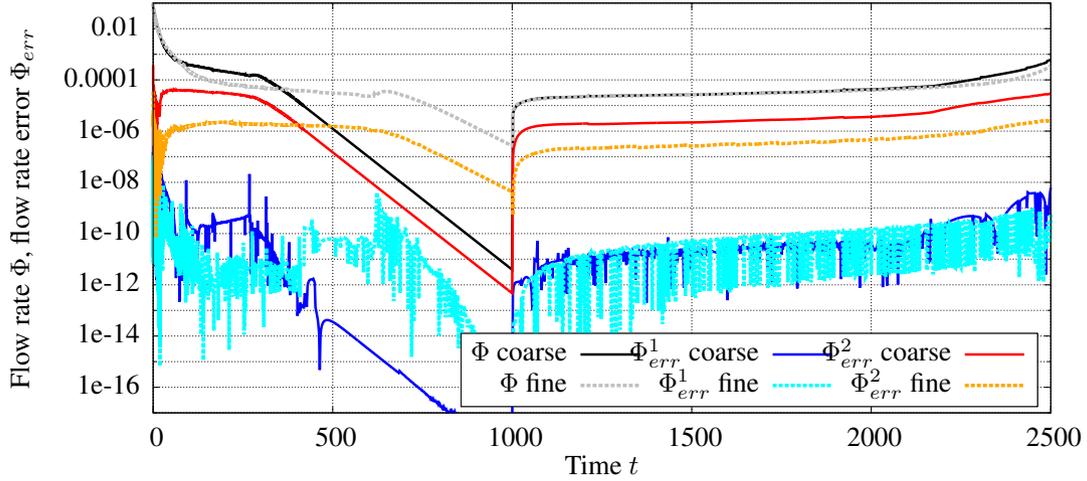
\end{minipage}
\caption{Comparison of computed flow rates and flow rate errors for the ``coarse'' and ``fine'' mesh resolution for the Navier slip interface condition. Herein, $\Phi$ is the flow rate on boundary $\nboundf$, $\Phi_{err}^1$ the overall flow rate error, and $\Phi_{err}^2$ the flow rate error on the interface $\fsiinterface$.}
\label{fig:ex1_massb_navslip}
\end{figure}

\begin{figure}[htbp]
\begin{minipage}[hbt]{1.0\textwidth}
\centering
\def\figscaling{0.75}
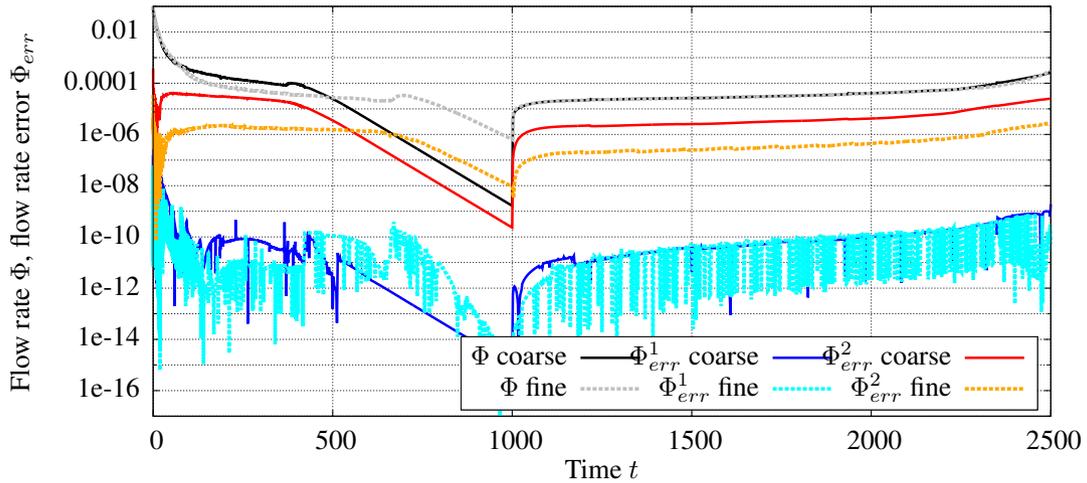
\end{minipage}
\caption{Comparison of computed flow rates and flow rate errors for the ``coarse'' and ``fine'' mesh resolution for the no-slip interface condition. Herein, $\Phi$ is the flow rate on boundary $\nboundf$, $\Phi_{err}^1$ the overall flow rate error, and $\Phi_{err}^2$ the flow rate error on the interface $\fsiinterface$.}
\label{fig:ex1_massb_noslip}
\end{figure}

The overall flowrate on boundary $\nboundf$ and two different flow rate errors are visualized in Figure \ref{fig:ex1_massb_navslip} including relaxation by the Navier slip interface condition and in Figure \ref{fig:ex1_massb_noslip} applying the no-slip interface condition.
Herein, the flow rate $\Phi$ through boundary $\nboundf$, the fluid displacement rate on the interface $\fsiinterface$ given by the fluid velocity $\velf$ or the solid velocity $\vels$ is computed as:
\begin{align}
\label{eq:flowrate}
\Phi = \left|\int_{\nboundf}{\velf \cdot \normal\, \text{d} \nboundf}\right|,\quad
\Phi^F_{\fsiinterface} = \left|\int_{\fsiinterface}{\velf \cdot \normal\, \text{d} \fsiinterface}\right|,\quad
\Phi^S_{\fsiinterface} = \left|\int_{\fsiinterface}{\vels \cdot \normal\, \text{d} \fsiinterface}\right|.
\end{align}
Due to the fluid incompressibility, all three rates have to be equal when taking into account the exact solution of the underlying problem.
Analyzing the flow rates $\Phi$ in Figure \ref{fig:ex1_massb_navslip}, an initial decrease of the fluid flow due to the deceleration of the structure in domain $\domainstwo$ for the approaching bodies can be observed. After a short-time raise at the point of first contact (at $t=268.2$ for the coarse mesh and $t=616$ for the fine mesh), the flow rate decreases to small magnitudes. At $t=1000$, the fluid load at $\nboundf$ starts linearly increasing, which results in a quick rise in the flow rate. As soon as contact is released (at $t=2166$ for the coarse mesh and $t=2290$ for the fine mesh), the structure in $\domainstwo$ moves in positive $y$-direction and so the flow increases.
To quantify the numerical error, two flow rate errors are considered:
\begin{align}
\label{eq:flowrate_errors}
\Phi^1_{err} = \left|\Phi^S_{\fsiinterface}-\Phi\right|,\quad
\Phi^2_{err} = \left|\Phi^S_{\fsiinterface} - \Phi^F_{\fsiinterface}\right|.
\end{align}
Herein, $\Phi^1_{err}$ indicates errors in the overall mass balance, and $\Phi^2_{err}$ characterizes the mass balance errors due to the weak imposition of the interface condition by the Nitsche method. 
When analyzing the overall mass balance $\Phi^1_{err}$, an unexpectedly small error for this mesh resolution can be observed. An explanation to this effect is given in the following.
The discrete fluid mass balance is comprised of the divergence term in \eqref{eq:w_fluid}, the weakly consistent face-oriented stabilization operators \eqref{eq:w_CIP} and ``ghost-penalty'' stabilization operators \eqref{eq:w_ghostpenalty}, and the skew-symmetric adjoint consistency term on the interface \eqref{eq:num_adj_n}.
Partial integration of the divergence term in \eqref{eq:w_fluid} for the fluid balance of mass is performed  and the resulting terms are combined with adjoint consistency term the \eqref{eq:num_adj_n} in \eqref{eq:fconti_partint}.
\begin{align}
 \text{Discrete fluid balance of mass:}\quad\underbrace{\innerp{\testpf}{\div \velf}{\domainf}}_{\text{from \eqref{eq:w_fluid}}} - \underbrace{\innerpb{\testpf\normal}{\velreln \normal}{\partial \domainf}}_{\text{from \eqref{eq:num_adj_n}}} 
 +\underbrace{\mathcal{W}^F }_{\text{from \eqref{eq:w_CIP}}}
 +\underbrace{\mathcal{W}^F_{\mathcal{G}}}_{\text{from \eqref{eq:w_ghostpenalty}}} = \nonumber\\
-\innerp{\grad \testpf}{\velf}{\domainf} + \innerpb{\testpf}{\velf \cdot \normal}{\partial \domainf} -\innerpb{\testpf}{\velreln \normal \cdot \normal}{\partial \domainf} + \mathcal{W}^F + \mathcal{W}^F_{\mathcal{G}} = \nonumber\\
-\innerp{\grad \testpf}{\velf}{\domainf} + \innerpb{\testpf}{\velf \cdot \normal}{\partial \domainf \setminus \fsiinterface} + \innerpb{\testpf}{\vels \cdot \normal}{\fsiinterface} + \mathcal{W}^F + \mathcal{W}^F_{\mathcal{G}}.
\label{eq:fconti_partint}
\end{align}
It can be observed that the fluid velocity in the boundary integral in the second line is replaced by the solid velocity on the interface $\fsiinterface$. The skew-symmetric adjoint consistency term \eqref{eq:num_adj_n} acts therefore as a compensation term for the violation of the balance of mass on the fluid-structure interface. 
Hence, the error $\Phi^1_{err}$ is not influenced by the accuracy of the FSI constraint enforcement but is solely attributed to the stabilization terms from the CIP and the GP stabilization. In addition, the finite convergence tolerance of the nonlinear solution procedure yields perturbations in the error level depending on the remaining residual.
Finally, the interface error $\Phi^2_{err}$ is observed to be significantly larger than the overall error $\Phi^1_{err}$. Comparing the ``coarse'' and the ``fine'' mesh resolution allows the analysis of the spatial convergence of this error. For the time range with similar flow rates ($\Phi$ coarse $\approx$ $\Phi$ fine), a reduction in the error, approximately of second order with respect to the fluid mesh element size $h$, can be observed.
To give a comprehensive view of the balance of mass for this FSCI formulation, the results for the application of the no-slip condition on the entire interface are also given in Figure \ref{fig:ex1_massb_noslip}. 
No significant difference between both results can be observed. Due to the logarithmic axis scaling, a deviation for the small flow rates ($600\leq t \leq 1000$) after contact established can be observed.
As this difference does not essentially influence the lift-off procedure afterwards, the principal discussion done for the Navier slip condition holds also for the no-slip condition.

\begin{figure}[htbp]
\begin{center}
\begin{tabular}{rccc}
Time \hspace{0.05cm}\vline& Overall traction & FSI traction & Contact traction\\\hline
{\rotatebox[origin=l]{90}{$\quad\, t=100$}}\hspace{0.3cm}\vline&
\hspace{-0.4cm}\includegraphics[width=0.3\textwidth]{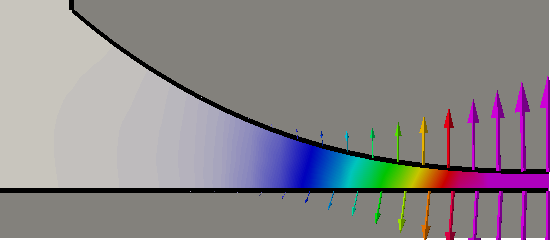} &
\hspace{-0.4cm}\includegraphics[width=0.3\textwidth]{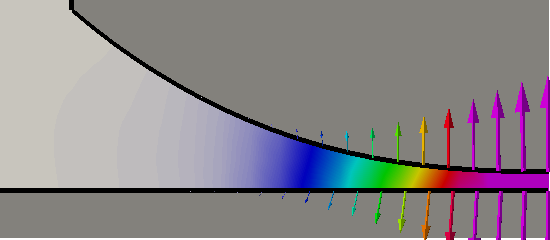} &
\hspace{-0.4cm}\includegraphics[width=0.3\textwidth]{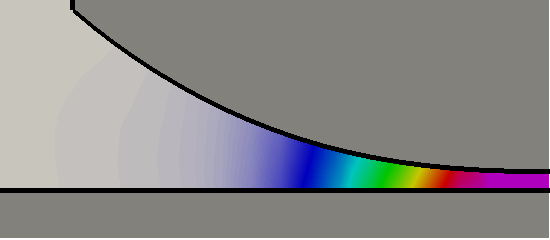}\\ 
{\rotatebox[origin=l]{90}{$\quad\,t=340$}}\hspace{0.3cm}\vline&
\hspace{-0.4cm}\includegraphics[width=0.3\textwidth]{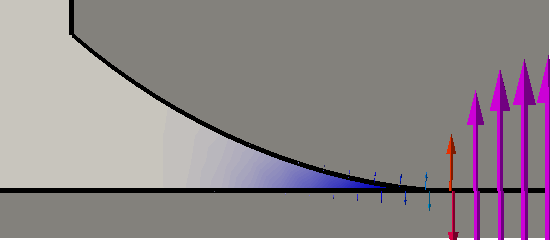}&
\hspace{-0.4cm}\includegraphics[width=0.3\textwidth]{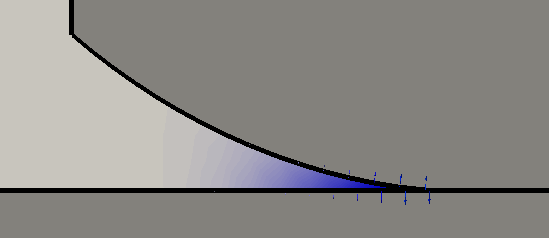}&
\hspace{-0.4cm}\includegraphics[width=0.3\textwidth]{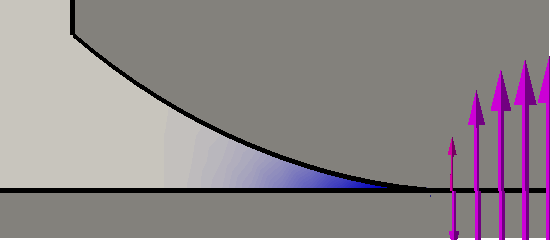}\\ 
{\rotatebox[origin=l]{90}{$\quad\,t=2020$}}\hspace{0.3cm}\vline&
\hspace{-0.4cm}\includegraphics[width=0.3\textwidth]{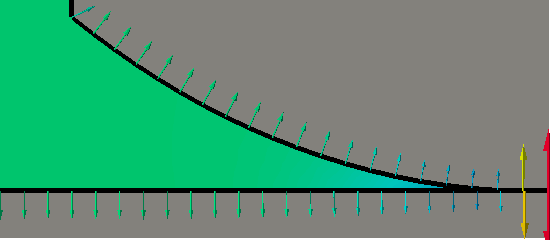}&
\hspace{-0.4cm}\includegraphics[width=0.3\textwidth]{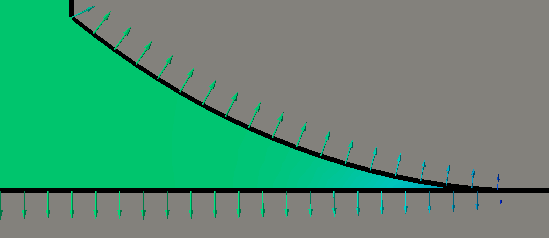}&
\hspace{-0.4cm}\includegraphics[width=0.3\textwidth]{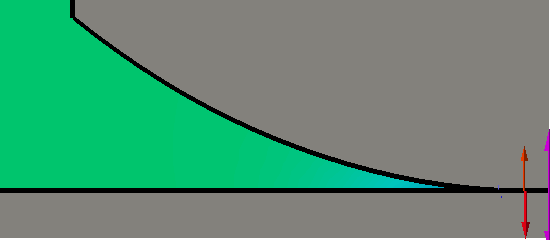}\\ 
{\rotatebox[origin=l]{90}{$\quad\,t=2420$}}\hspace{0.3cm}\vline&
\hspace{-0.4cm}\includegraphics[width=0.3\textwidth]{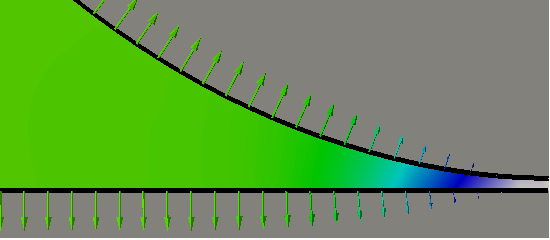}&
\hspace{-0.4cm}\includegraphics[width=0.3\textwidth]{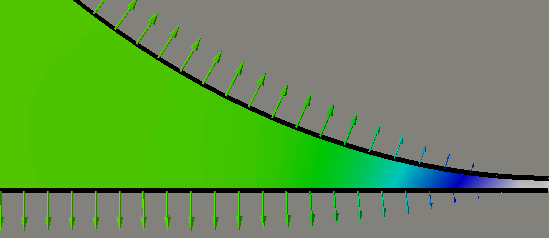}&
\hspace{-0.4cm}\includegraphics[width=0.3\textwidth]{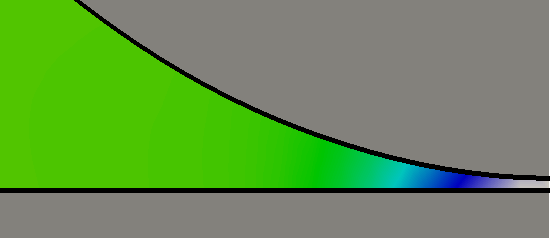}
  \end{tabular}
\end{center}
\begin{minipage}{1\textwidth}
\input{fig/ex1_det/pl.tex}
\end{minipage}
\caption{
Visualization of the computed falling, contacting and lifting process of the rounded stamp for the Navier slip interface condition at four instances in time. The color code in the fluid domain visualizes the computed fluid pressure and the color code of the arrows the respective traction magnitude. The arrows visualize the interface traction separated in three groups.
The overall traction includes all contributing of cases $I-IV$, the FSI traction includes case $I$, and the contact traction includes cases $II-IV$ (specified in the interface contributions \eqref{eq:num_caseI}-\eqref{eq:num_caseIII_alt}).
The visualization of the traction is reconstructed from the nodal interface contributions of \eqref{eq:num_caseI}-\eqref{eq:num_caseIII_alt}) to the overall weak form on the solid mesh.
}
\label{fig:ex1_detail}
\end{figure}

In Figure \ref{fig:ex1_detail}, a detailed view of the contacting zone for different points in time is given. Three different types of traction are visualized by arrows, namely the overall traction, the FSI traction, and the contact traction. At $t=100$ (first row in Figure \ref{fig:ex1_detail}), the body $\domainstwo$ approaches $\domainsone$ and as a result a high pressure peak occurs in the smallest constriction.
This peak is almost equal to the FSI traction concluding that viscous traction is not significant. 
At $t=340$ (second row in Figure \ref{fig:ex1_detail}), the majority of the external load is carried by the contact traction. For the overall traction, the continuous transition of FSI traction and contact traction can be seen.
An essential part of the external load at $t=2020$ (third row in Figure \ref{fig:ex1_detail}) is carried by the FSI traction, but due to the fluid inertia there is still contact at the area around $x=0$.
Finally, at $t=2420$ the structural body $\domainstwo$ completely lifted again and so the lowest pressure and FSI traction can be identified in the smallest constriction.

\subsection{Elastic Pump}
In the following example, an elastic fluid pump powered by an external load is analyzed. 
This configuration includes large deformation of the solid domain and a periodically changing topological connection of the fluid domain. Large fluid pressure discontinuities when crossing the valves occur which need to be represented properly by the fluid solution space.

\begin{figure}[htbp]
\centering
\input{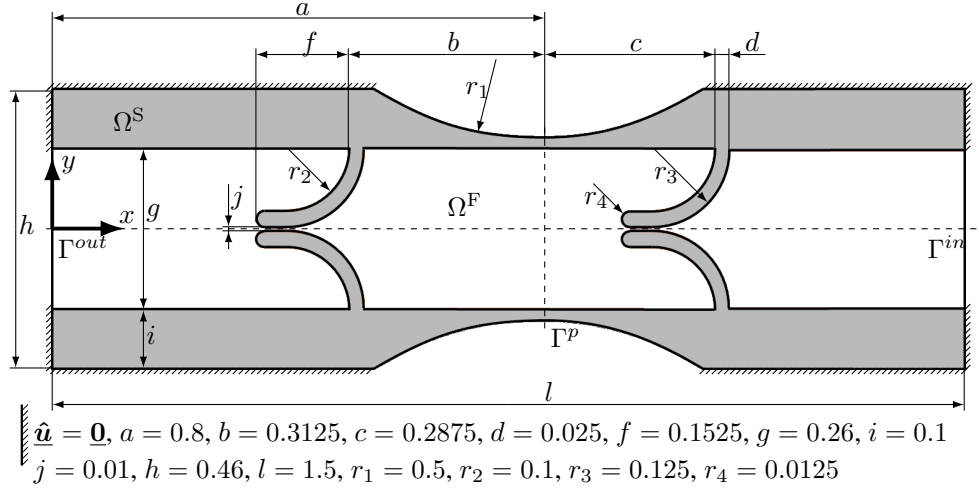}
\vspace{1.5cm}
\caption{Geometry and boundary conditions for the elastic pump.}
\label{fig:ex3_setup}
\end{figure}

\paragraph{Problem description}
The geometric setup and basic boundary conditions are depicted in Figure \ref{fig:ex3_setup}. 
The solid domain $\domains$ is designed to pump fluid in the domain $\domainf$ from the fluid inflow boundary $\Gamma^{in}$ to the fluid outflow boundary $\Gamma^{out}$.
The structural part includes two valves consisting of two flaps each to control the flow direction. The fluid flow is driven by the change of volume in the fluid chamber placed between the two valves.
The pump is powered by a time-dependent periodic traction in normal direction which is prescribed as Neumann condition on the circular solid boundary $\Gamma^p$ as
$\tractionsN  = -20 A (1-\text{cos}(40 \pi t)) \cdot \normal$, with $A=1.0$ for $t\in\left[0,0.15\right]$ and 
$A=1.5$ for $t\in\left[0.15,0.3\right]$. In the tangential plane of $\Gamma^p$, zero traction is prescribed.
Therefore, the pump is driven for three periodic cycles with a constant amplitude of the external load, followed by three periodic cycles with an external load increased by $50\%$.
Both the solid and the fluid are subject to a gravitational body force in negative $y$-direction: $\refdensitys \refbodyfs = \densityf \bodyff = \transpose{\left[0,-1\right]}$.
On the fluid boundaries $\Gamma^{in}$ and $\Gamma^{out}$, the hydrostatic pressure is prescribed by a Neumann boundary condition in $x$-direction ($\tractionfN \cdot \normal = y$), whereas zero velocity in $y$-direction is prescribed by a Dirichlet type boundary condition.

As material parameters, the fluid density is $\densityf=10^{-3}$ and the dynamic viscosity is $\viscf=10^{-4}$. The material behavior of the solid continuum is given by the Neo-Hookean model with the strain energy function \eqref{ex1:strainenergysp} and a Young's modulus $E=2000$ and Poisson's ration $\nu=0.3$. The initial density in $\domains$ equals the fluid density $\refdensitys=\densityf=10^{-3}$.

\begin{figure}[htbp]
\hspace{0.1cm}\includegraphics[width=0.4975\textwidth]{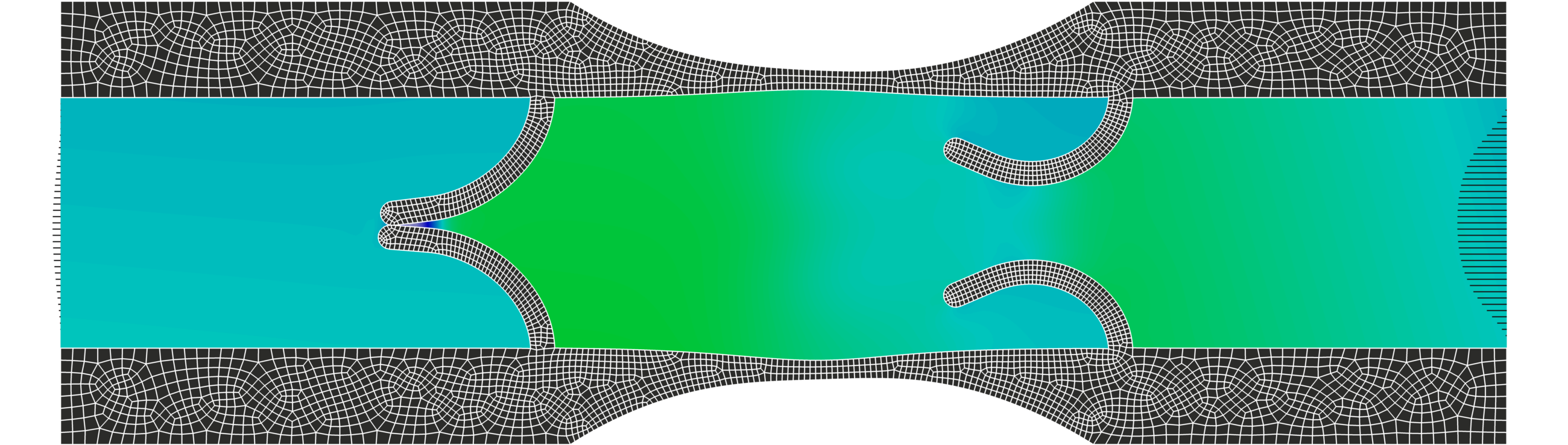}
\includegraphics[width=0.5\textwidth]{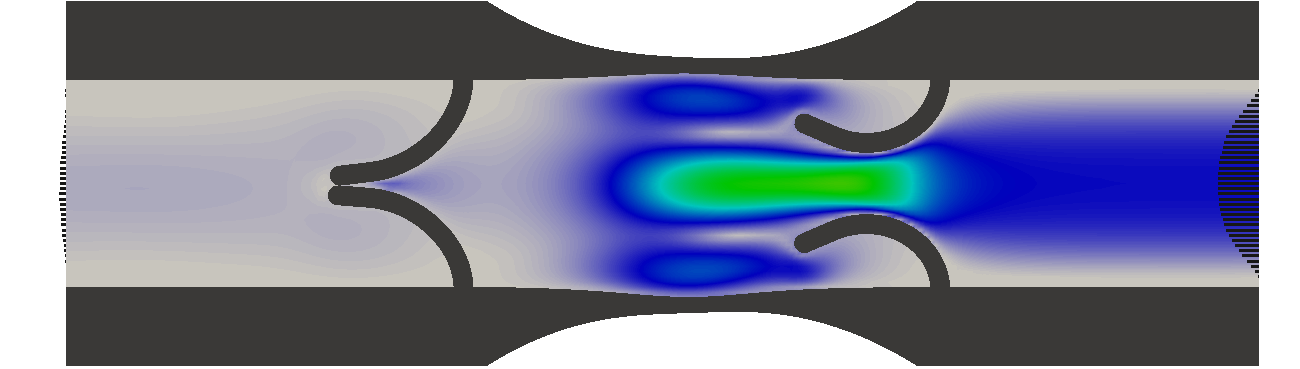}
\includegraphics[width=0.5\textwidth]{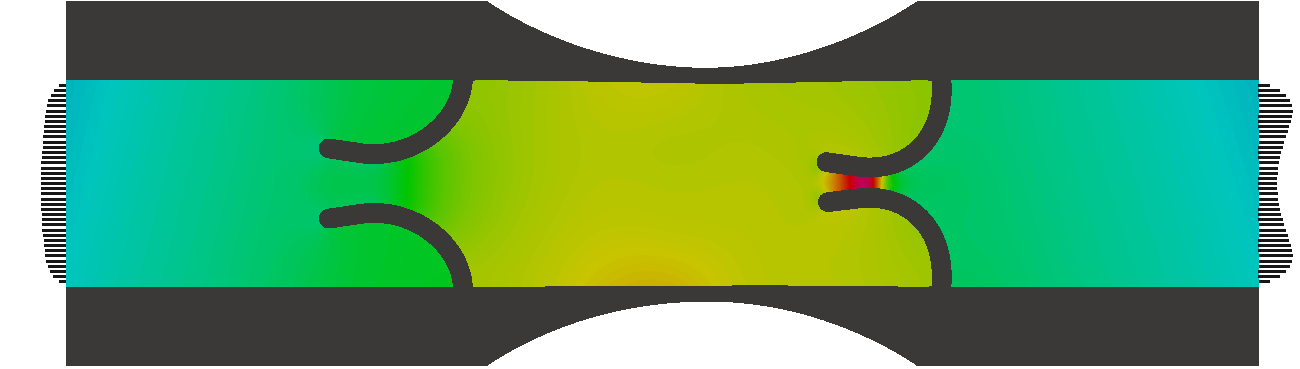}
\includegraphics[width=0.5\textwidth]{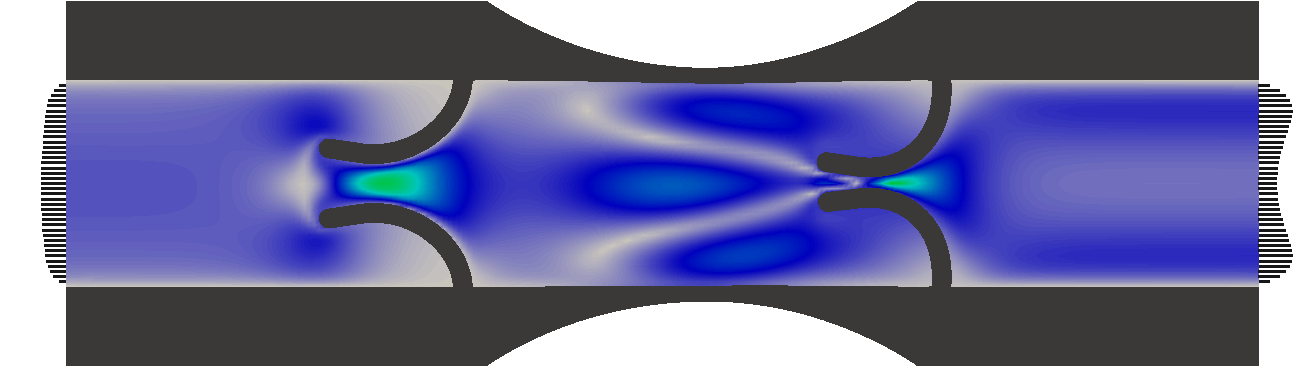}
\includegraphics[width=0.5\textwidth]{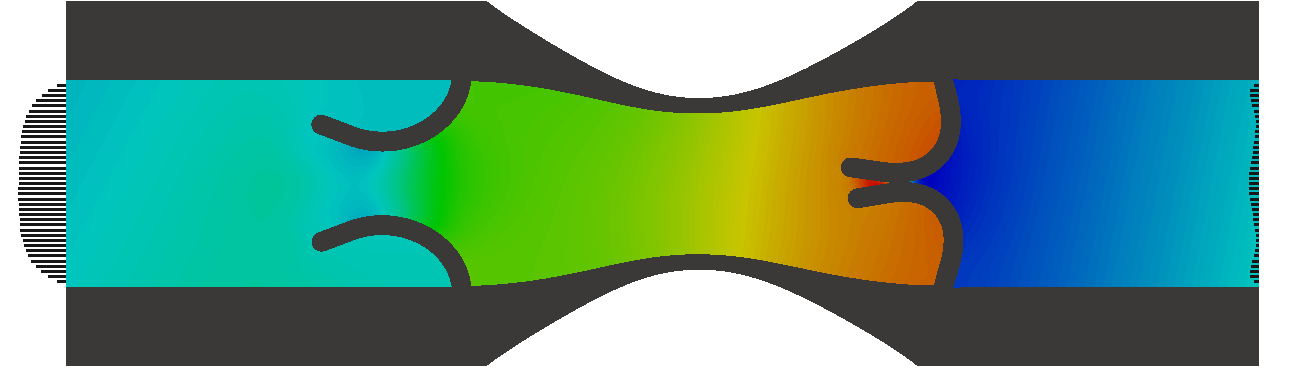}
\includegraphics[width=0.5\textwidth]{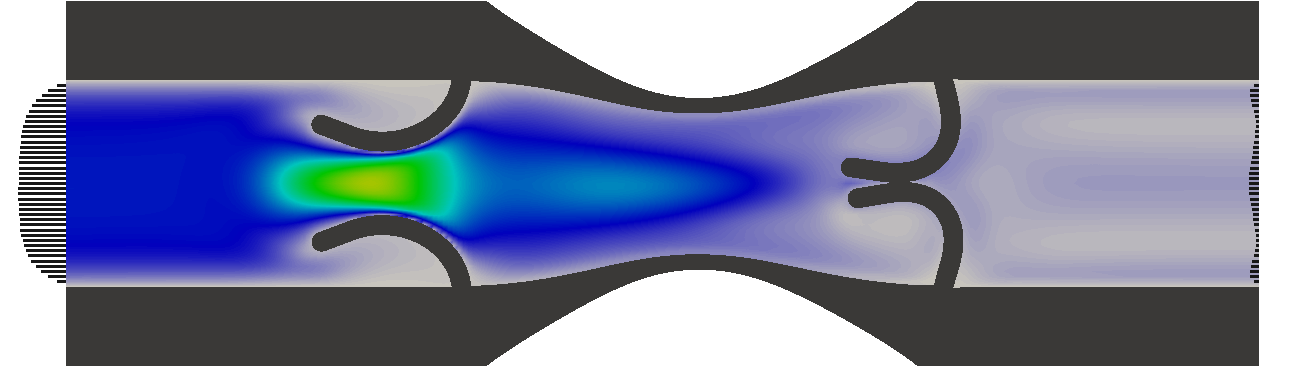}
\includegraphics[width=0.5\textwidth]{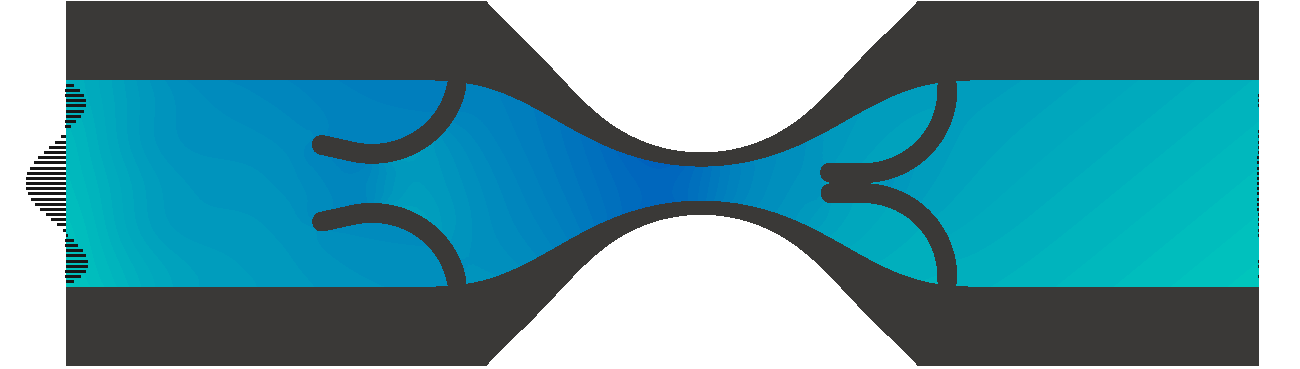}
\includegraphics[width=0.5\textwidth]{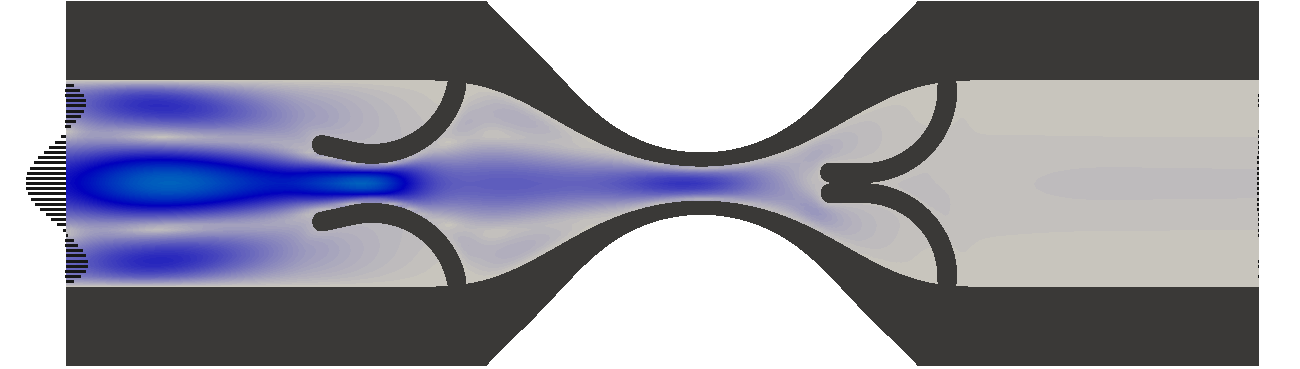}
\includegraphics[width=0.5\textwidth]{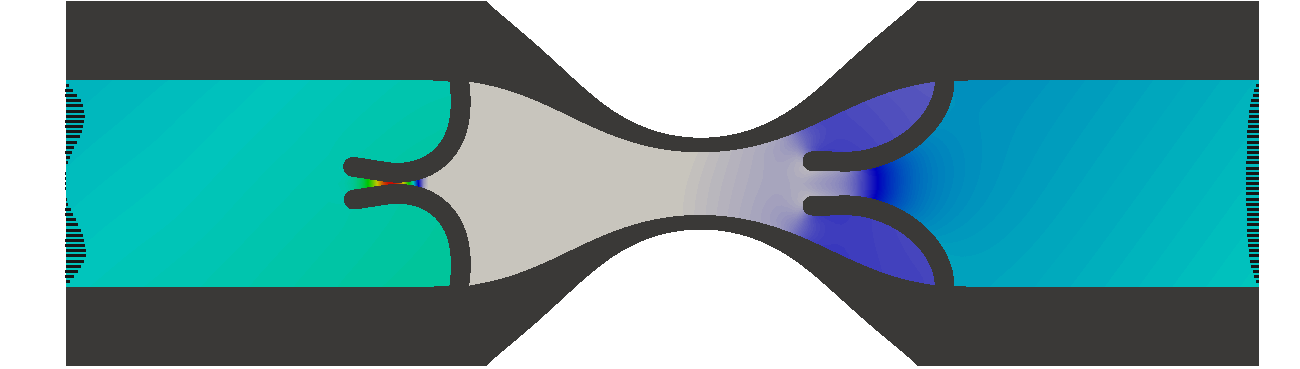}
\includegraphics[width=0.5\textwidth]{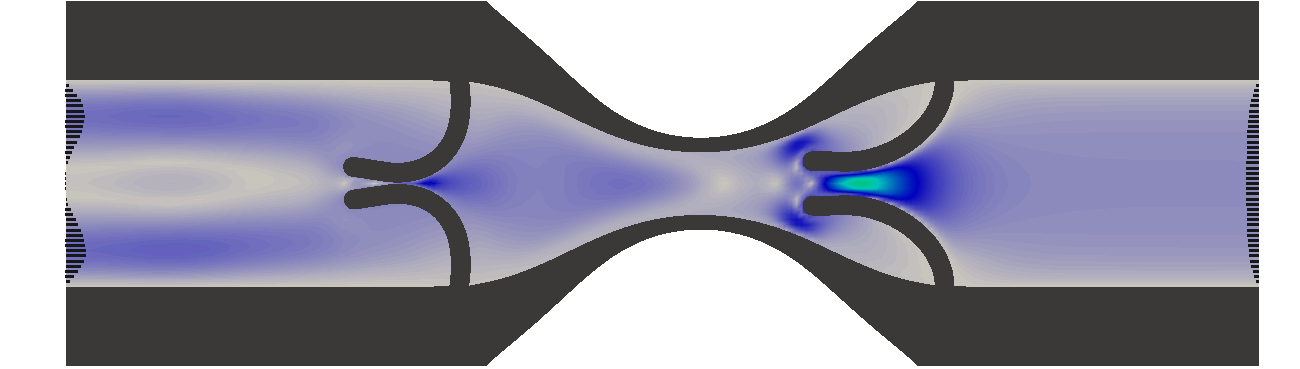}
\begin{minipage}{0.5\textwidth}

\input{fig/ex2_det/pl.tex}
\end{minipage}
\begin{minipage}{0.5\textwidth}
\input{fig/ex2_det/vl.tex}
\end{minipage}
\caption{Visualization of the computed fluid velocity and pressure and the computed deformation of solid domain for $t\in\left[0.1,0.15\right]$. In the left column, the color code represents the fluid pressure, whereas, on the right column, the fluid velocity magnitude is represented. Additionally, the black bars at the inflow boundary $\Gamma^{in}$ and outflow boundary $\Gamma^{out}$ indicate the computed fluid velocity at the corresponding boundary. Five points in time are represented by the rows, which are from top to bottom $t=0.1,t=0.106,t=0.11,t=0.125$, and $t=0.135$.}
\label{fig:ex2_detail}
\end{figure}

The fluid domain is discretized by a structured mesh consisting of $240\times54=12960$ (with $0.0\leq x \leq 1.5$ and $-0.1755 \leq y \leq 0.1755$) elements which is unfitted to the interface $\fsiinterface$.
The solid domain is discretized fitted to the interface $\fsiinterface$ by $4648$ elements (shown in Figure \ref{fig:ex2_detail} (upper left)).
A contact stress based on harmonic weighting between the stress representation of both solid domains, as discussed in Section \ref{sec:num_interface_n}, is applied. Due to the almost equal material parameters and mesh sizes of all contacting interfaces, this approach results approximately in a mean weighting $\omega\approx0.5$. The reference slip length is set to $\sliplengh_0=0.1$.
The temporal discretization is performed with $\theta = 1.0$ and a time step size of $\timestep=0.0002$ for $t \in \left[0,0.1698\right]$ and 
$\timestep=0.0001$ for $t \in \left[0.1698,0.3\right]$, to account for changing system dynamics, is applied.

\paragraph{Numerical results and discussion}
In Figure \ref{fig:ex2_detail}, the computed fluid velocity and pressure as well as the computed deformation for $t\in\left[0.1,0.15\right]$ are presented. These results correspond with the third load cycle and already exhibit a periodic response to the periodic external load with an amplitude of $A=1.0$.
Starting with $t=0.1$, where no external load on $\interface^P$ is applied, the left valve is closed and, due to the pressure gradient in the right valve, a flow into the fluid chamber occurs.
In the next point in time $t=0.106$, a compression of the fluid chamber resulting in an increasing pressure due to the external load is observable.
Due to the geometry of the two valves, an opening motion of the left valve and a closing motion of the right valve is induced. As both valves are still open at this point in time fluid, mass leaves the chamber through both valves and finally leads to an back flow at the inflow boundary. This behavior has changed at $t=0.11$, where the right valve prevents fluid flow as it is closed. It can be seen that the occurring pressure jump between both sides of the right valve can be well represented by the provided fluid function space. The resulting force of the discontinuous fluid pressure leads to a deformation of the right valve into positive $x$-direction. At the same time, the flaps of the left valve are opened by the fluid pressure and allow for a large fluid flow which finally leads to a high flow rate at the boundary $\Gamma^{out}$. At $t=0.125$, the volume in the chamber is almost minimal and as a consequence the structural velocity on $\interface^p$ nearly vanishes. Therefore, the fluid pressure gradients 
decrease and both valves relax towards the initial geometry. At $t=0.135$, the external load reduces and leads to an increasing volume in the fluid chamber. Consequently, the pressure in the chamber drops and induces a closing motion of the left valve.
A peak of the fluid pressure between the two left flaps occurs due to the acceleration of fluid mass.
The closed left valve prevents flow through the left valve, and the discontinuous pressure is carried elastically by the flaps. The right valve is opened by the pressure difference on both sides of the flaps and allows for fluid flow into the chamber. As the pumping motion is almost periodical, the results computed for $t=0.15$ are not distinguishable from the solution at $t=0.1$ and thus are not shown.

\begin{figure}[htbp]
\hspace{-0.4cm}
\begin{minipage}[hbt]{0.5\textwidth}
\centering
\def\figscaling{0.6}
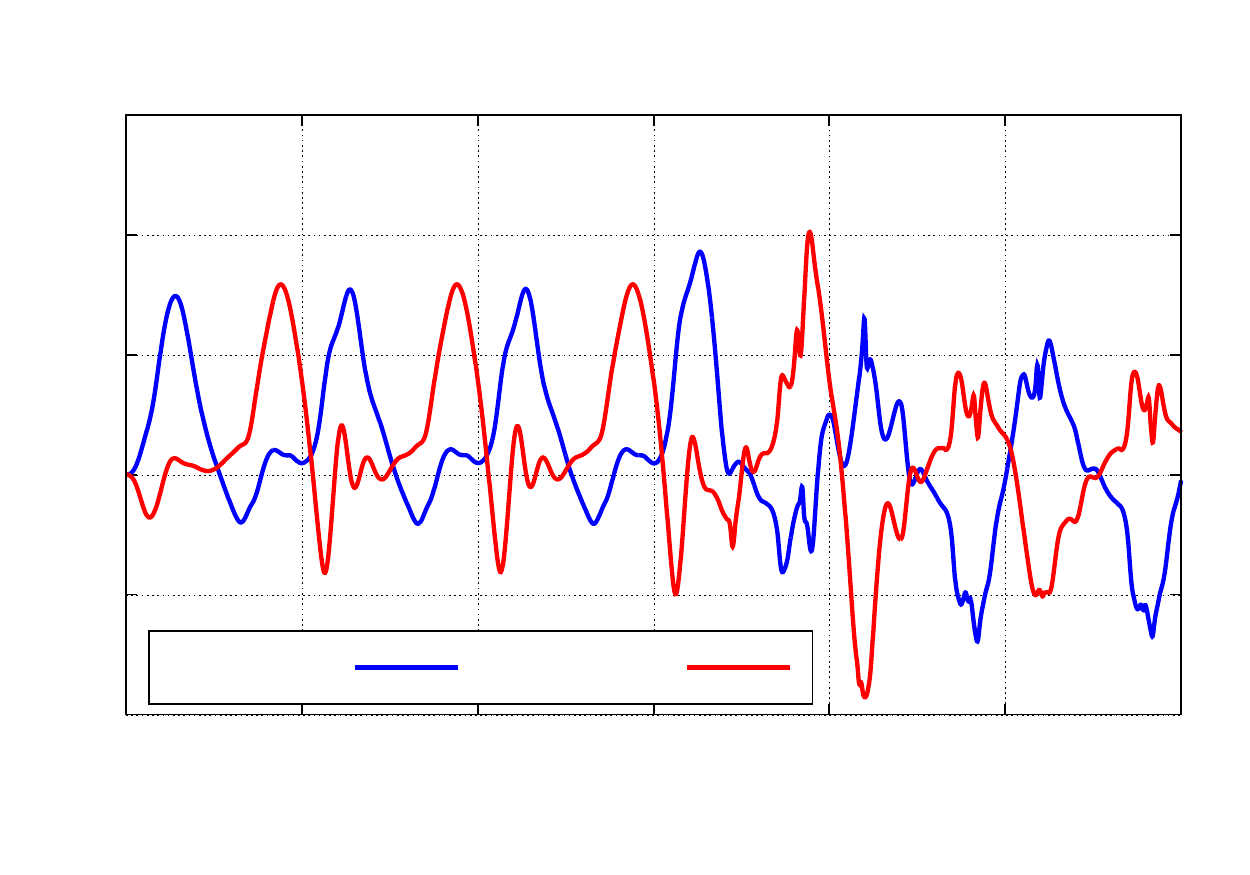
\end{minipage}
\hspace{-0.2cm}
 \begin{minipage}[hbt]{0.5\textwidth}
\centering
\def\figscaling{0.6}
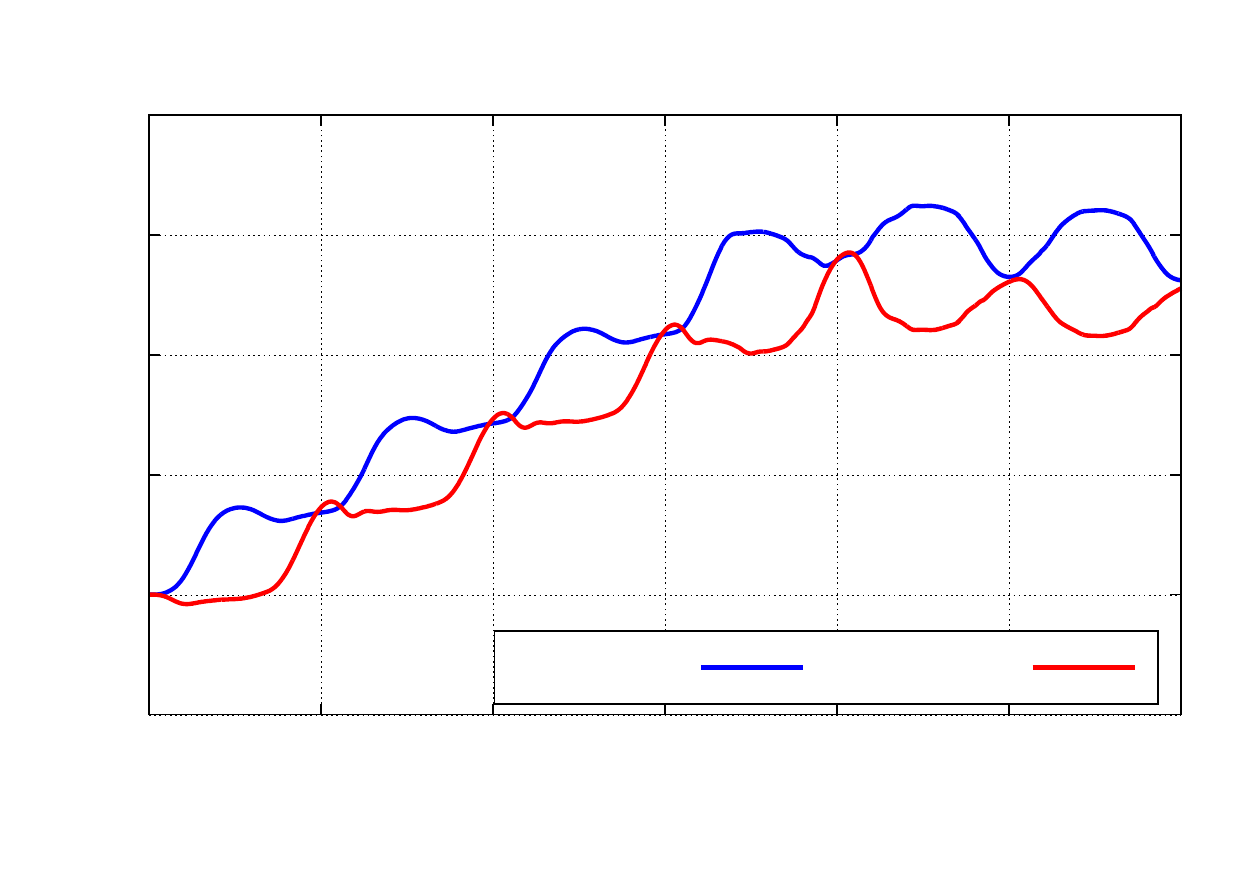
\end{minipage}
\caption{Computed flow rates at the inflow boundary $\Gamma^{in}$ and the outflow boundary $\Gamma^{out}$.
The normal vector therein is oriented in negative $x$-direction, which is the design flow direction of the pump (left).
Transported volume through the inflow boundary $\Gamma^{in}$ and the outflow boundary $\Gamma^{out}$ computed in a post-processing step where an integration in time of the flow rates is performed (right).}
\label{fig:ex2_flowrates}
\end{figure}

To quantify the output of the examined pump, the computed flow rates at the inflow boundary $\Gamma^{in}$ and outflow boundary $\Gamma^{out}$ are presented in Figure \ref{fig:ex2_flowrates} (left). First, the time interval $t\in\left[0.1,0.15\right]$, with a periodic external load of amplitude $A=1$, is analyzed. While the first cycle is still dominated by the start-up process from a system initially in rest, the flow rates of the second and third cycle are very similar.
Therefore, the cycle $t\in\left[0.1,0.15\right]$ can be classified as the periodic response to the periodic load with $A=1$ and was already discussed in detail previously. Now, analyzing the subsequent interval $t\in\left[0.15,0.3\right]$ with $A=1.5$, after a transition phase in the fourth load cycle the pump exhibits again an almost periodic behavior for the last two load cycles. It can be seen that oscillations with higher frequencies occur than for the smaller load amplitude, which is tackled by a reduced time step size in the time integration scheme.

To make a statement on the performance of the pump, the volume transported through the pump is presented in Figure \ref{fig:ex2_flowrates} (right).
It can be seen that in each cycle the transported volume through $\Gamma^{in}$ at first is smaller than through $\Gamma^{out}$ mainly due to the volume change in the fluid chamber. The difference in the transported volume is smaller for $A=1$ than for $A=1.5$ as larger deformation occurs. Analyzing the transported volume per cycle, it can be seen that, for the smaller amplitude, each cycle transports approximately $0.074$, whereas the higher load amplitude leads to a slight transport opposite to the design flow direction.

\begin{figure}[htbp]
\includegraphics[width=0.5\textwidth]{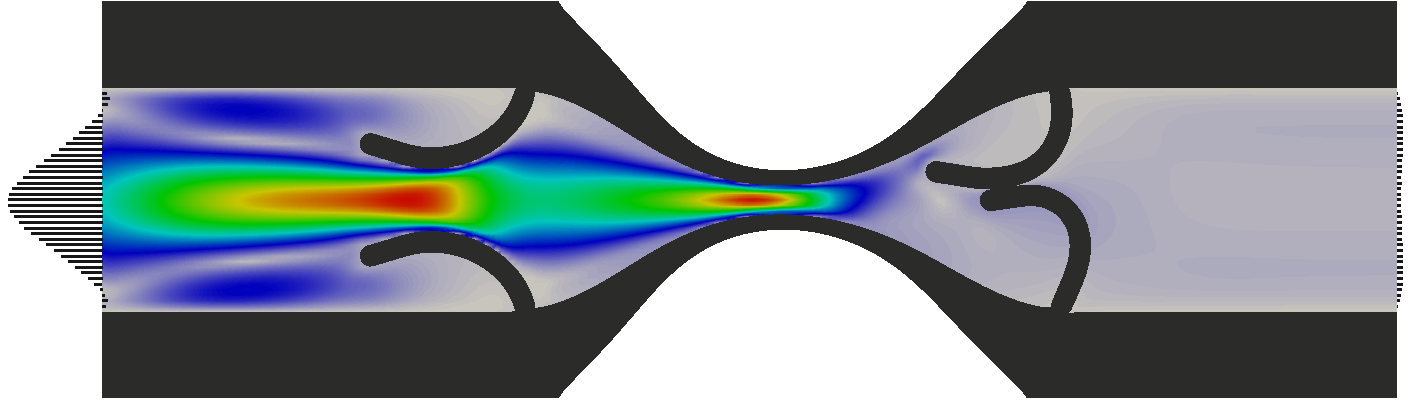}
\includegraphics[width=0.5\textwidth]{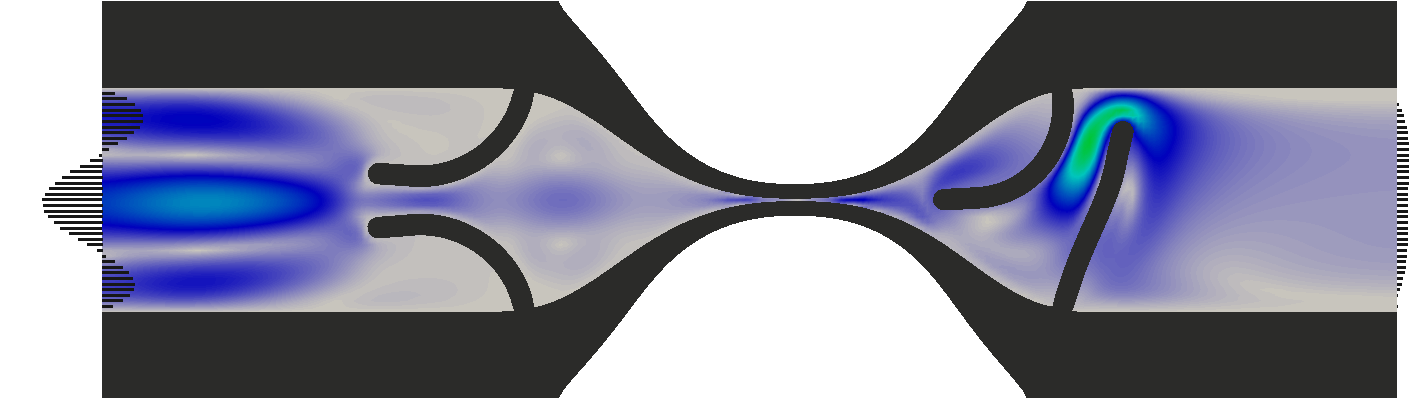}
\includegraphics[width=0.5\textwidth]{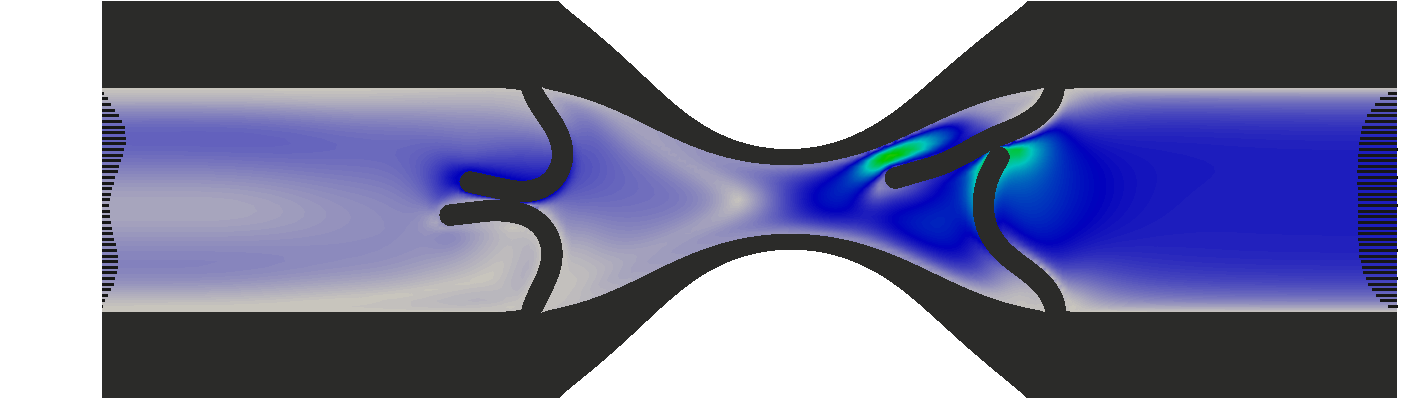}
\includegraphics[width=0.5\textwidth]{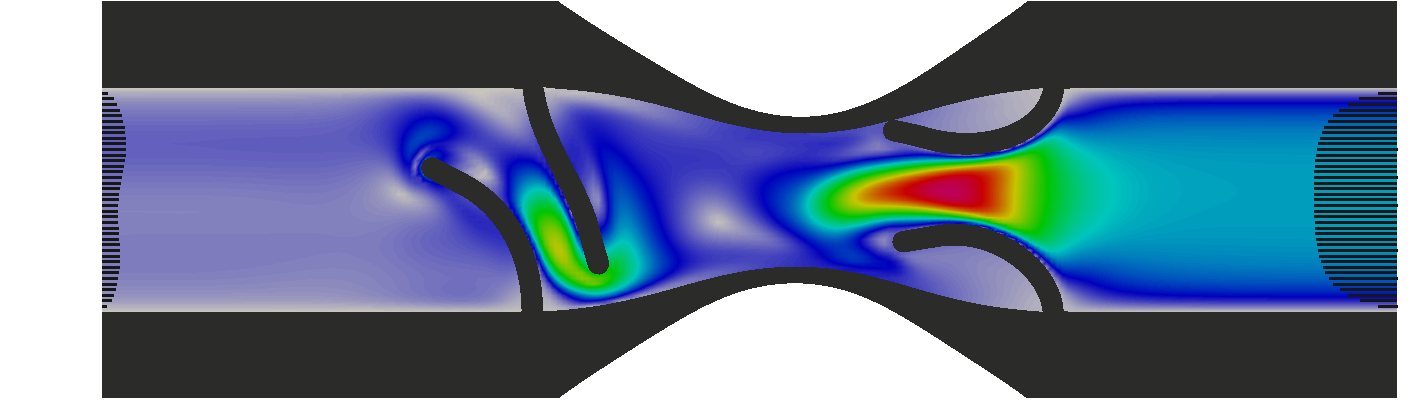}
\begin{minipage}{1\textwidth}
\input{fig/ex2_det/vl2.tex}
\end{minipage}
\caption{Visualization of the computed fluid velocity and pressure and the computed deformation of solid domain for $t\in\left[0.15,0.2\right]$. The color code represents the fluid velocity magnitude, and the black bars at the inflow boundary $\Gamma^{in}$ and outflow boundary $\Gamma^{out}$ indicate the computed fluid velocity at the corresponding boundary. Four points in time are represented from top-left to bottom-right $t=0.167,t=0.174,t=0.19$ and $t=0.194$.}
\label{fig:ex2_detail_2}
\end{figure}

To generate understanding for this phenomenon, four exemplary points in time with load amplitude $A=1.5$ are selected and shown in Figure \ref{fig:ex2_detail_2}. Compared to the load with amplitude $A=1.0$, higher fluid velocities occur leading to higher pressures and finally an increase of the interface traction, at $t=0.167$. This fluid state leads to a non-symmetric deformation of the flaps in the right valve. As it can be seen at $t=0.174$, finally the lower flap snaps through and as a result the right valve does not prevent flow properly anymore. For $t\in\left[0.1791,0.1822\right]$, contact between the upper and lower part of the fluid chamber occurs, prohibiting the flow in the chamber. At $t=0.19$, the lower flap of the right valve starts to snap back, whereas the left valve is exposed to large non-symmetric deformation. Finally at $t=0.194$, the left valve has snap-through, allowing for flow opposite to the design flow direction. In short, a load amplitude of $A=1.5$ is beyond the maximal load 
resulting in a proper operation of the elastic pump.

Nevertheless, from a computational point of view, it is noteworthy that the presented formulation demonstrates to be applicable also for these rather complex scenarios and hence promises to be a rather general tool.
Processes beyond the intended design can be computed without requiring changes to the problem setup. In this example, unexpected deformation and topological changes to the fluid domain were handled without any modifications to the problem setup.

\subsection{Flow-driven squeezed elastic structure}
In the following, a configuration is considered where an initially cylindrical elastic body $\domainstwo$ is squeezed through an elastic structure $\domainsone$ by the load of the surrounding fluid flow.
This configuration is designed to test the formulation's capability to handle frequent changes between the fluid-structure interface and the contact interface including large contacting areas and essential topological changes.

\paragraph{Problem description}
\begin{figure}[htbp]
\hspace*{1cm}
\begin{minipage}{0.45\textwidth}
\centering
\input{fig/ex2}
\end{minipage}
\begin{minipage}{0.45\textwidth}
\centering
\includegraphics[width=0.6\textwidth]{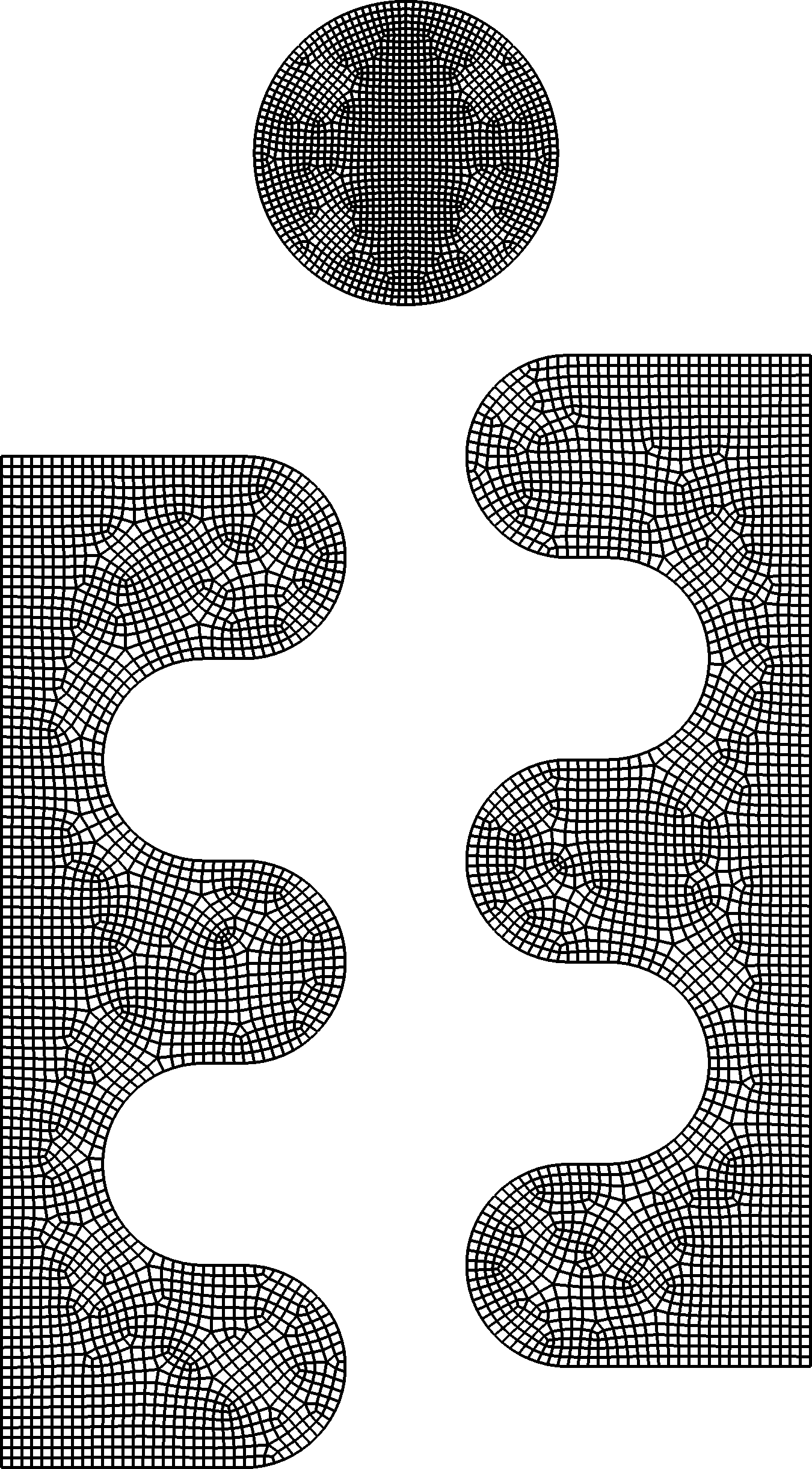}
\end{minipage}
\caption{Geometry and boundary conditions for the flow driven squeezed elastic structure (left). Visualization of the discretization for the solid domains $\domainsone$ and $\domainstwo$ with black lines indicating element boundaries (right).}
\label{fig:ex3_tbn}
\end{figure}

The problem setup of the this example, including all dimensions and basic boundary conditions, is visualized in Figure \ref{fig:ex3_tbn} (left).
All dimensions of solid body $\domainsone$, which are not explicitly indicated in this figure, are defined by symmetry and replication of the given dimensions (e.g.\ all unspecified radii are equal to $r_2$).

On the inflow boundary $\Gamma^{in}$, a time-dependent, parabolic velocity profile 
$\velfD = \left[0,-100 (1-x^2) 4000 t\right]^T$ for $t \in \left[0.0,0.00025\right]$ and
$\velfD = \left[0,-100 (1-x^2) \right]^T$ for $t \in \left[0.00025,0.016\right]$ is prescribed as Dirichlet boundary condition.
On the outflow boundary $\Gamma^{out}$, a zero traction Neumann boundary condition is prescribed.

The material properties of the incompressible fluid are specified by the density $\densityf = 10^{-6}$ and the dynamic viscosity $\viscf = 10^{-5}$. 
The initial density in both solid domains equals the fluid density $\refdensitys = \densityf = 10^{-6}$.
Similar to the numerical examples presented previously, a Neo-Hookean material model with strain energy function \eqref{ex1:strainenergysp} is considered for both solids. The parameters of the material model in the squeezed domain $\domainstwo$ are given by $E^{\structletterr_2}=100$ and $\nu^{\structletterr_2} = 0.3$, whereas the outer domain $\domainsone$ has an increased stiffness by $E^{\structletterr_1}=200$ and $\nu^{\structletterr_1} = 0.3$.

The structured computational mesh applied for the discretization of the fluid domain consists of $120\times300=36000$ bilinear elements. The solid domain is discretized fitted to the interface $\fsiinterface$ by 4890 elements in domain $\domainsone$ and by 1562 elements in domain $\domainstwo$. The solid discretization is given in Figure \ref{fig:ex3_tbn} (right).  Compared to the examples presented previously, the penalty parameters constants $\penfluid_0$ and $\penfluid_{t,0}$ are divided by a factor of $7$, in order to relax the kinematic constraints and thus support the nonlinear solution procedure (see Remark \ref{rem:fluidpen}). With this modification, the penalty parameters are still large enough to provide discrete stability of the formulation. A contact stress based on harmonic weighting between the stress representation of both solid domains, as discussed 
in Section \ref{sec:num_interface_n}, is applied. The reference slip length is set to $\sliplengh_0=0.1$.  The temporal discretization is preformed with $\theta=1$ and a time step size of $\Delta t = 0.00002$ for $t\in[0,0.0056]$ and $\Delta t = 0.000005$ for $t\
\in[0.0056,0.016]$ to account for the varying dynamic of the 
coupled system.

\begin{figure}[htbp]
\begin{minipage}{1\textwidth}
\includegraphics[width=0.12\textwidth]{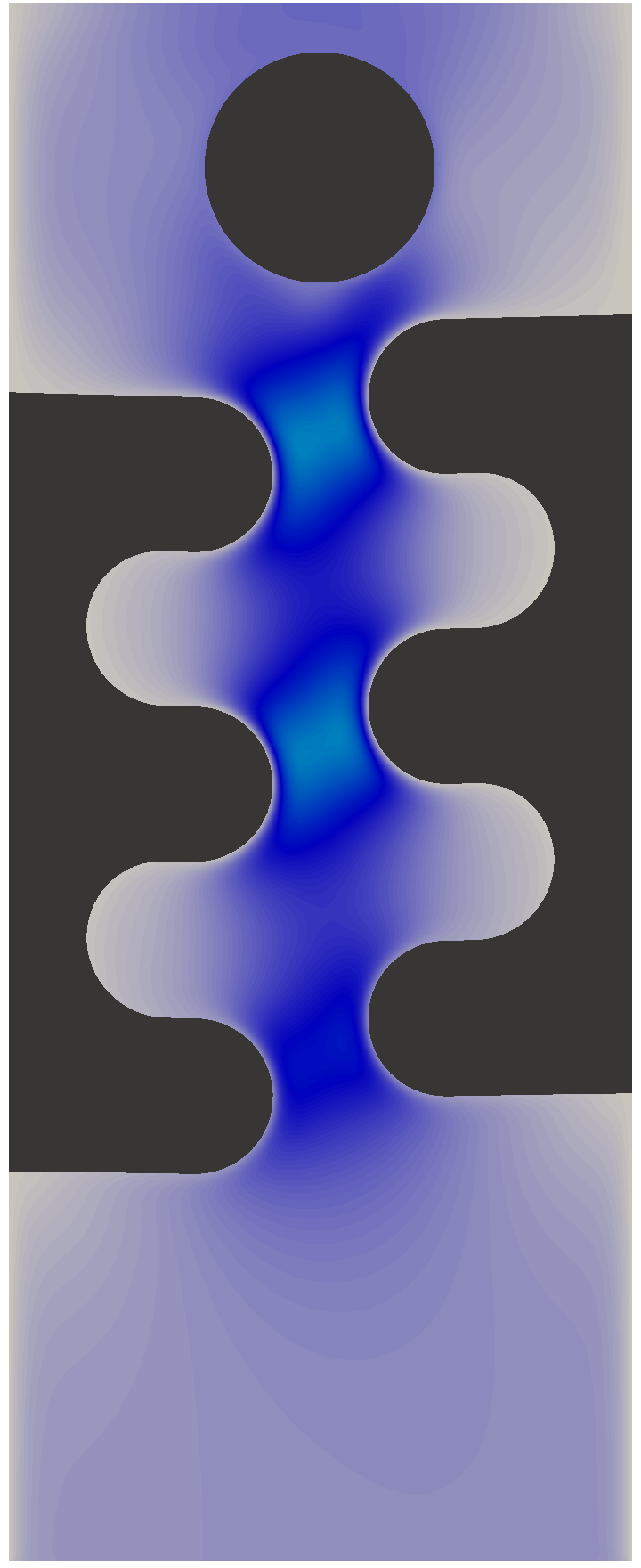}
\includegraphics[width=0.12\textwidth]{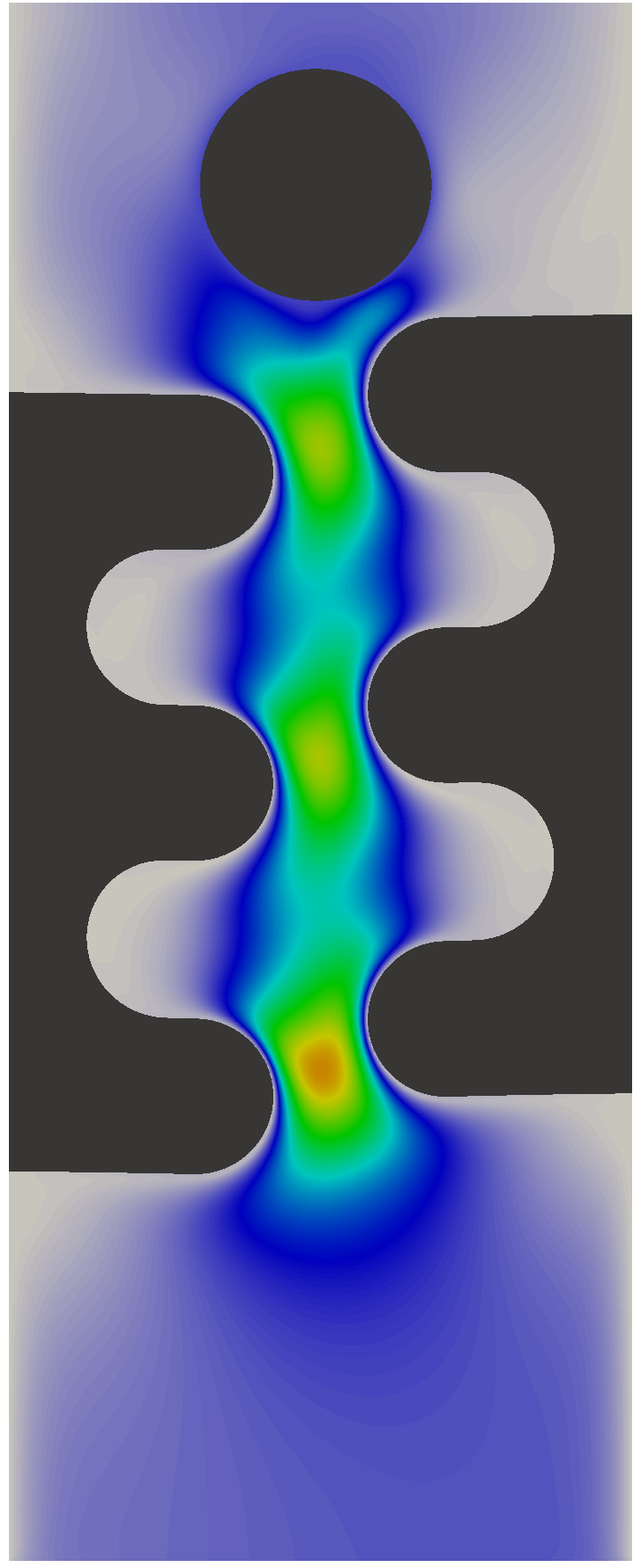}
\includegraphics[width=0.12\textwidth]{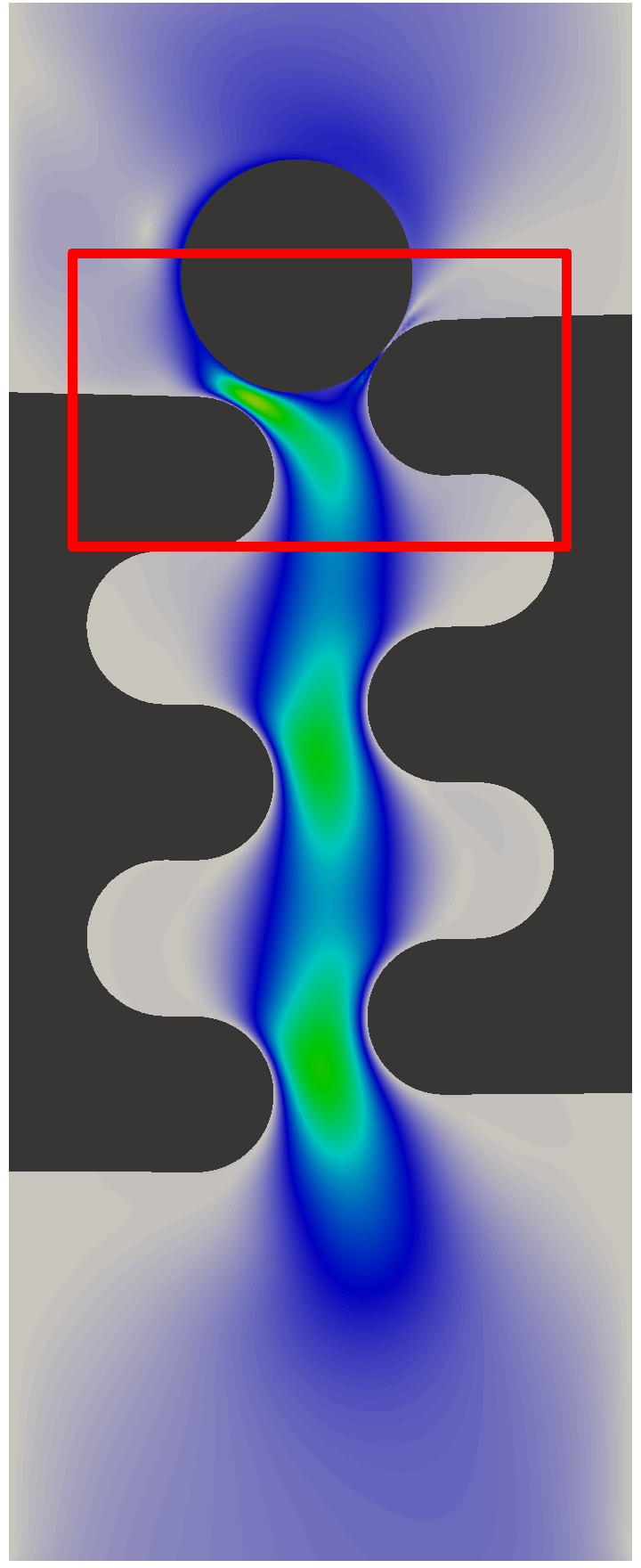}
\includegraphics[width=0.12\textwidth]{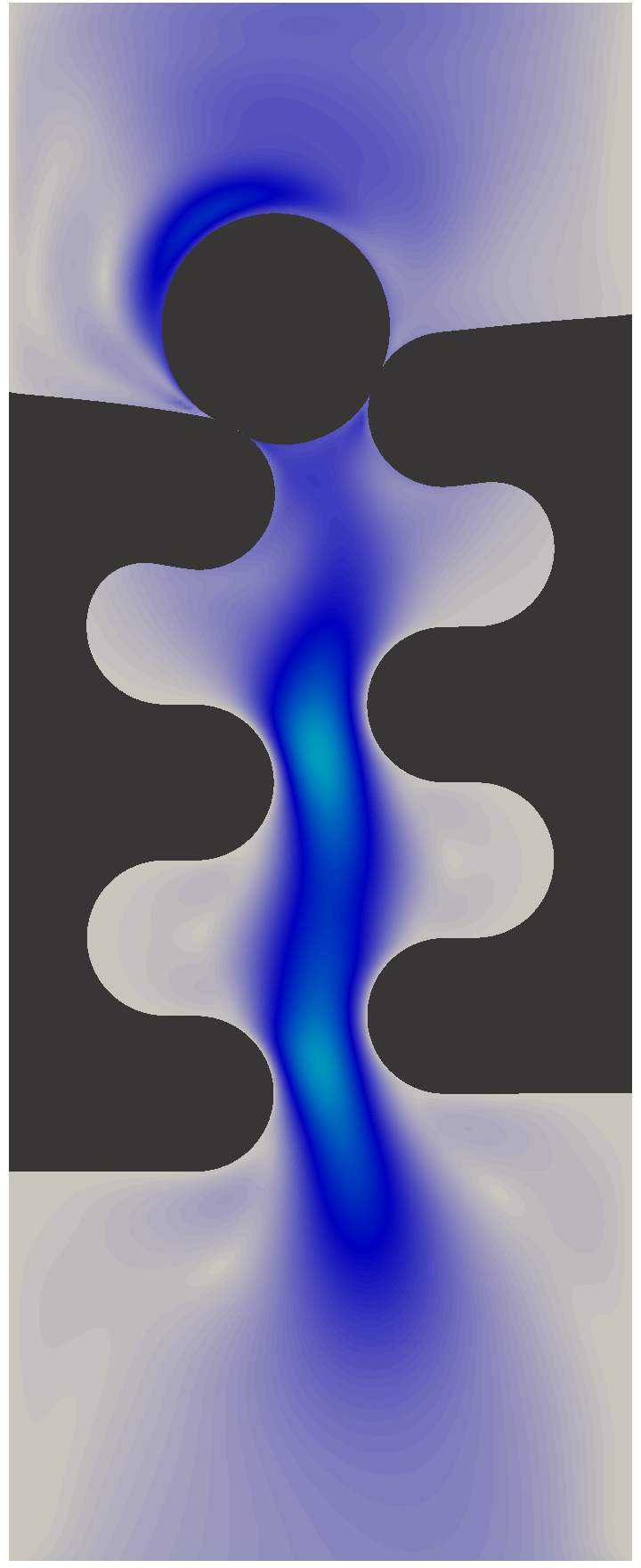}
\includegraphics[width=0.12\textwidth]{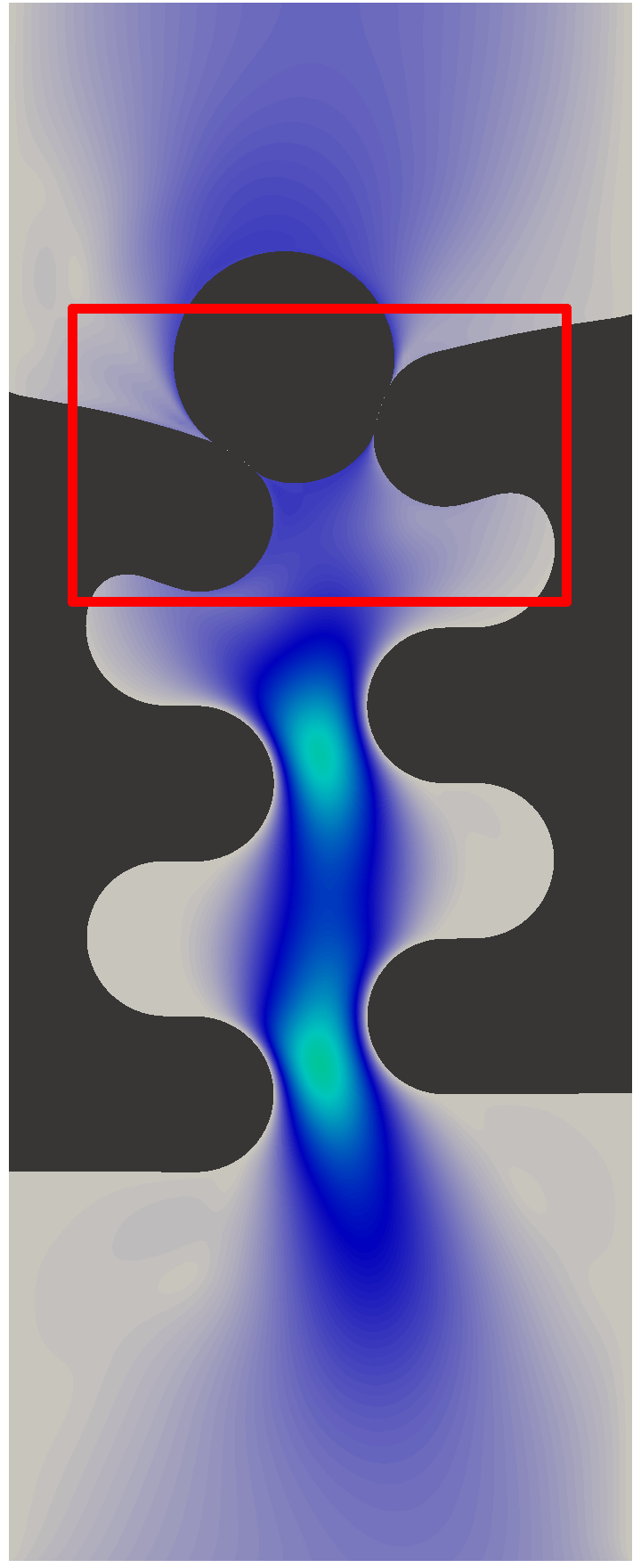}
\includegraphics[width=0.12\textwidth]{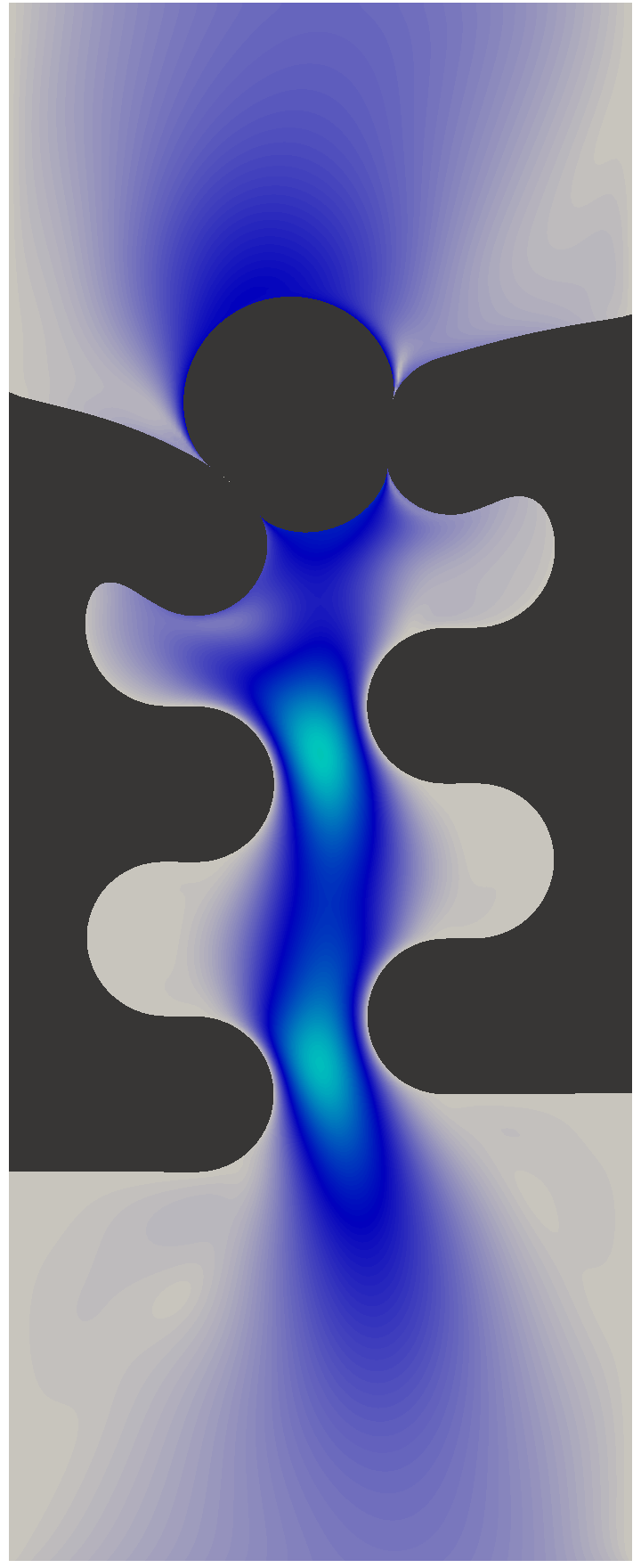}
\includegraphics[width=0.12\textwidth]{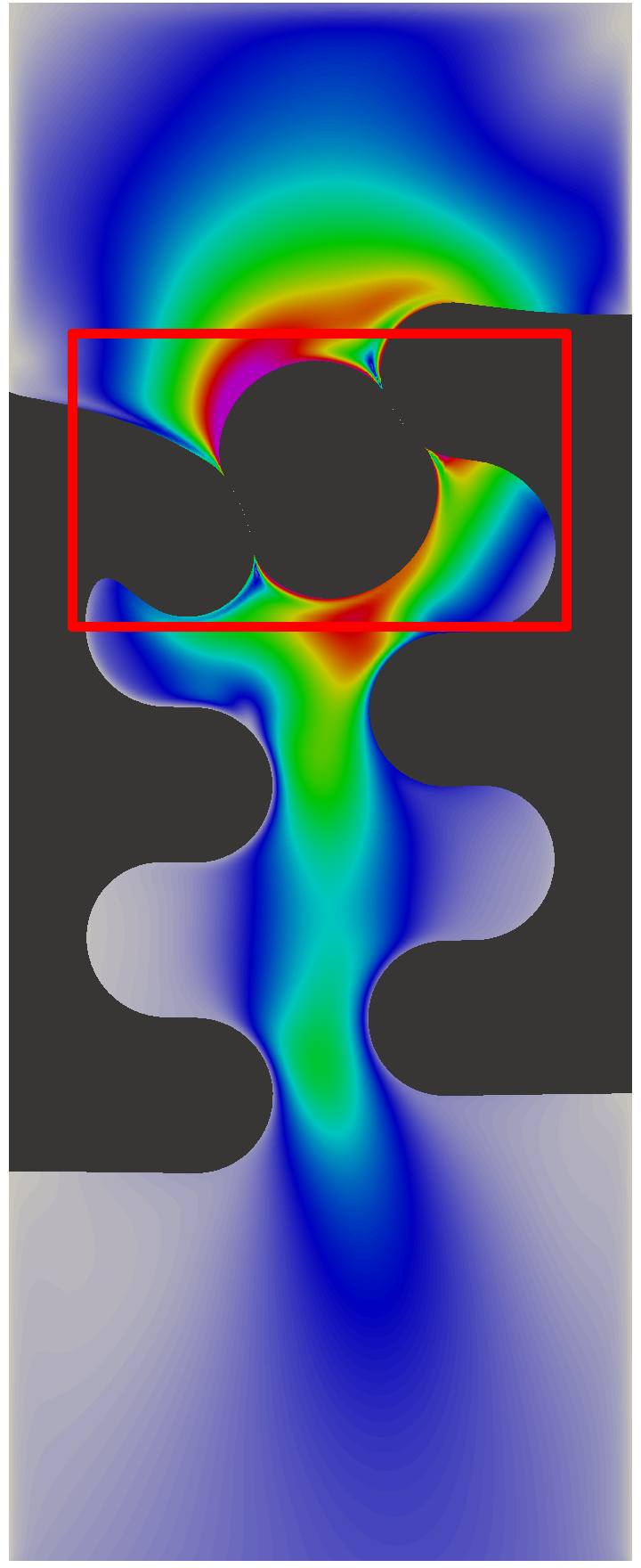}
\includegraphics[width=0.12\textwidth]{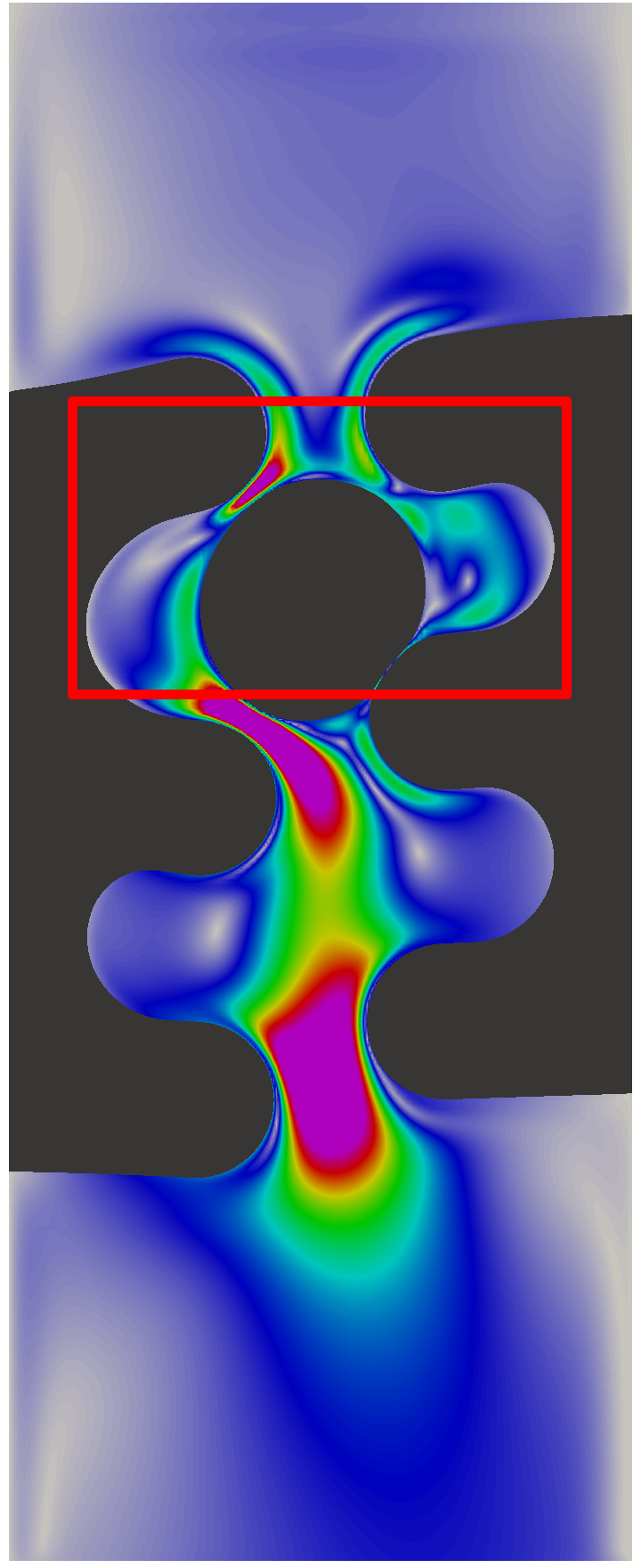}
\end{minipage}
\begin{minipage}{1\textwidth}
\includegraphics[width=0.12\textwidth]{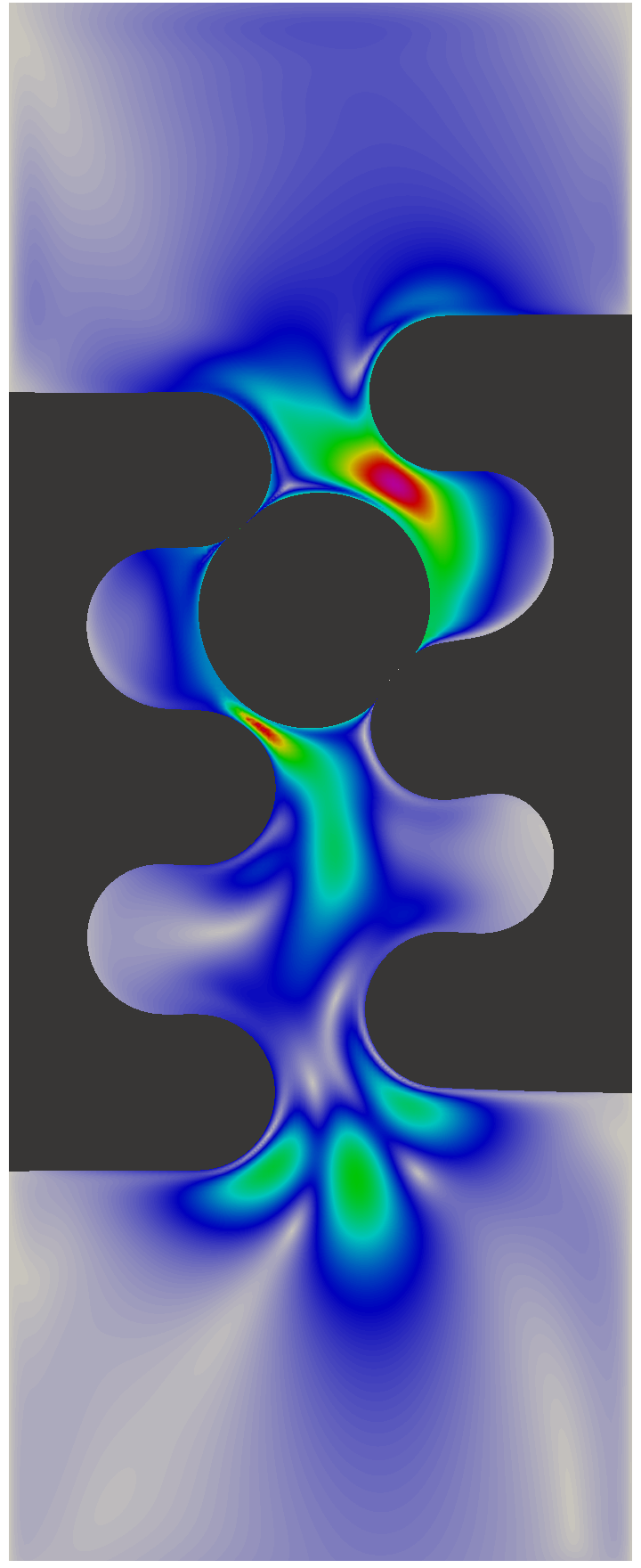}
\includegraphics[width=0.12\textwidth]{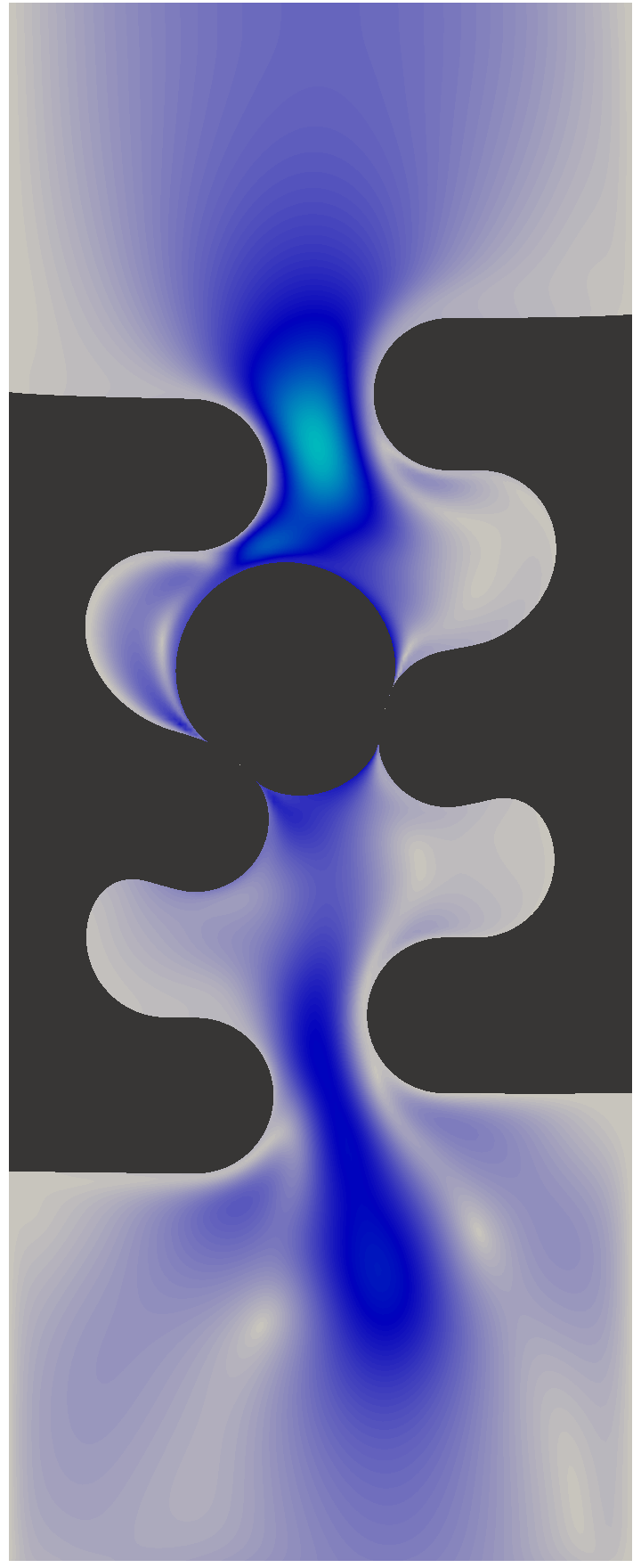}
\includegraphics[width=0.12\textwidth]{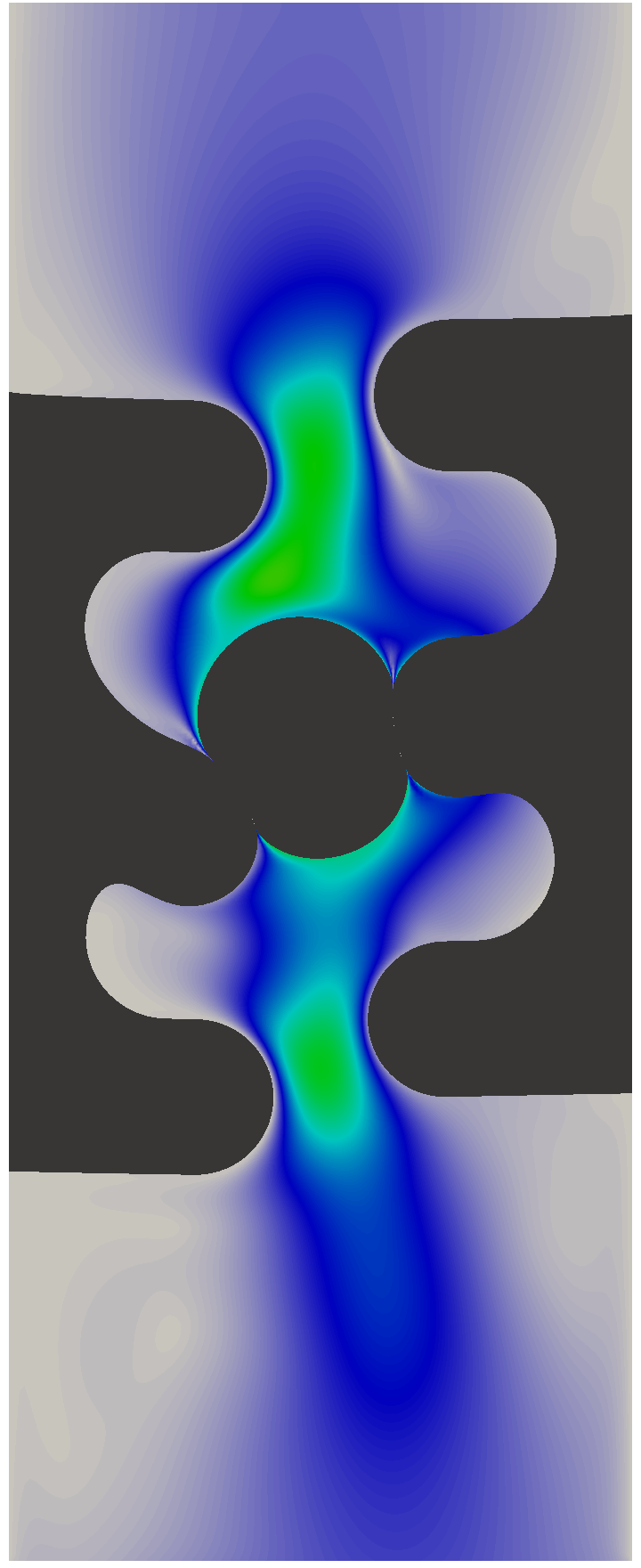}
\includegraphics[width=0.12\textwidth]{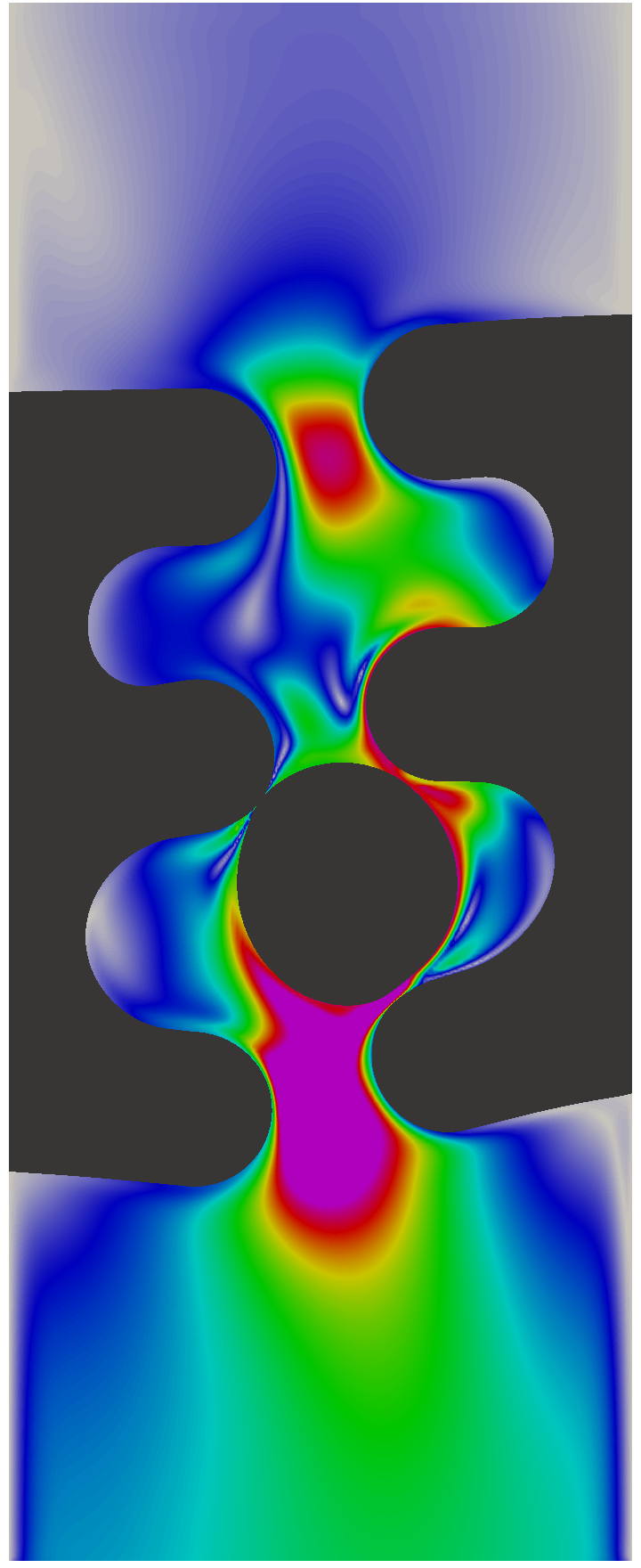}
\includegraphics[width=0.12\textwidth]{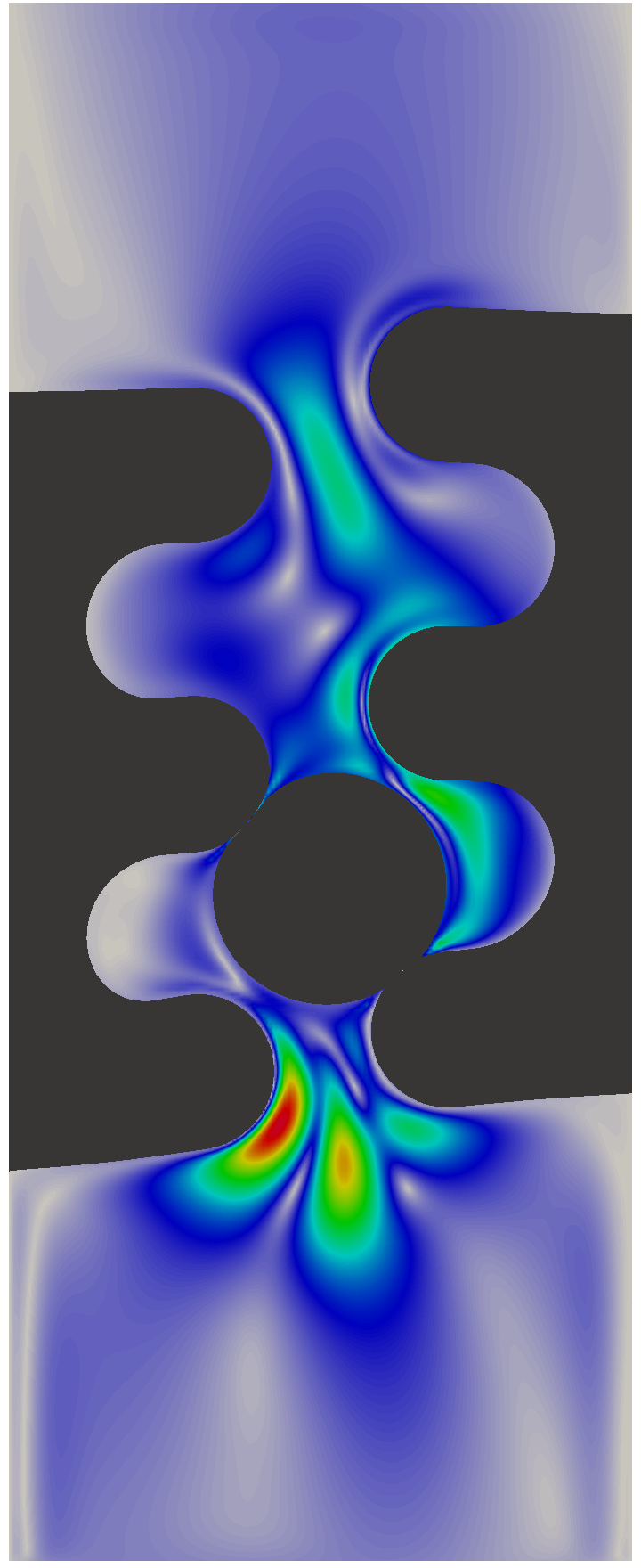}
\includegraphics[width=0.12\textwidth]{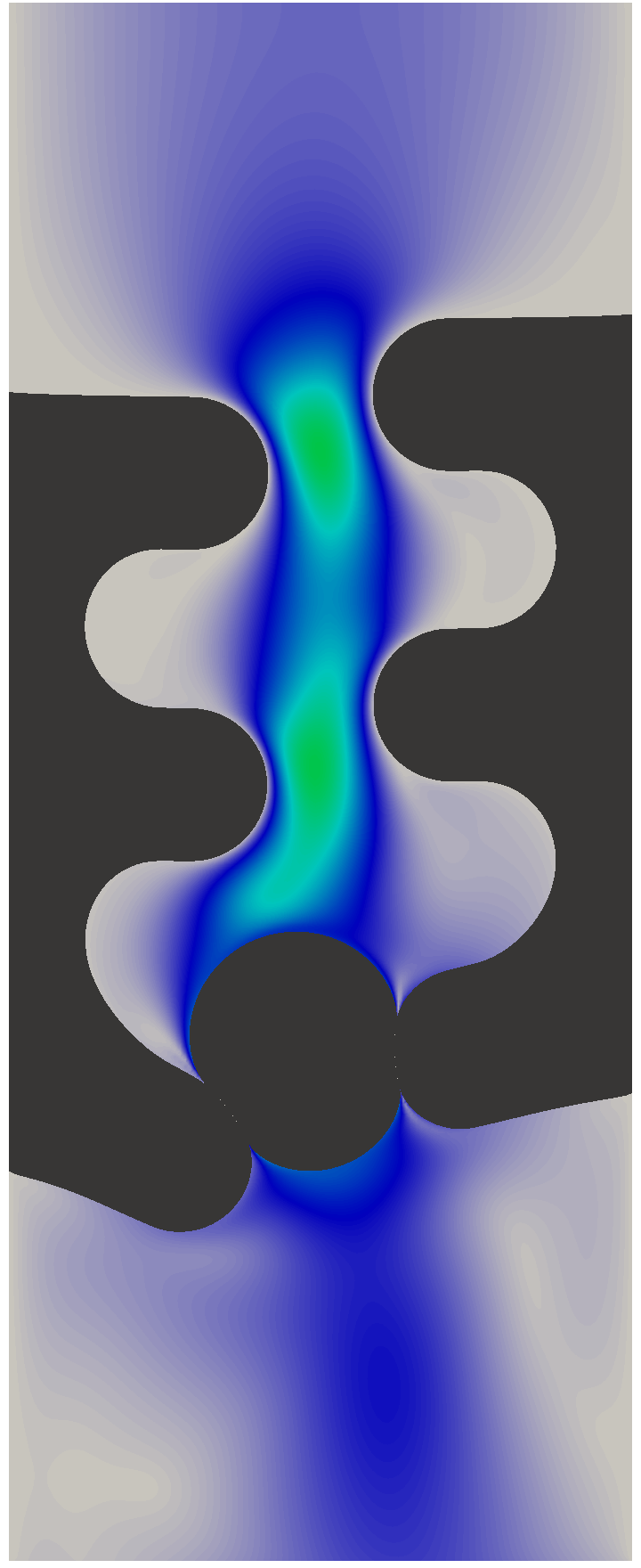}
\includegraphics[width=0.12\textwidth]{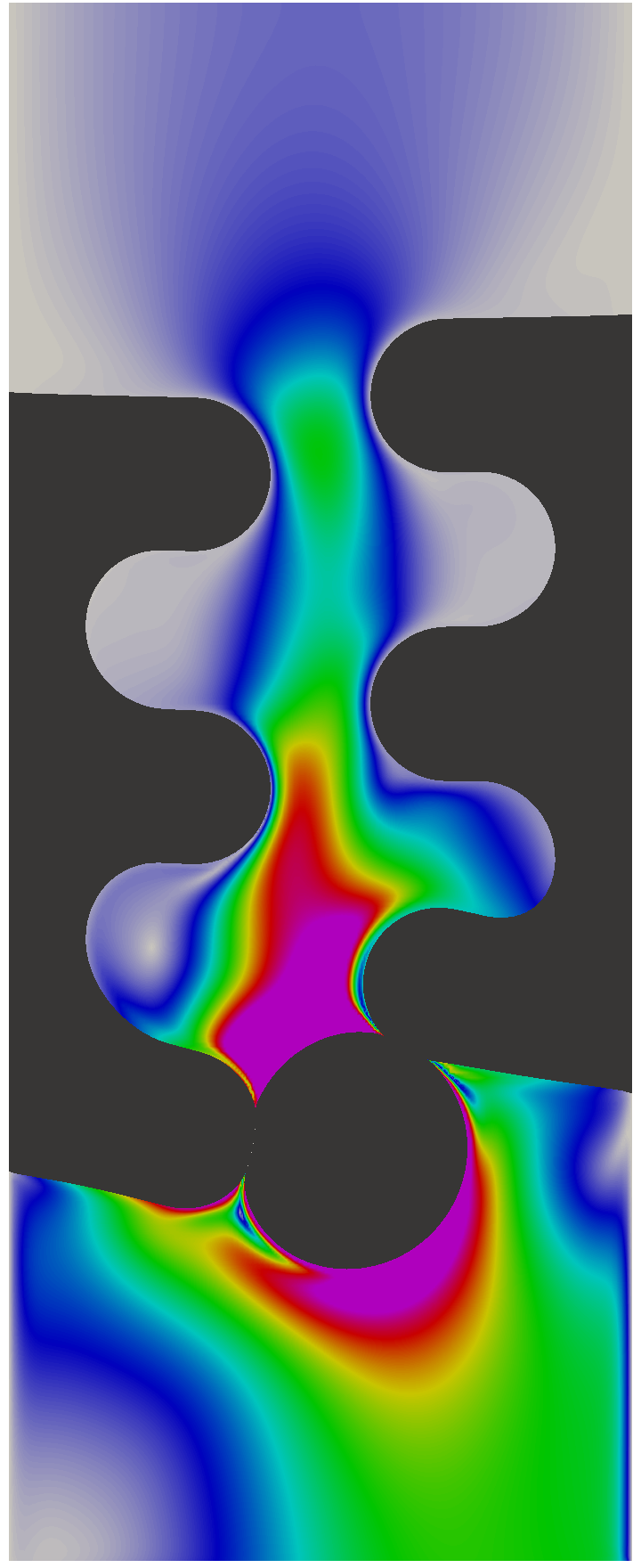}
\includegraphics[width=0.12\textwidth]{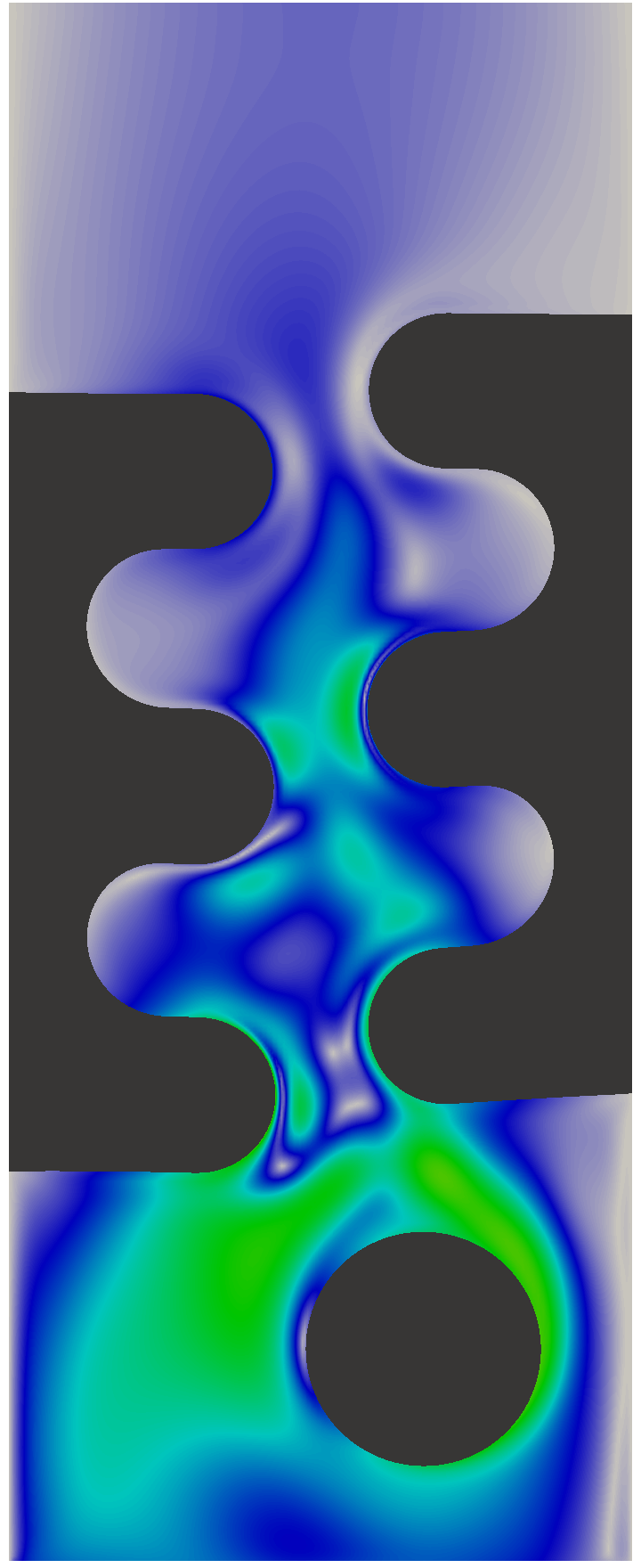}
\end{minipage}
\begin{minipage}{1\textwidth}
\input{fig/ex3/vl.tex}
\end{minipage}
\caption{Visualization of the computed fluid velocity and pressure and the computed deformation of solid domains. The color code represents the fluid velocity magnitude. Different points in time are represented from top-left to bottom-right
with $t=0.0005,t=0.001,t=0.003,t=0.004,t=0.005,t=0.006,t=0.0065,t=0.007$ in the first row and 
$t=0.0075,t=0.009,t=0.01,t=0.0105,t=0.011,t=0.0145,t=0.015,t=0.016$ in the second row. The red frame indicates the area of the detailed views in Figure \ref{fig:ex3_detail}.}
\label{fig:ex3_process}
\end{figure}

\paragraph{Numerical results and discussion}
In Figure \ref{fig:ex3_process}, the computed fluid velocity and the computed deformation of solid domains are presented for different points in time. Following the different snapshots, the motion of the solid domain $\domainstwo$ can be observed. In the initial phase ($0<t<0.00324$), a vertical motion of $\domainstwo$ is induced by the fluid flow. At $t=0.00324$ contact between $\domainstwo$ and the right part of $\domainsone$ occurs. Starting from $t=0.00386$, additional contact between $\domainstwo$ and the left part of $\domainsone$ establishes. Therefore, the topology of the fluid domain changes from one connected domain, to two separated fluid domains. 
In the subsequent phase ($0.00386<t<0.006$), the pressure in the upper fluid domain increases, which leads to a squeezing process of $\domainstwo$ and a deformation of $\domainsone$ and thus a storage of elastic energy. 
For $t>0.0065$, an acceleration in vertical direction of $\domainstwo$ can be observed by the transfer of the elastic energy via contact forces. Finally at $t=0.00668$, contact between both solid bodies is released and a single connected fluid domain reoccurs. Reestablishing contact at $t=0.00713$ of $\domainstwo$ and the right part of $\domainsone$, this principal process repeats for two additional cycles. Nevertheless, due to the varying geometric setup around the three smallest constrictions, the physical process is not repeated exactly and thus the robustness of the algorithm is tested for this challenging configuration.
Finally, at $t=0.015155$, both solid domains separate for the last time and the motion of $\domainstwo$. In the remaining period, the fluid traction is exclusively acting on the interface $\partial \domainstwo$.

\begin{figure}[htbp]
\begin{minipage}{1\textwidth}
\includegraphics[width=0.5\textwidth]{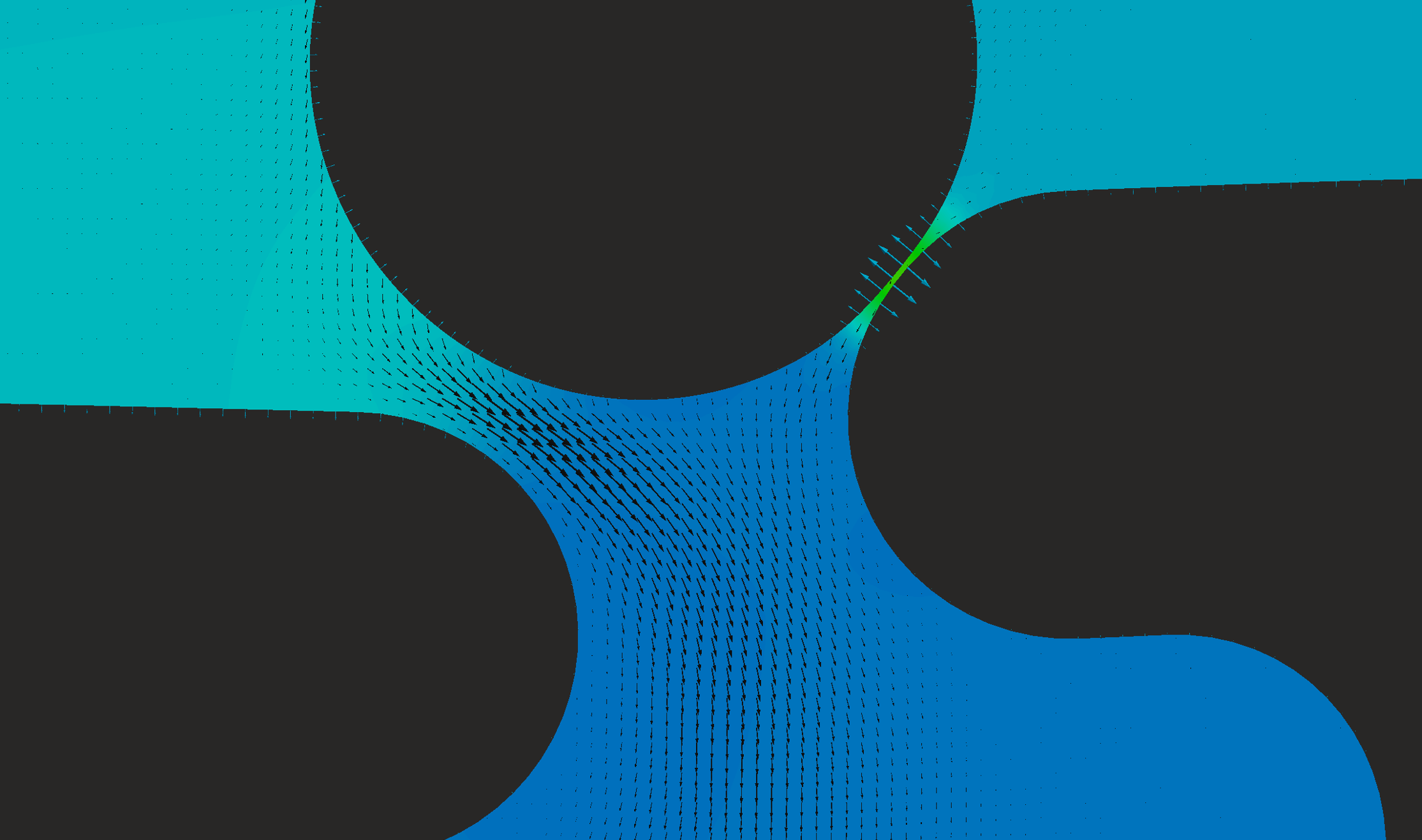}
\includegraphics[width=0.5\textwidth]{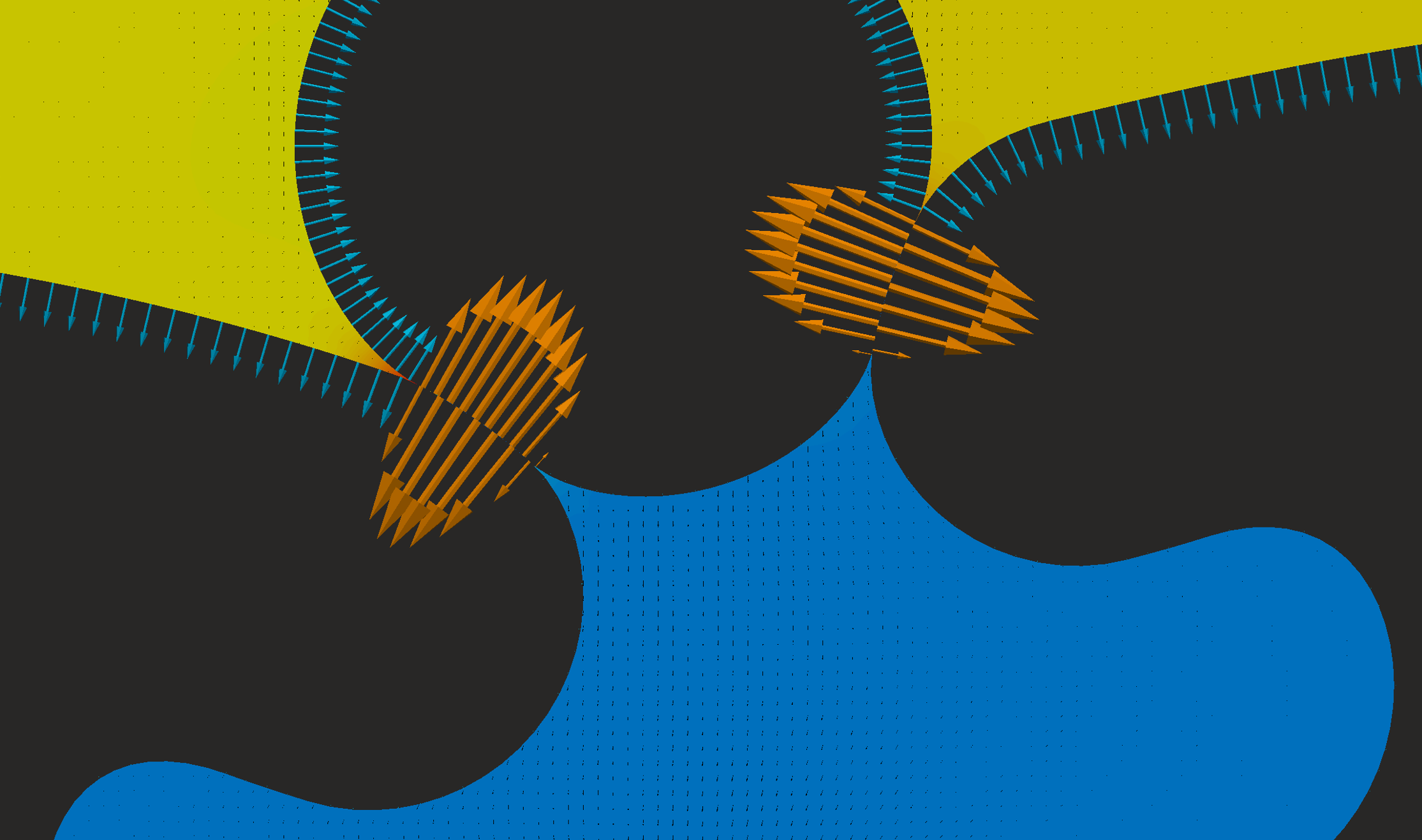}
\end{minipage}
\begin{minipage}{1\textwidth}
\includegraphics[width=0.5\textwidth]{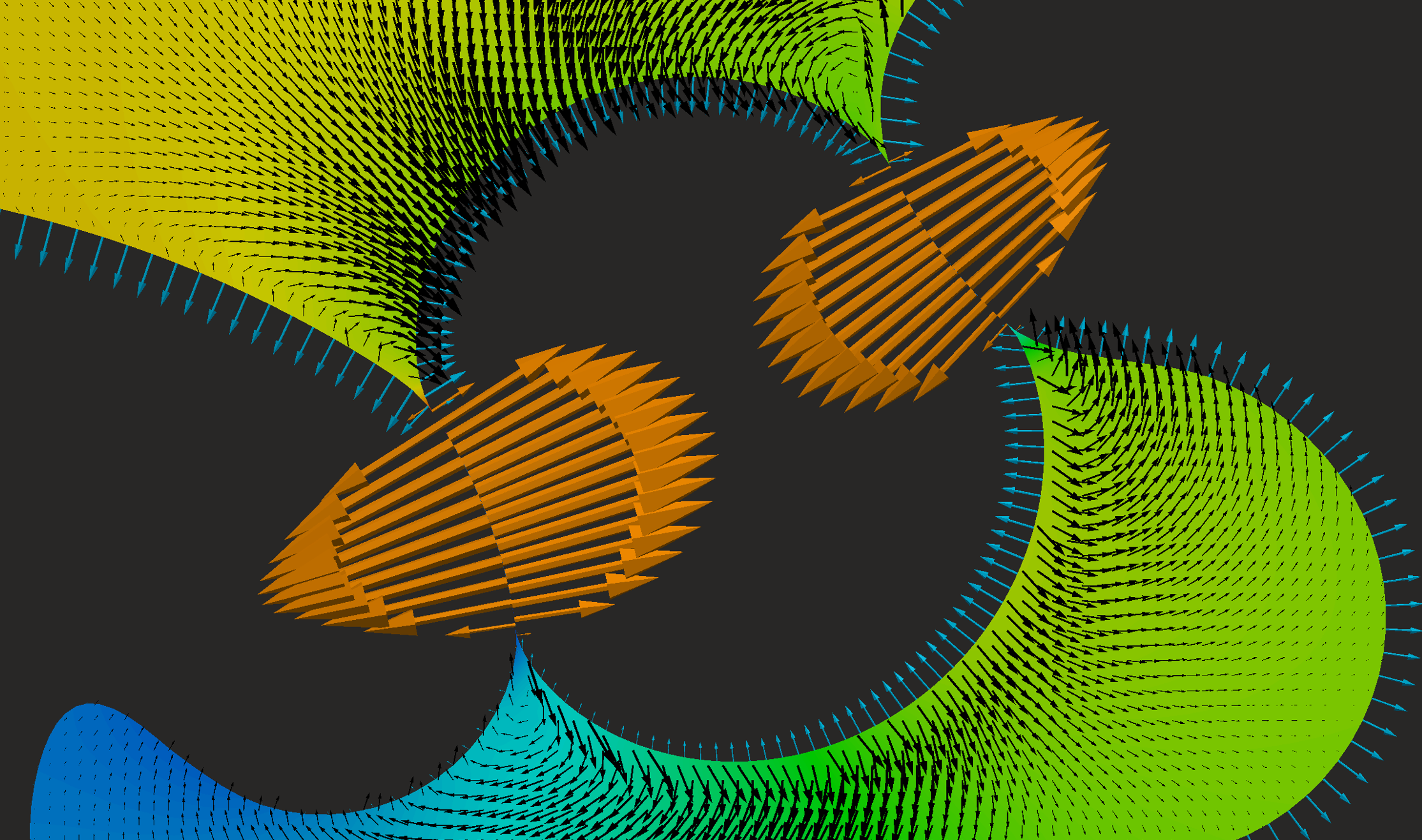}
\includegraphics[width=0.5\textwidth]{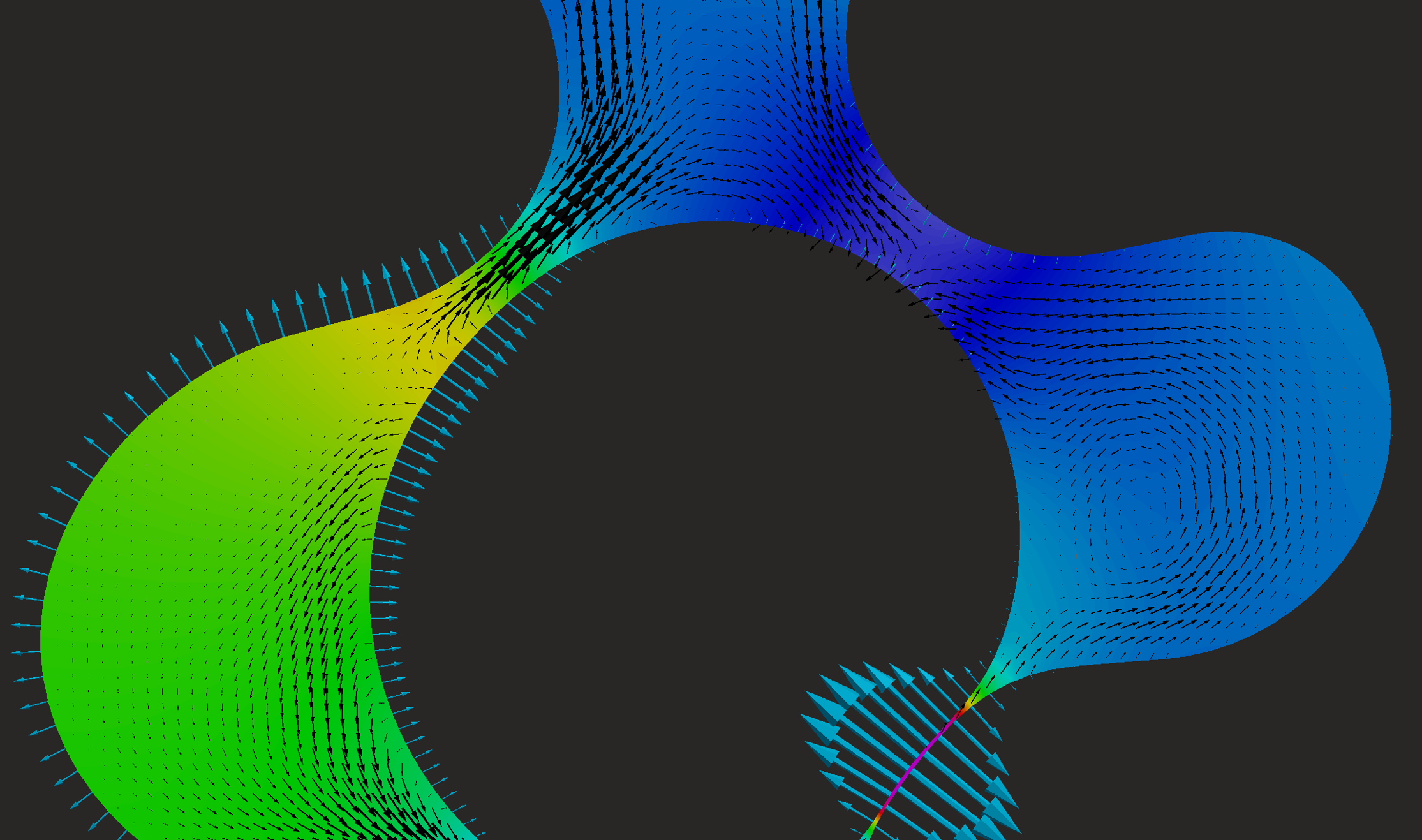}
\end{minipage}
\begin{minipage}{1\textwidth}
\input{fig/ex3/pl.tex}
\end{minipage}
\caption{Visualization of the computed fluid velocity and pressure, the computed deformation of solid domains, and the interface traction. 
The color code represents the fluid pressure and the black arrows in domain $\domainf$ indicate the fluid velocity. 
The blue arrows on $\Gamma$ represent the FSI traction (case $I$) and the orange arrows  on $\Gamma$ visualize the contact traction (case $II-IV$).
The visualization of the traction is reconstructed from the nodal interface contributions of \eqref{eq:num_caseI}-\eqref{eq:num_caseIII_alt}) to the overall weak form on the solid mesh.
Four points in time are represented from top-left to bottom-right $t=0.003,t=0.005,t=0.0065$ and $t=0.007$.
The position of each detailed view in the overall problem is marked in Figure \ref{fig:ex3_process} by a red frame.}
\label{fig:ex3_detail}
\end{figure}

To give a more detailed view of the computed process, the fluid solution as well as the interface traction for four exemplary points in time are shown in Figure \ref{fig:ex3_detail}. 
First, the point in time just before contact occurs $t=0.003$, is discussed. Due to the small cross-section of the connection between the upper and lower part of the fluid domain, the pressure in the upper part is already increased. Therefore, an essential fluid flow can be observed between $\domainstwo$ and the left part of $\domainsone$. The distance in the smallest constriction for the right part leads to an increased fluid pressure compared to the ambient pressure and thus an FSI traction separating the two bodies occurs. At $t=0.005$, contact between both solid domains is established in two positions. Due to the inflow on $\Gamma^{in}$, the pressure in the upper part of the fluid domain is increased, which leads to an increased FSI traction on the affected part of the interface. Although the maximal contact traction is significantly higher than the FSI traction, there is a continuous transition along the interface. The $y$-components of the resulting FSI force and contact force are almost in balance, 
and 
as a result only a very 
slow motion of the system (see fluid velocity) is observed, continuously adapting  to the increasing pressure difference.
At $t=0.0065$ this state changed fundamentally. Due to the deformation based change of the contact interface orientation, the resulting contact force accelerates the solid body in $\domainstwo$, and with it the surrounding fluid, in negative $y$-direction. The fluid pressure in the upper part of the flow domain drops, whereas the pressure in the lower part increases resulting in an almost constant FSI traction acting on $\partial \domainstwo$. Finally at $t=0.007$, contact is released and the structural body in $\domainstwo$ approaches the second barrier. This process leads again to an increased local fluid pressure and thus a growth of the related FSI traction. Due to the structural motion the pressure in the left chamber is raised, which results in a fluid flow out of the fluid chamber.
This description of the computed physical process highlights the capabilities of the presented formulation to predict the physical processes in FSCI without requiring a specific treatment whenever topological changes occur.

\begin{figure}[htbp]
\centering
\hspace{-0.5cm}
\begin{minipage}[hbt]{0.49\textwidth}
\centering
\def\figscaling{0.6}
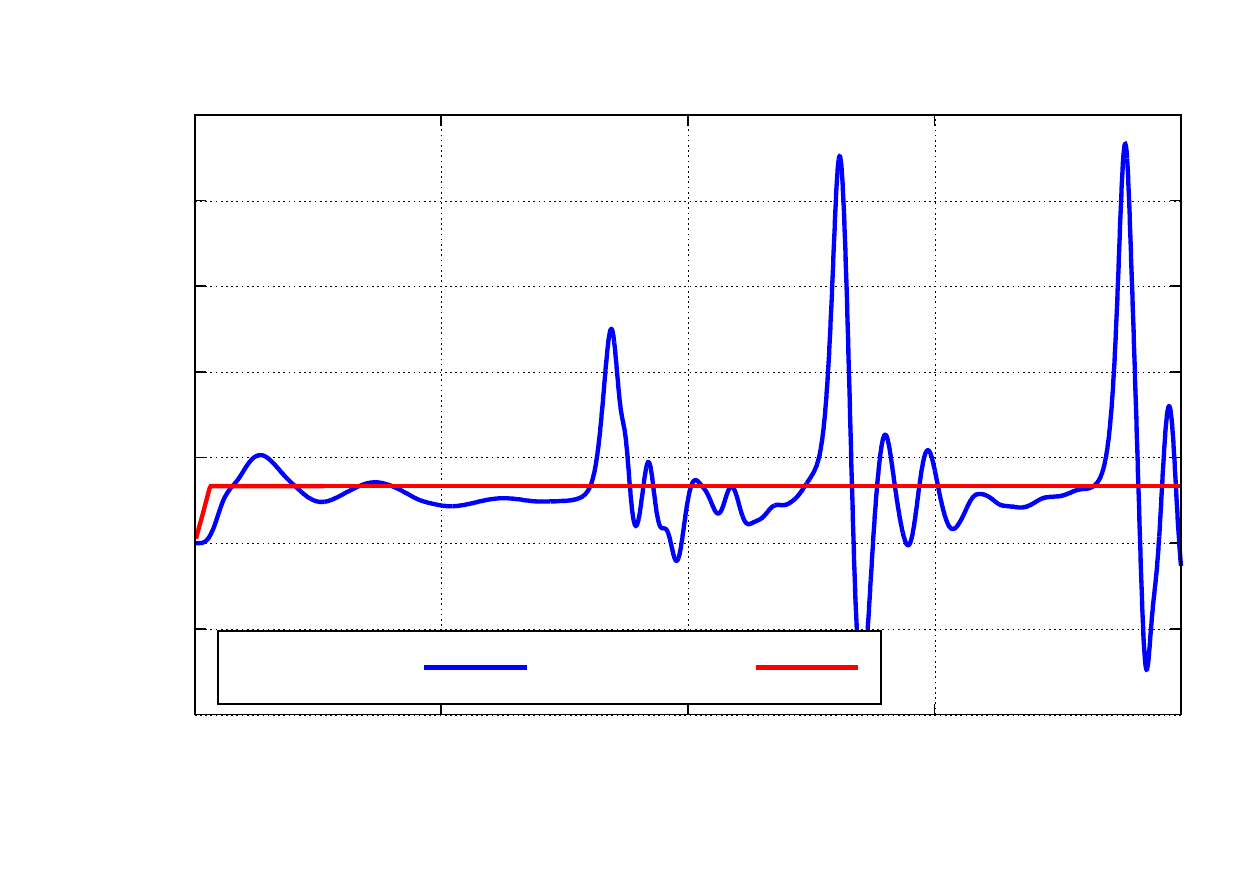
\end{minipage}
 \begin{minipage}[hbt]{0.49\textwidth}
\centering
\def\figscaling{0.6}
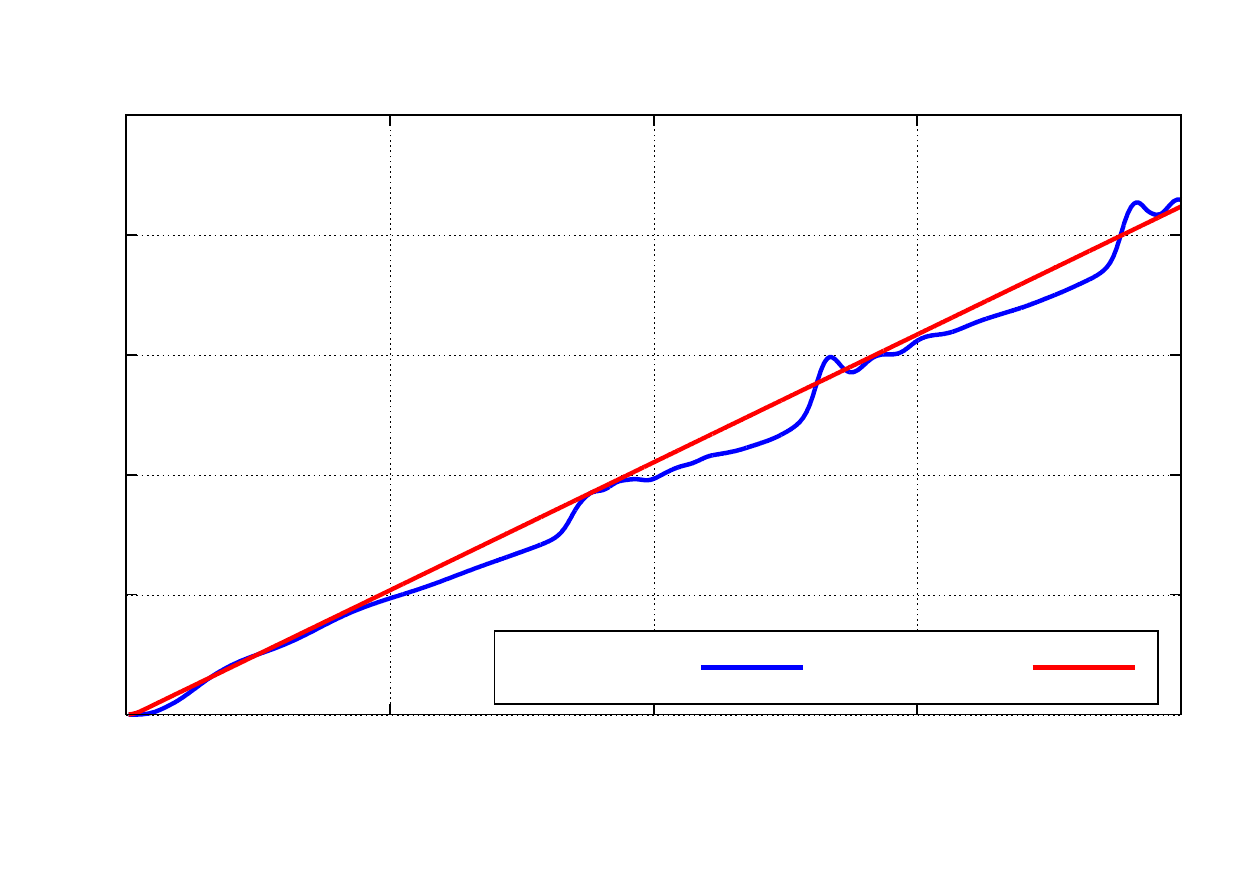
\end{minipage}
\caption{Computed flow rates at the inflow boundary $\Gamma^{in}$ and the outflow boundary $\Gamma^{out}$.
The normal vector therein is oriented in negative $y$-direction, which is the main flow direction of the overall configuration (left).
Transported volume through the inflow boundary $\Gamma^{in}$ and the outflow boundary $\Gamma^{out}$ computed in a post-processing step where an integration in time of the flow rates is performed (right).}
\label{fig:ex3_flowrate}
\end{figure}

In Figure \ref{fig:ex3_flowrate} (left), the computed flow rates at the inflow boundary and outflow boundary are presented. While the prescribed flow rate at the inflow is constant in time after the initial ramp up phase, the flow rate at the outflow boundary is massively influenced by the overall system. Three phases can be observed where a lower outflow rate (than the inflow rate) is followed by a peak of the flow rate. These can be identified as the phases where the solid domains are compressed due to increasing pressure as $\domainstwo$ blocks the flow. These phases are always followed by the highly dynamic process of squeezing through. To analyze the overall balance of mass, the transported volume through the inflow- and outflow- boundary is given in Figure \ref{fig:ex3_flowrate} (right). The difference between the transported volume of outflow and inflow results from the compression or expansion of the solid domains. As no systematical increase of this difference in time can be recognized, no relevant 
loss in mass occurs. 
This behavior is expected as discussed in the first presented numerical example in Section \ref{sec:ex1}.
\section{Conclusion}
\label{sec:conclusion}
In this contribution, we presented a consistent formulation to solve general fluid-structure-contact interaction (FSCI) problems numerically. 
Topological changes of the fluid domain are enabled by the CutFEM with non-interface fitted discretization. A weak incorporation of the governing interface conditions by approaches based on Nitsche's method allows the formulation of a continuous discrete problem even for changing interface conditions. 
To specify the fluid stress in the region of closed contact, we propose and apply an extension approach.
The continuous transition between the ``no-slip'' and frictionless contact condition in tangential interface orientation is enabled by a general Navier interface condition with a specific law for the slip length.

In a first numerical example, the fundamental process in FSCI problems, the contacting and lifting of a convex elastic structure in fluid is analyzed.
Therein, the suitability of applying the general Navier interface condition in comparison to a ``no-slip'' interface condition is evaluated.
The positive effect of a skew-symmetric fluid adjoint consistency interface term to the fluid mass conservation is observed and discussed.
In two more general examples, the treatment of challenging aspects by the formulation is demonstrated.
This includes the representation of large discontinuities of the fluid stress between opposite sides of the structure.
Further, large deformation and essential topological changes of the fluid domain as well as large contacting areas are considered.

Still, some aspects for solving general FSCI problems are left to future work.
This includes strategies for improving the spatial resolution close to the interface, for examining physically more sophisticated fluid stress representations in the contact zone compared to the simple extension based approach used so far, and for extending the formulation to frictional contact.

\bibliography{bib.bib}
\bibliographystyle{wileyj.bst}

\end{document}